\documentclass[final, 3p, times, 12pt, number]{elsarticle}

\usepackage{amsmath, amssymb, amsthm}
\usepackage{bm}
\usepackage{mathtools}
\usepackage{color, soul}
\usepackage{graphicx}
\usepackage{natbib}
\usepackage{makecell}

\usepackage{subcaption}
\usepackage{float}
\usepackage{booktabs}
\usepackage{multirow}
\usepackage{array}
\usepackage{tabularx}

\usepackage[dvipsnames, table]{xcolor}

\usepackage{siunitx}

\usepackage{algorithm}
\usepackage{algpseudocode}

\usepackage{microtype}
\usepackage{enumitem}

\usepackage[colorlinks=true,
            linkcolor=NavyBlue,
            citecolor=NavyBlue,
            urlcolor=NavyBlue]{hyperref}
\usepackage{cleveref}   

\journal{Computer Methods in Applied Mechanics and Engineering}

\begin{document}

\begin{frontmatter}

  \title{Solution Strategies for Dissolution-Driven Phase-Field Fracture in Geomaterials}

  \author[inst1]{Maithah M. Alaleeli}\corref{cor1}
  \ead{maithah.alaleeli@nyu.edu}
 
  \author[inst1]{Mohamed Elkhodbia}
  \author[inst1]{Mostafa E. Mobasher}\corref{cor1}
  \ead{mostafa.mobasher@nyu.edu}
 
  \cortext[cor1]{Corresponding authors}

  \affiliation[inst1]{
    organization={Civil and Urban Engineering Department, New York University},
    addressline={P.O. Box 129188},
    city={Abu Dhabi},
    country={United Arab Emirates}}

 \begin{abstract}

Dissolution-driven fracture governs the long-term integrity of geological CO$_2$ storage, geothermal reservoirs, and nuclear waste repositories. Modeling it requires solving strongly coupled mechanical, chemical, and phase-field equations. The solver choice controls their accuracy, cost, and convergence. This study investigates the role of solution strategies in dissolution-driven fracture in geomaterials. The coupled model is implemented in Abaqus user-element (UEL) subroutines, with all tangent blocks derived analytically. Five solvers are compared: (1) block-diagonal Newton monolithic, (2) BFGS quasi-Newton monolithic, (3) single-pass fully staggered, (4) single-pass partially staggered, and (5) hybrid multi-pass fully staggered. Performance is quantified by \textit{accuracy} (errors in peak load, peak and failure displacements, and local CO$_2$ concentration), \textit{cost} (CPU and wall-clock times), and \textit{robustness}, measured by new factors that assess convergence over the whole simulation and capture iteration spikes. The solvers are tested on mechanical and chemo-mechanical tension, shear, and multiple-crack benchmarks, including CO$_2$ diffusivities raised up to a thousandfold. The results reveal a clear trade-off between accuracy, cost, and robustness. The quasi-Newton solver matches the Newton reference at a fraction of its cost but can fail under intense transport. The single-pass fully staggered solver is inexpensive and always completes, but its post-peak response depends strongly on the time step. The hybrid multi-pass solver also always completes, retains near-reference accuracy, and avoids iteration spikes at moderate cost. The partially staggered solver is the least robust. These findings guide practical selection: the single-pass staggered solver suits rapid screening, quasi-Newton offers accuracy when convergent, and hybrid multi-pass provides the best balance.
 \end{abstract}


\end{frontmatter}

\section{Introduction}
\label{sec:intro}
The mechanical integrity of subsurface geomaterials is critical to several energy technologies, including geological CO$_2$ storage, geothermal energy extraction, and long-term waste containment~\citep{viswanathan2022fluid}. These systems involve interacting mechanical, hydraulic, thermal, and chemical processes. In CO$_2$ storage, the injected fluid acidifies the pore fluid and promotes dissolution of reactive minerals. Carbonates such as calcite dissolve orders of magnitude faster than most silicates~\citep{Holzheid2016}, so dissolution-driven weakening on injection time scales is most pronounced in carbonate rocks and carbonate-cemented sandstones. Dissolution increases porosity and degrades the mechanical properties of the solid, while deformation and cracking alter transport pathways and hence where further reaction occurs. These processes can reinforce one another, producing localized alteration and time-dependent deformation~\citep{IlgenNewell2019}. Predicting fracture under these conditions therefore requires a coupled framework in which dissolution and cracking evolve together. The resulting nonlinear partial differential equations are difficult to solve~\citep{Wick2020}, and the numerical strategy used to solve them affects both simulation accuracy and computational cost.

Coupled multiphysics problems are solved with either monolithic or staggered schemes~\citep{Wick2020,JinSvolos2026}. Monolithic schemes solve for all fields simultaneously. Newton-type variants differ in which inter-field coupling blocks the tangent retains, in how accurately the tangent is computed (consistent, inexact, or secant), and in how often it is updated (every iteration, or periodically as in quasi-Newton methods)~\citep{Nocedal2006}. Staggered schemes split the fields into fully or partially partitioned sub-problems that are solved in sequence, either once per increment (single-pass) or repeatedly until the fields agree (multi-pass)~\citep{Kim2011,FarrellMaurini2017,JinSvolos2026}. Strategies can also be combined or switched during an analysis~\citep{Dujc2022}. Each choice trades computational cost against coupling accuracy and robustness.

Phase-field methods are now widely used to model brittle fracture. Building on the variational theory of fracture~\citep{FrancfortMarigo1998}, \citet{Bourdin2000} adapted the Ambrosio--Tortorelli regularization~\citep{Ambrosio1990} to represent a sharp crack by a continuous damage field over a width set by the length scale $l_0$. The approach has since become a standard computational framework~\citep{Miehe2010a}, largely because crack propagation, branching, and coalescence emerge from energy minimization without explicit crack tracking. However, its numerical solution is demanding. Strong stress localization, coupled with phase-field growth, produces a highly nonlinear discrete system whose response changes abruptly as the stiffness is lost. After discretization of the weak form, the coupled displacement--phase-field system is non-convex when solved monolithically, although each sub-system is convex when the other field is held fixed~\citep{MieheHofacker2010,FarrellMaurini2017}. This structure shapes the choice of solver. 

Multi-pass staggered (alternate-minimization) schemes exploit the convexity of the sub-systems but can require many coupling iterations near fracture~\citep{FarrellMaurini2017}. Monolithic Newton schemes avoid these iterations but are challenged by the non-convexity of the coupled system and often fail to converge during unstable crack growth~\citep{Gerasimov2016,Wick2017,Bharali2022}. Proposed remedies include line search~\citep{Gerasimov2016,Heinzmann2025}, modified-Newton methods~\citep{Wick2017}, arc-length control with under-relaxation~\citep{Bharali2022}, and quasi-Newton BFGS and L-BFGS schemes~\citep{Wu2020,Kristensen2020,Jin2024,Ramos2025}. Viscous regularization~\citep{Miehe2010a} has also been used, but it is purely numerically motivated. A finite viscosity adds pseudo-viscous dissipation that smooths the post-peak response~\citep{Dammass2023}, and the rate-independent evolution is recovered only in the limit of vanishing viscosity~\citep{Almi2020}. Robust implementations have since converged without it~\citep{Kristensen2020,Navidtehrani2021}.

As phase-field models are extended to further physical processes, the coupled problems become larger and more strongly nonlinear, and the choice of solution strategy therefore becomes more consequential. For thermo-mechanical fracture, \citet{Ruan2023} solved displacement and temperature together, then ran a separate phase-field solve, repeating the sequence until convergence. In contrast, \citet{Mandal2021} solved all three fields simultaneously using a monolithic BFGS approach. \citet{JinSvolos2026} compared L-BFGS monolithic, fully staggered, and partially staggered thermo-mechanical schemes, and showed that grouping fields into one sub-problem does not necessarily make that sub-problem convex. For hydraulic fracture, \citet{MieheMauthe2016} solved displacement and pressure together and the phase field separately, either once per increment with a lagged crack-driving force or repeatedly until convergence. \citet{Chukwudozie2019} instead solved pressure, displacement, and the phase field separately in a multi-pass scheme with a fixed-stress split. \citet{navidtehrani2025phase} presented a framework that accommodates monolithic, staggered, and mixed schemes, but did not benchmark them against each other. Most of these studies adopt one strategy without comparing it to alternatives. Systematic comparisons, such as that of \citet{JinSvolos2026}, remain rare and do not yet cover chemically driven fracture.

Chemo-mechanical phase-field models add coupling through species transport, reaction, and changes in material properties. For hydrogen-assisted fracture, \citet{MartinezPaneda2018} used an Abaqus user element with a single-pass scheme in which the phase field and hydrostatic stress were lagged by one increment. For battery-electrode fracture, \citet{MieheBattery2016} formulated a multi-pass partially staggered scheme but applied a single pass per increment in their examples. For stress-corrosion cracking, \citet{Cui2021} solved all fields monolithically using Newton--Raphson with a block-diagonal tangent.

Mineral dissolution adds a further feedback because the reaction progressively changes the porosity and stiffness of the solid. For CO$_2$-driven calcite dissolution, \citet{Schuler2020} solved each field once per increment using cross-field quantities from the previous increment, without checking convergence between the fields. \citet{GuoNa2024} added a hydraulic equation and used a multi-pass partially staggered scheme with a coupled displacement--pressure solve. \citet{Mollaali2025} used a multi-pass fully staggered scheme with a fixed-stress split and evaluated the reactions separately in PHREEQC~\citep{Parkhurst2013,Clavijo2022}. In contrast, \citet{VenturaBorja2026} solved displacement, phase field, and porosity monolithically with the full coupled tangent, as \citet{Borja2023} did earlier for a gradient-enhanced damage formulation. Each study adopts a single strategy and assesses it only through its own benchmarks, without reporting cost or robustness against alternatives. Because the formulations and implementations also differ, the effect of the solution strategy cannot be separated from that of the model. No study has yet compared monolithic and staggered strategies for dissolution-driven fracture within a single framework. There are also no established measures of robustness beyond whether a simulation completes. Guidance on selecting a solution strategy is therefore lacking.

This study addresses the question: \textbf{\textit{what is the most effective and robust strategy for solving a strongly coupled mechanical--chemical--phase-field fracture system for dissolution-driven fracture in geomaterials?}} We answer it within a commercial finite-element environment, Abaqus. Phase-field fracture has been implemented in Abaqus before with monolithic and staggered schemes~\citep{Liu2016}, with layered user elements in a staggered solution~\citep{MolnarGravouil2017}, and with quasi-Newton methods compared against staggered schemes~\citep{Kristensen2020}. These studies address mechanical fracture only. The present work extends the comparison to a three-field chemo-mechanical model, adds a hybrid multi-pass scheme, and introduces numerical evaluation metrics.  

We implement the mechanical, dissolution--transport, and phase--field equations, along with their coupling, in custom user-element (UEL) subroutines. Each subroutine defines the element residual vector, tangent matrix, and state variables. The UELs also control explicitly how and when coupling quantities are exchanged between the fields. We assess five solution strategies: 
\begin{itemize}[noitemsep]
  \item Case~1: block-diagonal inexact-Newton monolithic,
  \item Case~2: quasi-Newton (BFGS) monolithic,
  \item Case~3: single-pass fully staggered,
  \item Case~4: single-pass partially staggered, with displacement and concentration grouped,
  \item Case~5: hybrid multi-pass fully staggered.
\end{itemize}
We compare the strategies on mechanical and chemo-mechanical tension, shear, and multiple-crack benchmarks. Solution accuracy is measured by errors in peak load, peak and failure displacements, and the local concentration ahead of the notch. Computational cost is measured by CPU and wall-clock times. Convergence robustness is measured by the completion ratio and by iteration-burst factors, introduced here to capture localized spikes in iteration demand. A transport-intensity study tests how each strategy responds as transport becomes more severe. In this study, we scale up the diffusivity and characterize transport intensity by a local Fourier number.

The remainder of the paper is organized as follows. Section~\ref{sec:theory} presents the coupled formulation, and Section~\ref{sec:implementation} describes the discretization and the solution strategies. Section~\ref{sec:benchmarks} presents the evaluation framework, the verification, the benchmark results, and the cross-benchmark assessment. Finally, Section~\ref{sec:conclusions} summarizes the conclusions and limitations.

\section{Theoretical Framework}
\label{sec:theory}
All solution strategies solve the same coupled three-field problem for the displacement $\bm{u}(\bm{x},t)$, the phase-field damage $d(\bm{x},t)\in[0,1]$, and the dissolved CO$_2$ concentration $c(\bm{x},t)\geq0$. The formulation follows the dissolution-assisted fracture model of \citet{Schuler2020}. It combines an AT2 phase-field description of brittle fracture with stress-assisted transport of CO$_2$ and dissolution-induced degradation of a calcite-bearing solid. This section first states the governing equations and then defines the constitutive relations for fracture, reactive transport, and diffusivity.

\begin{figure}[htbp]
  \centering
  \includegraphics[width=\textwidth]{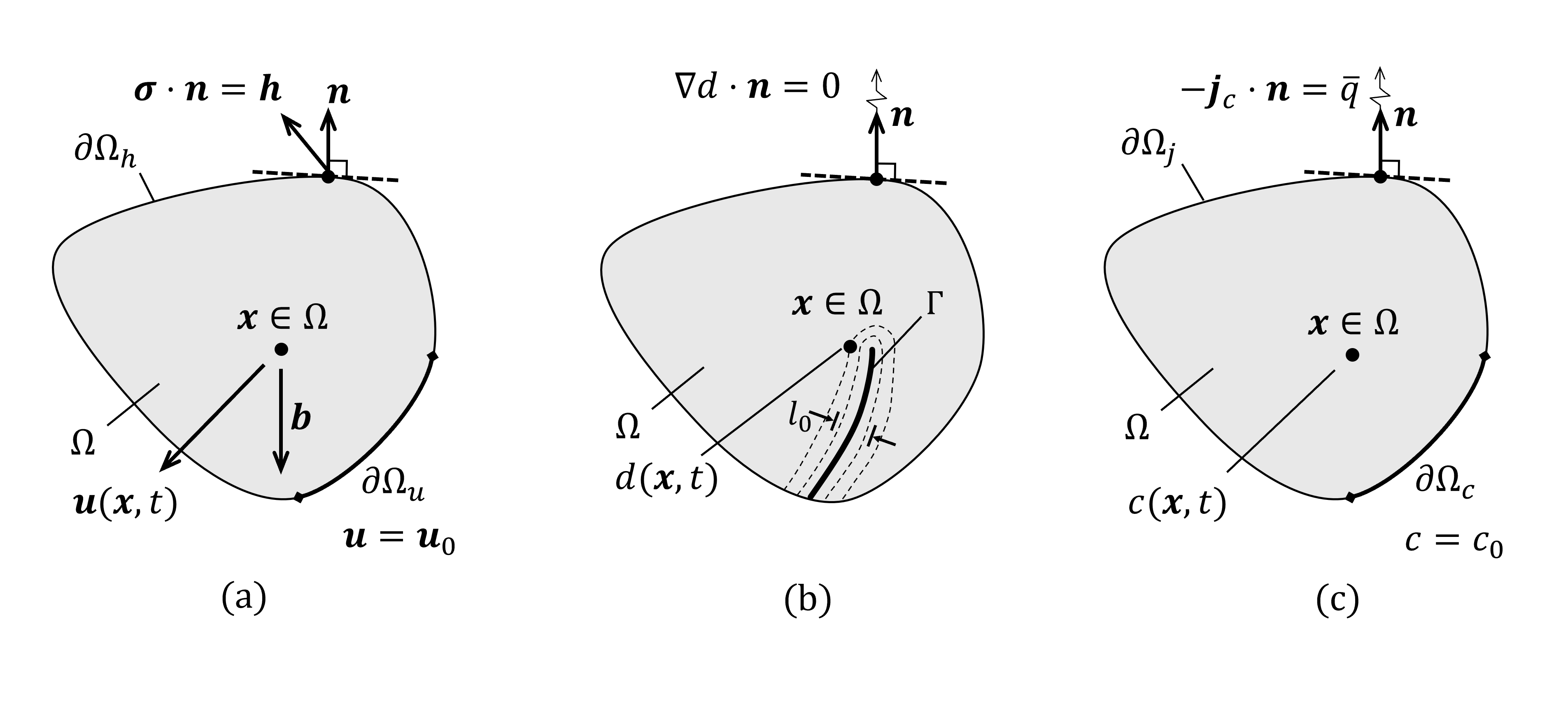}
  \caption{Schematic of the three coupled fields on a body $\Omega$ with outward unit normal $\bm{n}$. (a)~Mechanics: displacement $\bm{u}(\bm{x},t)$ under the body force $\bm{b}$, with $\bm{u}=\bm{u}_0$ prescribed on $\partial\Omega_u$ and the traction $\bm{\sigma}\cdot\bm{n}=\bm{h}$ on $\partial\Omega_h$. (b)~Phase field: the damage $d(\bm{x},t)$ regularizes the sharp crack $\Gamma$ over a band whose width scales with $l_0$, with $\nabla d\cdot\bm{n}=0$ on $\partial\Omega$. (c)~Transport: CO$_2$ concentration $c(\bm{x},t)$, with $c=c_0$ prescribed on $\partial\Omega_c$ and the inward flux $-\bm{j}_c\cdot\bm{n}=\bar{q}$ on $\partial\Omega_j$.}
  \label{fig:domain}
\end{figure}

Consider a deformable porous body occupying $\Omega\subset\mathbb{R}^{n_{\mathrm{dim}}}$ over $t\in[0,T]$, with boundary $\partial\Omega$ and outward unit normal $\bm{n}$ (\Cref{fig:domain}). The body may contain an initial crack $\Gamma_0\subset\Omega$. For a given chemical state, the mechanical and fracture behavior is governed by the energy functional:
\begin{align}
  \Pi[\bm{u},d]
  ={}&\int_{\Omega}\left[
    g(d,d_{\mathrm{chem}})\Psi_0(\bm{\varepsilon})
    +G_c\gamma_{l_0}(d,\nabla d)
  \right]\,\mathrm{d}\Omega \nonumber\\
  &-\int_{\Omega}\bm{b}\cdot\bm{u}\,\mathrm{d}\Omega
  -\int_{\partial\Omega_h}\bm{h}\cdot\bm{u}\,\mathrm{d}\Gamma,
  \label{eq:mechanical_fracture_functional}
\end{align}
where $\Psi_0$ is the undamaged elastic strain-energy density, $\gamma_{l_0}$ is the regularized crack-surface density, $G_c$ is the critical energy-release rate, and $g(d,d_{\mathrm{chem}})$ is the combined degradation function defined in Section~\ref{sec:theory:pf}. Stationarity of $\Pi$ with respect to $\bm{u}$ and $d$, with the crack-driving energy $\Psi_0$ replaced by the history field $H$ to enforce irreversibility, gives the first two equations below (see~\ref{app:strong_form}). The third is the mass balance of dissolved CO$_2$. The coupled strong form is:
\begin{subequations}
\begin{align}
  \nabla\cdot\bm{\sigma}+\bm{b}&=\bm{0}
  &&\text{in }\Omega,
  \label{eqn:strong_forms_a}\\
  G_c\left(\frac{d}{l_0}-l_0\nabla^2d\right)
  -2(1-d_{\mathrm{chem}})(1-d)H&=0
  &&\text{in }\Omega,
  \label{eqn:strong_forms_b}\\
  \dot{c}+\nabla\cdot\bm{j}_c&=0
  &&\text{in }\Omega,
  \label{eqn:strong_forms_c}
\end{align}
\label{eqn:strong_forms}
\end{subequations}
where $\bm{\sigma}$ is the degraded Cauchy stress and $\bm{j}_c$ is the dissolved-CO$_2$ flux. The associated boundary conditions are:
\begin{subequations}\label{eqn:bc}
\begin{align}
  \bm{u}&=\bm{u}_0 &&\text{on }\partial\Omega_u,
  \label{eqn:bc_a}\\
  \bm{\sigma}\cdot\bm{n}&=\bm{h} &&\text{on }\partial\Omega_h,
  \label{eqn:bc_b}\\
  d&=1 &&\text{on }\Gamma_0,
  \label{eqn:bc_c}\\
  \nabla d\cdot\bm{n}&=0 &&\text{on }\partial\Omega,
  \label{eqn:bc_d}\\
  c&=c_0 &&\text{on }\partial\Omega_c,
  \label{eqn:bc_e}\\
  -\bm{j}_c\cdot\bm{n}&=\bar{q} &&\text{on }\partial\Omega_j,
  \label{eqn:bc_f}
\end{align}
\end{subequations}
with $\partial\Omega_u\cup\partial\Omega_h=\partial\Omega_c\cup\partial\Omega_j=\partial\Omega$ and $\partial\Omega_u\cap\partial\Omega_h=\partial\Omega_c\cap\partial\Omega_j=\emptyset$. An initial crack may be prescribed through $d=1$ on $\Gamma_0$. The fields are coupled through dissolution, degradation, and transport, as summarized in Table~\ref{tab:couplings}.

\subsection{Dissolution-dependent Phase-field Fracture}
\label{sec:theory:pf}

Under the small-deformation assumption, the infinitesimal strain tensor and the undamaged elastic strain-energy density are:
\begin{equation}
  \bm{\varepsilon}(\bm{u})
  =\frac{1}{2}\left(\nabla\bm{u}+\nabla\bm{u}^{T}\right),
  \label{eq:strain}
\end{equation}
and
\begin{equation}
  \Psi_0(\bm{\varepsilon})
  =\frac{1}{2}\bm{\varepsilon}:\bm{C}_0:\bm{\varepsilon},
  \label{eq:psi0}
\end{equation}
where $\bm{C}_0$ is the fourth-order elastic stiffness tensor. The sharp crack surface $\Gamma$ is represented by the phase field:
\begin{equation}
  d=
  \begin{cases}
    0, & \text{intact material},\\
    1, & \text{fully fractured material},
  \end{cases}
  \label{eq:phase_field_limits}
\end{equation}
and regularized using the AT2 crack-surface approximation~\citep{Ambrosio1990,Miehe2010a}:
\begin{equation}
  \int_{\Gamma}\mathrm{d}S
  \approx\int_{\Omega}\gamma_{l_0}(d,\nabla d)\,\mathrm{d}\Omega,
  \label{eq:surface_regularisation}
\end{equation}
with
\begin{equation}
  \gamma_{l_0}(d,\nabla d)
  =\frac{d^2}{2l_0}+\frac{l_0}{2}\lvert\nabla d\rvert^2,
  \label{eq:crack_density}
\end{equation}
where $l_0$ controls the width of the diffuse fracture zone.

The phase field degrades the stiffness through the quadratic function $g_{\mathrm{m}}(d)=(1-d)^2$, and dissolution introduces the independent chemical factor $g_{\mathrm{c}}=1-d_{\mathrm{chem}}$. The two may be combined additively or multiplicatively~\citep{Pignatelli2013,LeBellego2003}. Following \citet{Schuler2020}, the multiplicative form is adopted:
\begin{equation}
  g(d,d_{\mathrm{chem}})
  =(1-d_{\mathrm{chem}})(1-d)^2+\kappa,
  \label{eq:degradation_combined}
\end{equation}
where $\kappa=10^{-6}$ is a numerical residual-stiffness parameter. The multiplicative form keeps the  two damage histories distinct and cannot exceed unity. For $d_{\mathrm{chem}}=0$, it recovers the standard quadratic degradation function of brittle phase-field fracture~\citep{Bourdin2000}. The degraded elastic energy and stress are:

\begin{equation}
  \Psi_{\mathrm{elastic}}
  =g(d,d_{\mathrm{chem}})\Psi_0(\bm{\varepsilon}),
  \label{eq:psi_elastic_mech}
\end{equation}
and
\begin{equation}
  \bm{\sigma}
  =\frac{\partial\Psi_{\mathrm{elastic}}}{\partial\bm{\varepsilon}}
  =g(d,d_{\mathrm{chem}})\bm{C}_0:\bm{\varepsilon}.
  \label{eq:stress_coupled}
\end{equation}

Crack irreversibility is enforced through the history-field construction of \citet{MieheHofacker2010}:
\begin{equation}
  H(\bm{x},t)=\max_{\tau\in[0,t]}
  \Psi_0\bigl(\bm{\varepsilon}(\bm{x},\tau)\bigr),
  \label{eq:history}
\end{equation}
which replaces $\Psi_0$ as the crack-driving energy in Eq.~\eqref{eqn:strong_forms_b}. No tension--compression split is used. This assumption is common to all solution strategies and should be considered when interpreting states involving substantial compression.

\subsection{Reactive Transport and Dissolution Kinetics}
\label{sec:theory:chem}

The flux in Eq.~\eqref{eqn:strong_forms_c} follows from the gradient of the chemical potential of dissolved CO$_2$, $\mu=\mu_0+RT\ln c-\bar{V}\sigma_H$, which decreases under hydrostatic tension. Taking $\bm{j}_c=-(D_{\mathrm{eff}}c/RT)\nabla\mu$ gives Fickian diffusion plus a stress-assisted term~\citep{MartinezPaneda2018,Schuler2020}:
\begin{equation}
  \bm{j}_c
  =-D_{\mathrm{eff}}\nabla c
  +\frac{D_{\mathrm{eff}}\bar{V}}{RT}\,c\,\nabla\sigma_H,
  \label{eqn:flux_final}
\end{equation}
where $\bar{V}$ is the partial molar volume of dissolved CO$_2$, $R$ is the universal gas constant, $T$ is the absolute temperature, and $\sigma_H=\operatorname{tr}(\bm{\sigma})/3$ is the hydrostatic stress under a tensile-positive convention. Following \citet{Schuler2020}, consumption of dissolved CO$_2$ by the reaction is neglected in Eq.~\eqref{eqn:strong_forms_c}.

The concentration $c$ controls dissolution through the pH of the pore fluid. Dissolved CO$_2$ hydrates to carbonic acid, whose dissociation releases H$^+$ and lowers the pore-fluid pH to about 3--5 under reservoir conditions~\citep{Gaus2010,Arshad2026}:

\begin{equation}
  \mathrm{H_2O+CO_2\rightleftharpoons H_2CO_3},\qquad
  \mathrm{H_2CO_3\rightleftharpoons HCO_3^-+H^+},\qquad
  \mathrm{HCO_3^-\rightleftharpoons CO_3^{2-}+H^+},
  \label{eq:carbonic_system}
\end{equation}
with the Henry's law constant $K_h$, first dissociation constant $K_{a1}$, and water ion product $K_w$:
\begin{equation}
  K_h=\frac{[\mathrm{H_2CO_3}]}{c},\qquad
  K_{a1}=\frac{[\mathrm{HCO_3^-}][\mathrm{H^+}]}{[\mathrm{H_2CO_3}]},\qquad
  K_w=[\mathrm{H^+}][\mathrm{OH^-}]=10^{-14}.
  \label{eq:equilibrium_constants}
\end{equation}
Charge neutrality requires $[\mathrm{H^+}]=[\mathrm{OH^-}]+[\mathrm{HCO_3^-}]+2[\mathrm{CO_3^{2-}}]$. The second dissociation ($\mathrm{p}K_{a2}\approx10.3$) is negligible under the acidic conditions considered, so substituting the equilibrium relations gives $[\mathrm{H^+}]^2=K_w+K_hK_{a1}c$~\citep{Schuler2020}:
\begin{equation}
  [\mathrm{H}^+]=\sqrt{K_w+K_hK_{a1}c},
  \qquad \mathrm{pH}=-\log_{10}[\mathrm{H}^+].
  \label{eq:pH_approx}
\end{equation}
Concentrations and equilibrium constants must use a consistent standard state and units. As in \citet{Schuler2020}, pH buffering by the dissolution products is neglected.

The released protons dissolve calcite, $\mathrm{CaCO_3+H^+\rightleftharpoons Ca^{2+}+HCO_3^-}$~\citep{Arshad2026}. Combined with the hydration and first dissociation in Eq.~\eqref{eq:carbonic_system}, this gives the overall dissolution reaction:
\begin{equation}
  \mathrm{CaCO_3\,(s)+CO_2\,(aq)+H_2O\,(l)
  \rightleftharpoons Ca^{2+}\,(aq)+2HCO_3^-\,(aq)}.
  \label{eq:calcite_reaction}
\end{equation}
Its rate per unit reactive surface area, $R_{\mathrm{Ca}}$ (mol\,m$^{-2}$\,s$^{-1}$), is taken from the quadratic fit of \citet{Schuler2020} to the experimental rates compiled by \citet{Holzheid2016}, with temperature and pressure held constant. It is converted to mol\,mm$^{-2}$\,s$^{-1}$ for consistency with the units of Table~\ref{tab:tensile_props}:
\begin{equation}
  \log_{10}(R_{\mathrm{Ca}})
  =0.056\,\mathrm{pH}^2-1.0218\,\mathrm{pH}-1.4878.
  \label{eq:dissolution_rate}
\end{equation}
Multiplying $R_{\mathrm{Ca}}$ by the reactive area per unit bulk volume, $A$, gives the moles of calcite dissolved per unit volume and time. Multiplying by the molar volume $M/\rho$ converts this to the rate of porosity increase~\citep{Schuler2020}:
\begin{equation}
  \dot{\Phi}=R_{\mathrm{Ca}}A\frac{M}{\rho}.
  \label{eq:porosity}
\end{equation}
The reactive area first grows as pores enlarge and then falls as they overlap and the calcite is consumed. This evolution is described by the modified random-pore model of \citet{Khinast1996}:
\begin{equation}
  \frac{A}{A_0}
  =(1-X)^{a_s}\left[1-\mathcal{C}\ln(1-X)\right]^{b_s},
  \qquad
  X=\frac{n_{\mathrm{diss}}}{n_0},
  \label{eq:surface_area_evolution}
\end{equation}
where $X\in[0,1]$ is the fraction of the initial calcite dissolved. The values $\mathcal{C}=37$, $a_s=1.7$, and $b_s=0.69$ were fitted by \citet{Khinast1996} to the reactive surface area inferred from thermogravimetric decomposition of limestone particles at 1053~K. \citet{Schuler2020} adopted them for aqueous calcite dissolution, and they are retained here for consistency with that model. They are therefore not calibrated for the present rock or reaction. The structural parameter $\mathcal{C}$ depends on the initial porosity and pore-size distribution~\citep{Khinast1996}, and reactive surface areas in real reservoirs remain highly uncertain~\citep{Arshad2026}. The predicted timing of dissolution, and hence of chemical damage, should therefore be regarded as indicative. Because all solution strategies use identical parameters, this limitation does not affect their relative comparison.

The porosity increase is mapped to chemical damage through:
\begin{equation}
  d_{\mathrm{chem}}=1-\exp(-r\Phi),
  \label{eq:dchem}
\end{equation}
which corresponds to an exponential decrease of the elastic modulus with porosity, i.e. $E=E_0\exp(-r\Phi)$, proposed for rocks undergoing dissolution~\citep{Wojtacki2015,Schuler2020}. Thus $d_{\mathrm{chem}}=1-E/E_0$ is the fractional loss of stiffness due to dissolution, and the material parameter $r$ sets how rapidly stiffness is lost as porosity increases.

\subsection{Damage- and Porosity-dependent Transport}
\label{Diffusion_coef}

Dissolution and fracture both increase diffusivity. The porosity-dependent contribution is~\citep{NavarreeSitchler2009}:
\begin{equation}
  D(\Phi_e)=D_p+D_0\Phi_e^2,
  \qquad
  \Phi_e=
  \begin{cases}
    0, & \Phi_T\leq\Phi_c,\\
    a(\Phi_T-\Phi_c)^\beta, & \Phi_T>\Phi_c,
  \end{cases}
  \label{eq:D_porosity}
\end{equation}
where $D_p$ is the intact-material diffusivity, $D_0$ is the diffusivity in water, $\Phi_T=\Phi_0+\Phi$ is the total porosity, and $\Phi_e$ is the connected porosity above the percolation threshold $\Phi_c$. Following \citet{Schuler2020}, $\Phi_c=\Phi_0$, so the connected porosity is governed by dissolution. The phase-field contribution follows \citet{Gerard1998}:
\begin{equation}
  D(d)=D_p+D_0\left(1-\frac{1}{1+(d/d_{\mathrm{cr}})^{n_D}}\right),
  \label{eq:D_damage}
\end{equation}
where $d_{\mathrm{cr}}$ is the damage level at which diffusivity rises steeply and $n_D$ controls how abruptly it does so. The effective diffusivity takes the locally dominant contribution~\citep{Gerard1998,Schuler2020}:
\begin{equation}
  D_{\mathrm{eff}}=\max\!\left(D(\Phi_e),D(d)\right).
  \label{eq:D_eff}
\end{equation}
The parameters $a=2.4$ and $\beta=0.9$ were obtained by \citet{NavarreeSitchler2009} from X-ray microtomography and diffusion measurements on a weathered basalt clast. The values $d_{\mathrm{cr}}=0.4$ and $n_D=5$ were proposed by \citet{Gerard1998} for damaged cementitious materials. Both sets were adopted unchanged by \citet{Schuler2020} and are retained here. As with the reactive-area parameters, they were not calibrated for the present rock. This limits the quantitative predictions of the model but not the comparison of solution strategies.

Table~\ref{tab:couplings} summarizes the resulting inter-field couplings. Each coupling corresponds to an off-diagonal block of the tangent. These blocks are defined in Section~\ref{sec:residual_tangent}, and the solution strategies differ in how they treat them.

\begin{table}[!ht]
  \centering
  \caption{Principal inter-field couplings of the model and the corresponding off-diagonal tangent blocks.}
  \label{tab:couplings}
  \small
  \renewcommand{\arraystretch}{1.2}
  \begin{tabularx}{\textwidth}{@{}l>{\raggedright\arraybackslash}Xl@{}}
    \toprule
    Coupling & Mechanism & Block \\
    \midrule
    $d\rightarrow\bm{u}$ & Stiffness degradation $g(d,d_{\mathrm{chem}})$, Eq.~\eqref{eq:stress_coupled} & $\bm{J}_{ud}$ \\
    $\bm{u}\rightarrow d$ & Crack-driving energy $H$, Eq.~\eqref{eq:history} & $\bm{J}_{du}$ \\
    $c\rightarrow\bm{u}$ & $c\rightarrow\mathrm{pH}\rightarrow R_{\mathrm{Ca}}\rightarrow\Phi\rightarrow d_{\mathrm{chem}}\rightarrow g$, Eqs.~\eqref{eq:pH_approx}--\eqref{eq:dchem} & $\bm{J}_{uc}$ \\
    $c\rightarrow d$ & Same chain, through the factor $(1-d_{\mathrm{chem}})$ in Eq.~\eqref{eqn:strong_forms_b} & $\bm{J}_{dc}$ \\
    $\bm{u}\rightarrow c$ & Stress-assisted flux through $\nabla\sigma_H$, Eq.~\eqref{eqn:flux_final} & $\bm{J}_{cu}$ \\
    $d\rightarrow c$ & Damage-enhanced diffusivity $D(d)$ and degraded $\sigma_H$, Eqs.~\eqref{eq:D_damage}--\eqref{eq:D_eff} & $\bm{J}_{cd}$ \\
    \bottomrule
  \end{tabularx}
\end{table}

\section{Numerical Implementation}
\label{sec:implementation}
This section describes how the coupled problem of Section~\ref{sec:theory} is discretized and solved in Abaqus. It presents the weak form and the temporal and spatial discretization, and then the element residuals and the tangent blocks that link the three fields. These blocks form the basis of the five solution strategies, which are defined next and compared side by side. The section closes with the metrics used to evaluate the strategies in Section~\ref{sec:benchmarks}.

\subsection{Weak Formulation}
\label{sec:weak_form}

Let $\delta\bm{u}$, $\delta d$, and $\delta c$ denote  test functions. Multiplying Eqs.~\eqref{eqn:strong_forms_a}--\eqref{eqn:strong_forms_c} by these functions, integrating by parts, and applying the natural boundary conditions gives:
\begin{equation}
    \int_\Omega \delta\bm{\varepsilon}:\bm{\sigma}\,\mathrm{d}\Omega
    -\int_\Omega \delta\bm{u}\cdot\bm{b}\,\mathrm{d}\Omega
    -\int_{\partial\Omega_h}\delta\bm{u}\cdot\bm{h}\,\mathrm{d}\Gamma=0,
    \label{eq:weak_u}
\end{equation}
\begin{equation}
    \int_\Omega\left[
        G_c l_0\nabla\delta d\cdot\nabla d
        +\delta d\left(
            \frac{G_c}{l_0}d
            -2(1-d_{\mathrm{chem}})(1-d)H
        \right)
    \right]\mathrm{d}\Omega=0,
    \label{eq:weak_d}
\end{equation}
and
\begin{equation}
    \int_\Omega\left[
        \delta c\,\dot{c}
        +D_{\mathrm{eff}}\nabla\delta c\cdot\nabla c
        -\frac{D_{\mathrm{eff}}\bar{V}}{RT}
        c\,\nabla\delta c\cdot\nabla\sigma_H
    \right]\mathrm{d}\Omega
    -\int_{\partial\Omega_j}\delta c\,\bar{q}\,\mathrm{d}\Gamma=0.
    \label{eq:weak_c}
\end{equation}
\noindent During the phase-field variation, $H$ and $d_{\mathrm{chem}}$ are treated as known state variables. Their update timing differs among the solution strategies described below.

\subsection{Temporal and Spatial Discretization}
\label{sec:discretisation}
\label{sec:time_disc}
\label{sec:space_disc}

Only the concentration balance contains a time derivative. It is integrated with the generalized trapezoidal rule:
\begin{equation}
    \dot{c}^{n+1}
    =\frac{c^{n+1}-c^n}{\theta\Delta t}
    -\frac{1-\theta}{\theta}\dot{c}^n,
    \qquad \theta\in(0,1],
    \label{eqn:general_rate_update}
\end{equation}
with the fully implicit choice $\theta=1$ in all simulations, for which the residuals below are written. The dissolution-induced porosity is updated as:
\begin{equation}
    \Phi^{n+1}=\Phi^n
    +R_{\mathrm{Ca}}\bigl(c^{n+1}\bigr)\,A^{n}\,\Delta t\,\frac{M}{\rho},
    \label{eq:porosity_update}
\end{equation}
with the reactive area held at its value at the start of the increment.

All analyses are two-dimensional under plane strain. Consider an element with $n_n$ nodes and $n_{\mathrm{dof}}$ displacement degrees of freedom per node. Its nodal unknowns are:
\begin{equation}
    \bm{u}_e\in\mathbb{R}^{n_n n_{\mathrm{dof}}},\qquad
    \bm{d}_e\in\mathbb{R}^{n_n},\qquad
    \bm{c}_e\in\mathbb{R}^{n_n}.
\end{equation}
All fields are interpolated with the same eight-node serendipity quadrilateral (Q8) shape functions and integrated with a $3\times3$ Gauss rule. Quadratic elements are used because the stress-assisted flux requires $\nabla\sigma_H$, which is obtained by extrapolating the integration-point values of $\sigma_H$ to the nodes and differentiating them with the shape functions. Eight-node elements give a more accurate approximation of this gradient than linear elements~\citep{MartinezPaneda2018,Reddy2025}. They also resolve the steep phase-field and concentration profiles near the crack with fewer elements across $l_0$.

The same Q8 interpolation is used for $\bm{u}$, $d$, and $c$. Unequal orders are needed only when one field acts as a constraint on another, as pressure does in incompressible or undrained poroelastic problems, where equal-order elements can produce spurious pressure oscillations. Here, $d$ and $c$ affect the momentum balance only by degrading the stiffness through $g$, so no such instability arises. All three fields also appear with first derivatives only in the weak form. Sharing the same nodes and integration points lets coupling quantities such as $d_{\mathrm{chem}}$ and $\sigma_H$ pass between fields without interpolation. With the standard interpolation and gradient matrices $\bm{N}_u$, $\bm{B}_u$, $\bm{N}_d$, $\bm{B}_d$, $\bm{N}_c$, and $\bm{B}_c$, the fields and their variations are approximated by:
\begin{subequations}\label{eqn:field_interpolation}
\begin{align}
    \bm{u}&=\bm{N}_u\bm{u}_e, &
    \bm{\varepsilon}&=\bm{B}_u\bm{u}_e, &
    \delta\bm{u}&=\bm{N}_u\delta\bm{u}_e, &
    \delta\bm{\varepsilon}&=\bm{B}_u\delta\bm{u}_e,
    \label{eqn:field_interpolation_u}\\
    d&=\bm{N}_d\bm{d}_e, &
    \nabla d&=\bm{B}_d\bm{d}_e, &
    \delta d&=\bm{N}_d\delta\bm{d}_e, &
    \nabla\delta d&=\bm{B}_d\delta\bm{d}_e,
    \label{eqn:field_interpolation_d}\\
    c&=\bm{N}_c\bm{c}_e, &
    \nabla c&=\bm{B}_c\bm{c}_e, &
    \delta c&=\bm{N}_c\delta\bm{c}_e, &
    \nabla\delta c&=\bm{B}_c\delta\bm{c}_e,
    \label{eqn:field_interpolation_c}
\end{align}
\end{subequations}
where $\bm{N}_d=\bm{N}_c$ and $\bm{B}_d=\bm{B}_c$ because of the common interpolation.
\subsection{Element Residuals and Tangent Structure}
\label{sec:residual_tangent}

Substituting Eq.~\eqref{eqn:field_interpolation} into Eqs.~\eqref{eq:weak_u}--\eqref{eq:weak_c} gives the element residuals:
\begin{subequations}\label{eqn:residual_blocks}
\begin{align}
    \bm{R}_u={}&
    \int_{\Omega^e}\bm{B}_u^T\bm{\sigma}\,\mathrm{d}\Omega
    -\int_{\Omega^e}\bm{N}_u^T\bm{b}\,\mathrm{d}\Omega
    -\int_{\partial\Omega_h^e}\bm{N}_u^T\bm{h}\,\mathrm{d}\Gamma,
    \label{eqn:Ru}\\[4pt]
    \bm{R}_d={}&
    \int_{\Omega^e}\Bigg[
        \bm{B}_d^T\bm{B}_d\bm{d}_e
        +\left(\frac{1}{l_0^2}
        +\frac{2(1-d_{\mathrm{chem}})H}{G_c l_0}\right)
        \bm{N}_d^T\bm{N}_d\bm{d}_e
        -\frac{2(1-d_{\mathrm{chem}})H}{G_c l_0}\bm{N}_d^T
    \Bigg]\mathrm{d}\Omega,
    \label{eqn:Rd}\\[4pt]
    \bm{R}_c={}&
    \int_{\Omega^e}\Bigg[
        \bm{N}_c^T\bm{N}_c
        \frac{\bm{c}_e^{n+1}-\bm{c}_e^n}{\Delta t}
        +D_{\mathrm{eff}}\bm{B}_c^T\bm{B}_c\bm{c}_e
        -\frac{D_{\mathrm{eff}}\bar{V}}{RT}
        \bm{B}_c^T\nabla\sigma_H\,(\bm{N}_c\bm{c}_e)
    \Bigg]\mathrm{d}\Omega
    -\int_{\partial\Omega_j^e}\bm{N}_c^T\bar{q}\,\mathrm{d}\Gamma.
    \label{eqn:Rc}
\end{align}
\end{subequations}

At nonlinear iteration $k$, the assembled system is written as:
\begin{equation}
    \bm{J}^{(k)}\Delta\bm{x}^{(k)}=-\bm{R}^{(k)},
    \qquad
    \bm{x}=\{\bm{u},\bm{d},\bm{c}\},
    \label{eqn:global_newton_system}
\end{equation}
with the formal block Jacobian
\begin{equation}
    \bm{J}=\frac{\partial\bm{R}}{\partial\bm{x}}
    =\begin{bmatrix}
        \bm{J}_{uu} & \bm{J}_{ud} & \bm{J}_{uc}\\
        \bm{J}_{du} & \bm{J}_{dd} & \bm{J}_{dc}\\
        \bm{J}_{cu} & \bm{J}_{cd} & \bm{J}_{cc}
    \end{bmatrix}.
    \label{eqn:global_jacobian}
\end{equation}
The diagonal blocks supplied by the UEL are:
\begin{subequations}\label{eqn:jacobian_diag}
\begin{align}
    \bm{J}_{uu}={}&
    \int_{\Omega^e}g(d,d_{\mathrm{chem}})
    \bm{B}_u^T\bm{C}_0\bm{B}_u\,\mathrm{d}\Omega,
    \label{eqn:Juu}\\
    \bm{J}_{dd}={}&
    \int_{\Omega^e}\left[
        \bm{B}_d^T\bm{B}_d
        +\left(\frac{1}{l_0^2}
        +\frac{2(1-d_{\mathrm{chem}})H}{G_c l_0}\right)
        \bm{N}_d^T\bm{N}_d
    \right]\mathrm{d}\Omega,
    \label{eqn:Jdd}\\
    \bm{J}_{cc}={}&
    \frac{1}{\theta\Delta t}
    \int_{\Omega^e}\bm{N}_c^T\bm{N}_c\,\mathrm{d}\Omega
    +\int_{\Omega^e}D_{\mathrm{eff}}\left[
        \bm{B}_c^T\bm{B}_c
        -\frac{\bar{V}}{RT}
         \bm{B}_c^T\nabla\sigma_H\,\bm{N}_c
    \right]\mathrm{d}\Omega.
    \label{eqn:Jcc}
\end{align}
\end{subequations}

With $\bm{\sigma}_0=\bm{C}_0\bm{B}_u\bm{u}_e$ the undegraded stress in Voigt form, the off-diagonal blocks, one for each coupling in Table~\ref{tab:couplings}, are:
\begin{subequations}\label{eqn:jacobian_offdiag}
\begin{align}
    \bm{J}_{ud}={}&
    \int_{\Omega^e}\frac{\partial g}{\partial d}\,
    \bm{B}_u^T\bm{\sigma}_0\,\bm{N}_d\,\mathrm{d}\Omega,
    \label{eqn:Jud}\\
    \bm{J}_{uc}={}&
    \int_{\Omega^e}\frac{\partial g}{\partial d_{\mathrm{chem}}}
    \frac{\partial d_{\mathrm{chem}}}{\partial c}\,
    \bm{B}_u^T\bm{\sigma}_0\,\bm{N}_c\,\mathrm{d}\Omega,
    \label{eq:Kuc}\\
    \bm{J}_{du}={}&
    -\int_{\Omega^e}\frac{2(1-d_{\mathrm{chem}})(1-d)}{G_cl_0}\,
    \chi_H\,\bm{N}_d^T\bm{\sigma}_0^T\bm{B}_u\,\mathrm{d}\Omega,
    \label{eqn:Jdu}\\
    \bm{J}_{dc}={}&
    \int_{\Omega^e}\frac{2(1-d)H}{G_cl_0}
    \frac{\partial d_{\mathrm{chem}}}{\partial c}\,
    \bm{N}_d^T\bm{N}_c\,\mathrm{d}\Omega,
    \label{eqn:Jdc}\\
    \bm{J}_{cu}={}&
    -\int_{\Omega^e}\frac{D_{\mathrm{eff}}\bar{V}}{RT}
    (\bm{N}_c\bm{c}_e)\,\bm{B}_c^T
    \frac{\partial\nabla\sigma_H}{\partial\bm{u}_e}\,\mathrm{d}\Omega,
    \label{eq:Kcu}\\
    \bm{J}_{cd}={}&
    \int_{\Omega^e}\left[
    \frac{\partial D_{\mathrm{eff}}}{\partial d}
    \left(\bm{B}_c^T\bm{B}_c\bm{c}_e
    -\frac{\bar{V}}{RT}(\bm{N}_c\bm{c}_e)\,\bm{B}_c^T\nabla\sigma_H\right)\bm{N}_d
    -\frac{D_{\mathrm{eff}}\bar{V}}{RT}
    (\bm{N}_c\bm{c}_e)\,\bm{B}_c^T
    \frac{\partial\nabla\sigma_H}{\partial\bm{d}_e}
    \right]\mathrm{d}\Omega,
    \label{eqn:Jcd}
\end{align}
\end{subequations}
where $\partial g/\partial d=-2(1-d_{\mathrm{chem}})(1-d)$, $\partial g/\partial d_{\mathrm{chem}}=-(1-d)^2$, and $\chi_H=1$ when $\Psi_0>H^n$ (loading) and $\chi_H=0$ otherwise. Each diagonal block is the derivative of a residual with respect to its own field only. Coefficients set by the other fields ($H$, $d_{\mathrm{chem}}$, $D_{\mathrm{eff}}$, and $\nabla\sigma_H$) are treated as constants in this derivative, and so is the dependence of $D_{\mathrm{eff}}$ on $c$ through $\Phi$.

The concentration sensitivity of the chemical damage follows from the chain $c\rightarrow\mathrm{pH}\rightarrow R_{\mathrm{Ca}}\rightarrow\Phi\rightarrow d_{\mathrm{chem}}$ (Eqs.~\eqref{eq:pH_approx}, \eqref{eq:dissolution_rate}, \eqref{eq:porosity_update}, and~\eqref{eq:dchem}), with the reactive area fixed at $A^n$ within the increment:
\begin{equation}
    \frac{\partial d_{\mathrm{chem}}}{\partial c}
    =\underbrace{r\,e^{-r\Phi}}_{\partial d_{\mathrm{chem}}/\partial\Phi}\;
    \underbrace{A^n\,\Delta t\,\frac{M}{\rho}}_{\partial\Phi/\partial R_{\mathrm{Ca}}}\;
    \underbrace{\ln(10)\,R_{\mathrm{Ca}}\,(0.112\,\mathrm{pH}-1.0218)}_{\partial R_{\mathrm{Ca}}/\partial\mathrm{pH}}\;
    \underbrace{\left(-\frac{K_hK_{a1}}{2\ln(10)\,(K_w+K_hK_{a1}c)}\right)}_{\partial\mathrm{pH}/\partial c}.
    \label{eqn:ddchem_dc}
\end{equation}
The damage sensitivity of the diffusivity follows from Eqs.~\eqref{eq:D_damage} and~\eqref{eq:D_eff}:
\begin{equation}
    \frac{\partial D_{\mathrm{eff}}}{\partial d}=
    \begin{cases}
    \dfrac{n_DD_0}{d_{\mathrm{cr}}}\dfrac{(d/d_{\mathrm{cr}})^{n_D-1}}{\bigl[1+(d/d_{\mathrm{cr}})^{n_D}\bigr]^2}, & D(d)\geq D(\Phi_e),\\[8pt]
    0, & \text{otherwise}.
    \end{cases}
    \label{eqn:dDeff_dd}
\end{equation}

The hydrostatic-stress gradient is computed in four steps, following the nodal-extrapolation approach of \citet{MartinezPaneda2018}. First, the nodal displacements give the strain at each Gauss point $\bm{\xi}_q$, and hence the undegraded hydrostatic stress under plane strain, $\sigma_{H0,q}=K\,\bm{m}^T\bm{B}_u(\bm{\xi}_q)\bm{u}_e$, where $K$ is the bulk modulus and $\bm{m}^T=[1,\,1,\,0]$. Second, these values are extrapolated to the eight nodes. Only Gauss points $q=1,\ldots,8$ of the $3\times3$ rule are used, so the matrix $P_{qI}=N_I(\bm{\xi}_q)$, with $q,I=1,\ldots,8$, is square and its inverse $\bm{E}=\bm{P}^{-1}$ exists. Third, the nodal values are differentiated with the shape-function gradients $\bm{B}_c$ at each Gauss point. Fourth, the result is multiplied by $g$ at that Gauss point:
\begin{equation}
    \nabla\sigma_H=g\,\bm{B}_c\,\hat{\bm{\sigma}}_{H0},
    \qquad
    \hat{\bm{\sigma}}_{H0}=\bm{E}\,\bm{s}_{H0},
    \qquad
    \bm{s}_{H0}=K
    \begin{bmatrix}
    \bm{m}^T\bm{B}_u(\bm{\xi}_1)\\ \vdots\\ \bm{m}^T\bm{B}_u(\bm{\xi}_8)
    \end{bmatrix}\bm{u}_e,
    \label{eqn:gradsigmaH}
\end{equation}
so that
\begin{equation}
    \frac{\partial\nabla\sigma_H}{\partial\bm{u}_e}
    =g\,\bm{B}_c\,\bm{E}\,K
    \begin{bmatrix}
    \bm{m}^T\bm{B}_u(\bm{\xi}_1)\\ \vdots\\ \bm{m}^T\bm{B}_u(\bm{\xi}_8)
    \end{bmatrix},
    \qquad
    \frac{\partial\nabla\sigma_H}{\partial\bm{d}_e}
    =\frac{\partial g}{\partial d}\,\bm{B}_c\,\hat{\bm{\sigma}}_{H0}\,\bm{N}_d.
    \label{eqn:dgradsigmaH}
\end{equation}

\noindent The solution strategies in Section~\ref{sec:impl:overview} differ in which of these blocks they retain and in the state at which the coupling coefficients are evaluated.

\subsection{Solution Strategies}
\label{sec:impl:overview}

The coupled system can be solved monolithically, with all fields updated together, or with a staggered scheme, in which some coupling inputs are frozen so that the fields can be solved separately. Five solution strategies are compared:
\begin{itemize}
  \item \textbf{Case~1, \emph{Block-diagonal inexact-Newton monolithic}}: all fields are solved simultaneously with a block-diagonal tangent.
  \item \textbf{Case~2, \emph{Quasi-Newton (BFGS) monolithic}}: as Case~1, with the tangent updated by BFGS.
  \item \textbf{Case~3, \emph{Single-pass fully staggered}}: each field is solved separately, with coupling terms lagged from the last converged increment.
  \item \textbf{Case~4, \emph{Single-pass partially staggered $[\bm{u},c]$}}: $\bm{u}$ and $c$ are solved together, with the phase field lagged.
  \item \textbf{Case~5, \emph{Hybrid multi-pass fully staggered}}: three single-field elements are evaluated in a fixed order, each passing updated coupling quantities to the next within the same iteration.
\end{itemize}

The strategies differ in two respects. The first is the \textit{approximate} matrix $\tilde{\bm{J}}$, which can depart from the \textit{consistent tangent} $\bm{J}$ through three independent choices: which coupling blocks it retains, how accurately it is computed (consistent, inexact, or secant), and how often it is updated (every iteration, or periodically). The second is whether the quantities each residual takes from the other fields are evaluated at the current iterate $(k)$ or at the last converged increment $n$. To show the latter, each residual is written with its unknown first and these coupling inputs after the semicolon:
\begin{equation}
    \bm{R}_u\bigl(\bm{u};\,d,\,d_{\mathrm{chem}}\bigr),\qquad
    \bm{R}_d\bigl(d;\,H,\,d_{\mathrm{chem}}\bigr),\qquad
    \bm{R}_c\bigl(c;\,D_{\mathrm{eff}},\,\nabla\sigma_H\bigr).
    \label{eq:residual_arguments}
\end{equation}
Because all degrees of freedom are solved together in every case, a strategy is monolithic here when all coupling inputs are current, so that the converged solution satisfies the fully coupled equations. In Cases~3 and~4, enough inputs are lagged that the residuals decouple into sub-problems, so solving them together is equivalent to solving those sub-problems one after another. Case~5 instead exchanges coupling quantities between iterations. Chemical quantities with superscript $(k)$ are recomputed from $\Phi^n$ and $c^{(k)}$ through Eqs.~\eqref{eq:pH_approx}--\eqref{eq:D_eff} and \eqref{eq:porosity_update}.

\subsubsection{Case~1: Block-Diagonal Inexact-Newton Monolithic Scheme}
\label{sec:impl:newton}

All fields are updated together, with all coupling inputs at the current iterate:
\begin{equation}
  \tilde{\bm{J}}^{(k)}=
  \begin{bmatrix}
    \bm{J}_{uu} & \bm{0} & \bm{0} \\
    \bm{0} & \bm{J}_{dd} & \bm{0} \\
    \bm{0} & \bm{0} & \bm{J}_{cc}
  \end{bmatrix}^{(k)},
  \qquad
  \bm{R}^{(k)}=
  \begin{Bmatrix}
    \bm{R}_u\bigl(\bm{u}^{(k)};\,d^{(k)},\,d_{\mathrm{chem}}^{(k)}\bigr)\\
    \bm{R}_d\bigl(d^{(k)};\,H^{(k)},\,d_{\mathrm{chem}}^{(k)}\bigr)\\
    \bm{R}_c\bigl(c^{(k)};\,D_{\mathrm{eff}}^{(k)},\,\nabla\sigma_H^{(k)}\bigr)
  \end{Bmatrix},
  \label{eq:newton_tangent}
\end{equation}
with $H^{(k)}=\max\bigl(H^n,\Psi_0(\bm{u}^{(k)})\bigr)$. In terms of the three tangent choices, Case~1 retains only the diagonal blocks, computes them approximately (Section~\ref{sec:residual_tangent}), and reassembles them at every iteration. This inexact Newton method avoids the indefinite coupled tangent but loses quadratic convergence. Because the residual is evaluated at the current iterate, the converged solution satisfies the fully coupled equations, and Case~1 is used as the reference.

\subsubsection{Case~2: Quasi-Newton (BFGS) Monolithic Scheme}
\label{sec:impl:monolithic}

Case~2 uses the same residual as Case~1 but replaces the reassembled tangent by a BFGS approximation~\citep{Nocedal2006,Wu2020,Kristensen2020}. Instead of rebuilding the tangent at every iteration, BFGS keeps the previous tangent and corrects it using how much the solution and the residual changed over the last iteration. Starting from $\tilde{\bm{J}}^{(0)}$, the block-diagonal tangent of Eq.~\eqref{eq:newton_tangent}, the tangent and residual take the form
\begin{equation}
  \tilde{\bm{J}}^{(k)}=
  \begin{bmatrix}
    \tilde{\bm{J}}_{uu} & \tilde{\bm{J}}_{ud} & \tilde{\bm{J}}_{uc} \\
    \tilde{\bm{J}}_{ud}^T & \tilde{\bm{J}}_{dd} & \tilde{\bm{J}}_{dc} \\
    \tilde{\bm{J}}_{uc}^T & \tilde{\bm{J}}_{dc}^T & \tilde{\bm{J}}_{cc}
  \end{bmatrix}^{(k)},
  \qquad
  \bm{R}^{(k)}\ \text{as in Eq.~\eqref{eq:newton_tangent}},
  \label{eq:bfgs_tangent}
\end{equation}
where the off-diagonal blocks are zero at $k=0$ and are built up by the update in Eq.~\eqref{eq:bfgs_update}. Each iteration computes:
\begin{equation}
  \tilde{\bm{J}}^{(k)}\bm{p}^{(k)}=-\bm{R}^{(k)},\qquad
  \bm{x}^{(k+1)}=\bm{x}^{(k)}+\alpha^{(k)}\bm{p}^{(k)},
  \label{eq:bfgs_step}
\end{equation}
with step length $\alpha^{(k)}$ from a line search. The tangent is then updated with $\bm{s}^{(k)}=\bm{x}^{(k+1)}-\bm{x}^{(k)}$ and $\bm{y}^{(k)}=\bm{R}^{(k+1)}-\bm{R}^{(k)}$:
\begin{equation}
    \tilde{\bm{J}}^{(k+1)}
    =
    \tilde{\bm{J}}^{(k)}
    -
    \frac{
      \tilde{\bm{J}}^{(k)}\bm{s}^{(k)}
      (\bm{s}^{(k)})^{T}
      \tilde{\bm{J}}^{(k)}
    }{
      (\bm{s}^{(k)})^{T}
      \tilde{\bm{J}}^{(k)}\bm{s}^{(k)}
    }
    +
    \frac{
      \bm{y}^{(k)}(\bm{y}^{(k)})^{T}
    }{
      (\bm{y}^{(k)})^{T}\bm{s}^{(k)}
    },
    \label{eq:bfgs_update}
\end{equation}
which enforces the secant condition $\tilde{\bm{J}}^{(k+1)}\bm{s}^{(k)}=\bm{y}^{(k)}$. The tangent is reassembled from Eq.~\eqref{eq:newton_tangent} after every eight updates. Because $\bm{y}^{(k)}$ is the change in the full coupled residual, it includes the response of each field equation to changes in the other fields. Through the secant condition, the updated tangent therefore acquires an approximation of the off-diagonal coupling along the search directions, even though $\tilde{\bm{J}}^{(0)}$ is block-diagonal. The BFGS update is designed for symmetric tangents~\citep{Nocedal2006}, and Case~2 uses symmetric matrix storage. As a result, the solver works with the symmetric part of the tangent, including $\tfrac{1}{2}(\bm{J}_{cc}+\bm{J}_{cc}^T)$ in $\tilde{\bm{J}}^{(0)}$. The coupled Jacobian is not symmetric, because its off-diagonal blocks arise from different mechanisms, such as chemical softening in $\bm{J}_{uc}$ and the stress-assisted flux in $\bm{J}_{cu}$. Case~2 can therefore approximate only the symmetric part of the coupling. This does not change the converged solution, because the residual is the full coupled residual of Eq.~\eqref{eq:newton_tangent}. It affects only the search directions and hence the convergence rate (Section~\ref{sec:tensile:transport}).

\subsubsection{Case~3: Single-Pass Fully Staggered Scheme}
\label{sec:impl:fullystag}

Case~3 uses the block-diagonal tangent of Case~1, but lags the coupling inputs to the last converged increment:
\begin{equation}
  \tilde{\bm{J}}^{(k)}=
  \begin{bmatrix}
    \bm{J}_{uu} & \bm{0} & \bm{0} \\
    \bm{0} & \bm{J}_{dd} & \bm{0} \\
    \bm{0} & \bm{0} & \bm{J}_{cc}
  \end{bmatrix}^{(k)},
  \qquad
  \bm{R}^{(k)}=
  \begin{Bmatrix}
    \bm{R}_u\bigl(\bm{u}^{(k)};\,d^n,\,d_{\mathrm{chem}}^n\bigr)\\
    \bm{R}_d\bigl(d^{(k)};\,H^{(k)},\,d_{\mathrm{chem}}^n\bigr)\\
    \bm{R}_c\bigl(c^{(k)};\,D_{\mathrm{eff}}^n,\,\nabla\sigma_H^n\bigr)
  \end{Bmatrix},
  \label{eq:case3}
\end{equation}
with $H^{(k)}=\max\bigl(H^n,\Psi_0(\bm{u}^{(k)})\bigr)$.Because the inputs are lagged, the three residuals are solved as independent sub-problems for $\bm{u}$, $d$ and $c$ in turn.
Each sub-problem is linear in its own field, with tangent $\bm{J}_{uu}$, $\bm{J}_{dd}$, or $\bm{J}_{cc}$. The sequence is not repeated within the increment, so the accuracy of Case~3 depends on the time step (Section~\ref{sec:tensile}).

\subsubsection{Case~4: Single-Pass Partially Staggered Scheme}
\label{sec:impl:partialstag}

A partially staggered scheme groups two fields and solves the third separately. The three possible partitions compare as follows:
\begin{itemize}[noitemsep]
  \item $[d,c]+\bm{u}$ gains little, because $d$ and $c$ interact mainly through the displacement-dependent stress, so the grouped block is nearly block-diagonal.
  \item $[\bm{u},d]+c$ groups the two fields that make the energy non-convex, and its tangent becomes indefinite after the onset of softening~\citep{JinSvolos2026}.
  \item $[\bm{u},c]+d$ retains the chemo-mechanical blocks $\bm{J}_{uc}$ and $\bm{J}_{cu}$ and lags the non-convex displacement--phase-field coupling.
\end{itemize}
Therefore, the last partition is adopted. Within each increment, two sub-problems are solved once, in sequence:
\begin{subequations}\label{eq:case4}
\begin{align}
  &\text{I:}\quad
  \begin{bmatrix}
    \bm{J}_{uu} & \bm{J}_{uc} \\
    \bm{J}_{cu} & \bm{J}_{cc}
  \end{bmatrix}^{(k)}
  \begin{Bmatrix}
    \Delta\bm{u} \\ \Delta\bm{c}
  \end{Bmatrix}^{(k)}
  =-
  \begin{Bmatrix}
    \bm{R}_u\bigl(\bm{u}^{(k)};\,d^n,\,d_{\mathrm{chem}}^{(k)}\bigr) \\
    \bm{R}_c\bigl(c^{(k)};\,D_{\mathrm{eff}}(d^n,\Phi^{(k)}),\,\nabla\sigma_H^{(k)}\bigr)
  \end{Bmatrix}
  &&\rightarrow\ \bm{u}^{n+1},\ c^{n+1},
  \label{eq:ps4_tangent}\\
  &\text{II:}\quad \bm{R}_d\bigl(d^{n+1};\,H^{n+1},\,d_{\mathrm{chem}}^{n+1}\bigr)=\bm{0},
  \quad H^{n+1}=\max\bigl(H^n,\Psi_0(\bm{u}^{n+1})\bigr)
  &&\rightarrow\ d^{n+1},
  \label{eq:ps4_II}
\end{align}
\end{subequations}
where $\bm{J}_{uc}$ and $\bm{J}_{cu}$ are given by Eqs.~\eqref{eq:Kuc} and~\eqref{eq:Kcu}, evaluated with $d=d^n$. Sub-problem~I is nonlinear, because $d_{\mathrm{chem}}$ and $D_{\mathrm{eff}}$ depend on $c$, and is iterated to convergence. Sub-problem~II is linear in $d$, with the positive-definite tangent $\bm{J}_{dd}$. Unlike in Case~3, the chemo-mechanical coupling between $\bm{u}$ and $c$ is current, and only the phase field is lagged. The sequence is not repeated within the increment, so the scheme is single-pass.

\subsubsection{Case~5: Hybrid Multi-pass Fully Staggered Scheme}
\label{sec:impl:multipass}

Case~5 uses separate mechanical, transport, and phase-field elements that share the same nodes. Each supplies only the tangent of its own field, so $\tilde{\bm{J}}$ is the block-diagonal matrix of Eq.~\eqref{eq:newton_tangent}. At every iteration $k$, the elements are evaluated in this order:
\begin{subequations}\label{eq:case5}
\begin{align}
  &\bm{R}_u^{(k)}=\bm{R}_u\bigl(\bm{u}^{(k)};\,d^n,\,d_{\mathrm{chem}}^{(k-1)}\bigr)
  &&\rightarrow\ \Psi_0^{(k)},\ \nabla\sigma_H^{(k)},
  \label{eq:case5_u}\\
  &\bm{R}_c^{(k)}=\bm{R}_c\bigl(c^{(k)};\,D_{\mathrm{eff}}^{(k)},\,\nabla\sigma_H^{(k)}\bigr)
  &&\rightarrow\ d_{\mathrm{chem}}^{(k)},
  \label{eq:case5_c}\\
  &\bm{R}_d^{(k)}=\bm{R}_d\bigl(d^{(k)};\,H^{(1)},\,d_{\mathrm{chem}}^{(k)}\bigr),
  &&H^{(1)}=\max\bigl(H^n,\Psi_0^{(1)}\bigr),
  \label{eq:case5_d}
\end{align}
\end{subequations}
where the arrows indicate the quantities passed to the elements evaluated next. At the first iteration, $d_{\mathrm{chem}}^n$ is used in $\bm{R}_u$, and $H^{(1)}$ is the history field from that iteration. Before evaluating $\bm{R}_c$, the transport element computes $\Phi^{(k)}$, $d_{\mathrm{chem}}^{(k)}$, and $D_{\mathrm{eff}}^{(k)}=D_{\mathrm{eff}}(d^{(k)},\Phi^{(k)})$ from $c^{(k)}$. These chemical quantities are committed only when the increment converges. The scheme is multi-pass because every iteration repeats the ordered evaluation with the latest shared quantities. It is hybrid because the chemical coupling is refreshed at every pass, whereas the phase field in $\bm{R}_u$ stays at $d^n$ and $H$ is frozen after its first evaluation.

\subsection{Comparison of the Solution Strategies}
\label{sec:impl:summary}

Table~\ref{tab:solver_summary} summarizes the five strategies. They share the interpolation, integration rule, constitutive model, and time integration, and differ as follows:
\begin{itemize}[noitemsep]
  \item \textbf{Element implementation.} Cases~1--4 use one three-field element per mesh cell. Case~5 uses three single-field elements per cell, so each iteration involves more element evaluations.
  \item \textbf{Tangent.}  Retained blocks: diagonal only in Cases~1, 3, and~5 and in the initial tangent of Case~2, with $\bm{J}_{uc}$ and $\bm{J}_{cu}$ added in Case~4. Accuracy: approximate diagonal blocks in every case, replaced by a secant (BFGS) approximation in Case~2. Update frequency: every iteration, except Case~2, which reassembles the tangent after every eight BFGS updates.
  \item \textbf{Coupling update.} Cases~1 and~2 evaluate all coupling inputs at the current iterate. Case~3 lags all of them except $H$. Case~4 lags only the phase field. Case~5 refreshes the chemical coupling at every iteration but keeps $d^n$ in $\bm{R}_u$ and freezes $H$.
  \item \textbf{Inter-field convergence.} Cases~1 and~2 converge the coupled residual. Cases~3 and~4 are single-pass, so their accuracy depends on the time step. Cases~3 and~5 use the same three-field partition, and their main difference is how often they exchange coupling information: once per increment in Case~3 and at every iteration in Case~5. 
  \item \textbf{History field.} $H$ is evaluated from the current iterate $\bm{u}^{(k)}$ in Cases~1 and~2, from $\bm{u}^{n+1}$ in Cases~3 and~4, and frozen at $H^{(1)}$ in Case~5.
\end{itemize}

\begin{table}[htbp]
  \centering
  \caption{Summary of the five solution strategies. ``Current'' denotes evaluation at the current iterate, ``lagged'' evaluation at the last converged increment, and ``mixed'' the partially updated treatment of Case~5 (Section~\ref{sec:impl:multipass}).}
  \label{tab:solver_summary}
  \small
  \setlength{\tabcolsep}{4pt}
  \renewcommand{\arraystretch}{1.25}
  \begin{tabular}{@{}
    >{\centering\arraybackslash}m{0.045\textwidth}
    >{\centering\arraybackslash}m{0.17\textwidth}
    >{\centering\arraybackslash}m{0.08\textwidth}
    >{\centering\arraybackslash}m{0.14\textwidth}
    >{\centering\arraybackslash}m{0.07\textwidth}
    >{\centering\arraybackslash}m{0.15\textwidth}
    >{\raggedright\arraybackslash}m{0.25\textwidth}@{}}
    \toprule
    \multirow{2}{0.045\textwidth}{\centering\textbf{Case}} &
    \multirow{2}{0.17\textwidth}{\centering\textbf{Scheme}} &
    \multirow{2}{0.08\textwidth}{\centering\textbf{Field partitions}} &
    \multirow{2}{0.14\textwidth}{\centering\textbf{Grouped fields}} &
    \multicolumn{2}{c}{\textbf{Update timing}} &
    \multirow{2}{0.25\textwidth}{\centering\textbf{Tangent}} \\
    \cmidrule(lr){5-6}
    & & & & $d$ & $d_{\mathrm{chem}},\,D_{\mathrm{eff}},\,\sigma_H$ & \\
    \midrule
    1 & Block-diagonal inexact-Newton monolithic  & 1 & $[\bm{u},\,d,\,c]$ & Current & Current &
    Block-diagonal, reassembled every iteration \\
    2 & Quasi-Newton (BFGS) monolithic & 1 & $[\bm{u},\,d,\,c]$ & Current & Current &
    Block-diagonal initial tangent with BFGS updates \\
    3 & Single-pass fully staggered & 3 & $\bm{u}\,|\,d\,|\,c$ & Lagged & Lagged &
    Block-diagonal, one pass per increment \\
    4 & Single-pass partially staggered & 2 & $[\bm{u},\,c]+d$ & Lagged & Current &
    Coupled $[\bm{u},c]$ block with $\bm{J}_{uc}$, $\bm{J}_{cu}$ \\
    5 & Hybrid multi-pass fully staggered & 3 & $\bm{u}\,|\,c\,|\,d$ & Mixed & Mixed &
    Block-diagonal, three single-field elements, ordered evaluation every iteration \\
    \bottomrule
  \end{tabular}
\end{table}

\subsection{Evaluation Framework}
\label{sec:evaluation_framework}

The strategies in Section~\ref{sec:impl:summary} differ in their tangent, in how often they update the coupling, and in whether they converge the coupled residual. Each of these choices affects accuracy, cost, and robustness in a different way, so they are compared with a common set of metrics. These are summarized in Table~\ref{tab:evaluation_metrics}. Accuracy is compared at common physical states and cost at a common physical endpoint. Once its spatial and temporal discretizations have been verified, Case~1 serves as the reference for all comparisons. In what follows, the superscript $(s)$, $s\in\{1,\ldots,5\}$, denotes the case number.

\paragraph{Solution accuracy} For a scalar response quantity $Q\in\{F_{\max},\delta_{\mathrm{peak}},\delta_f\}$ (the peak load, the displacement at peak load, and the failure displacement), the error of Case~$s$ is:

\begin{equation}
\varepsilon_Q^{(s)}
=
100\,
\frac{\left|Q^{(s)}-Q^{(1)}\right|}
{\left|Q^{(1)}\right|},
\label{eq:scalar_accuracy_metric}
\end{equation}
where $\delta_f$ is the first post-peak displacement at which $F\leq0.05F_{\max}$. In the chemo-mechanical benchmarks, the local concentration error at a common prescribed displacement is:
\begin{equation}
\varepsilon_{c,\mathrm{tip}}^{(s)}
=
100\,
\frac{\left\|\bm{c}_{\mathrm{tip}}^{(s)}-\bm{c}_{\mathrm{tip}}^{(1)}\right\|_2}
{\left\|\bm{c}_{\mathrm{tip}}^{(1)}\right\|_2},
\label{eq:notch_tip_concentration_accuracy}
\end{equation}
where $\bm{c}_{\mathrm{tip}}$ collects the concentrations in the two elements adjacent to the pre-notch tip along the expected initial crack path. These elements lie directly ahead of the tip in the tensile and multiple-crack benchmarks, and directly below it in the shear benchmark. This region is sensitive to solver-dependent differences because the concentration prescribed on the crack faces produces steep gradients there, where stress-assisted transport and crack initiation interact.

\paragraph{Computational performance} Cost is measured by the total CPU time $t_{\mathrm{CPU}}$, the wall-clock time $t_{\mathrm{wall}}$, the total number of equilibrium iterations $N_{\mathrm{iter}}$, and the speed-up relative to Case~1:
\begin{equation}
S_{\mathrm{CPU}}^{(s)}
=
\frac{t_{\mathrm{CPU}}^{(1)}}
{t_{\mathrm{CPU}}^{(s)}},
\qquad
N_{\mathrm{iter}}^{(s)}
=
\sum_{i=1}^{M^{(s)}} n_i^{(s)},
\label{eq:cpu_speedup}
\end{equation}
where $n_i^{(s)}$ is the number of equilibrium iterations in accepted increment $i$ and $M^{(s)}$ is the number of accepted increments up to the endpoint. $S_{\mathrm{CPU}}^{(s)}>1$ indicates a lower CPU cost than Case~1.

\paragraph{Convergence robustness} Robustness is quantified by the iteration-burst factors $B_{95}$ and $B_{\max}$ and the completion ratio $R_{\mathrm{comp}}$. The burst factors are:
\begin{equation}
B_{95}^{(s)}
=
\frac{n_{95}^{(s)}}{\bar n^{(s)}},
\qquad
B_{\max}^{(s)}
=
\frac{n_{\max}^{(s)}}{\bar n^{(s)}},
\label{eq:iteration_burst_factors}
\end{equation}
where $\bar n^{(s)}$, $n_{95}^{(s)}$, and $n_{\max}^{(s)}$ are the mean, 95th-percentile, and maximum numbers of equilibrium iterations per accepted increment.They measure how strongly the iteration demand is concentrated in localized spikes, independently of the average cost per increment. For a prescribed set of simulations, the completion ratio is:
\begin{equation}
R_{\mathrm{comp}}^{(s)}
=
\frac{N_{\mathrm{completed}}^{(s)}}
{N_{\mathrm{run}}^{(s)}},
\label{eq:completion_ratio}
\end{equation}
where a simulation is complete only if it reaches its prescribed endpoint.

Within-increment residual histories provide an additional diagnostic. The maximum residuals of the displacement--phase-field equations and of the concentration equation are normalized by the convergence criterion:
\begin{equation}
\widehat r_{u,d}^{(k)}
=
\frac{\left|r_{u,d}^{\max,(k)}\right|}
{R_{\mathrm{tol}}\,\bar f_{u,d}^{(k)}},
\qquad
\widehat r_c^{(k)}
=
\frac{\left|r_c^{\max,(k)}\right|}
{R_{\mathrm{tol}}\,\bar f_c^{(k)}},
\qquad
R_{\mathrm{tol}}=5\times10^{-3},
\label{eq:criterion_normalised_residual}
\end{equation}
where $k$ is the equilibrium iteration and $\bar f_{u,d}^{(k)}$ and $\bar f_c^{(k)}$ are the time-averaged reference magnitudes of the corresponding residuals (\ref{app:abaqus}). Values below unity satisfy the residual criterion. For an increment with $N>1$ iterations, the histories are plotted against the normalized iteration progress $\xi_k=(k-1)/(N-1)$, which aligns the first and last checks across strategies.

\paragraph{Transport intensity} The diffusion over one increment relative to the element size is measured by the local Fourier number~\citep{Bergman2011}, the square of the ratio of the diffusion length $\sqrt{D_{\mathrm{eff}}\Delta t}$ to the element size:
\begin{equation}
\mathrm{Fo}_{q,i}
=
\frac{D_{\mathrm{eff},q,i}\,\Delta t_i}{h_q^2},
\qquad
\mathrm{Fo}_{95}
=
\operatorname{P}_{95}
\bigl(\{\mathrm{Fo}_{q,i}\}_{q,i}\bigr),
\label{eq:fourier_number}
\end{equation}
where $\operatorname{P}_{95}$ denotes the 95th percentile, $q$ is a material point, $i$ an accepted increment, and $h_q$ the size of the element containing $q$. Each simulation is characterized by $\mathrm{Fo}_{95}$, introduced here as a measure of transport intensity rather than as a universal convergence threshold.
\begin{table}[htbp]
\centering
\caption{Metrics used to compare the solution strategies.}
\label{tab:evaluation_metrics}
\small
\begin{tabularx}{\textwidth}{
@{}
>{\raggedright\arraybackslash}p{0.12\textwidth}
@{\hspace{1em}}
>{\raggedright\arraybackslash}X
@{\hspace{2em}}
>{\raggedright\arraybackslash}p{0.19\textwidth}
@{}}
\toprule
Symbol & Definition & Purpose \\
\midrule

\(\varepsilon_{F_{\max}}\)
& Error in peak load relative to Case~1

& \multirow{4}{0.20\textwidth}{Solution accuracy} \\

\(\varepsilon_{\delta_{\mathrm{peak}}}\)
& Error in displacement at peak load relative to Case~1
& \\

\(\varepsilon_{\delta_f}\)
& Error in failure displacement ($F\leq0.05F_{\max}$) relative to Case~1

& \\

\(\varepsilon_{c,\mathrm{tip}}\)
& Concentration error in the two elements at the pre-notch tip
& \\

\midrule

\(t_{\mathrm{CPU}}\), \(t_{\mathrm{wall}}\)
& CPU and wall-clock times to the endpoint or termination
& \multirow{3}{0.20\textwidth}{Computational performance} \\

\(S_{\mathrm{CPU}}\)
& Case~1 CPU time divided by that of Case~$s$
& \\

\(N_{\mathrm{iter}}\)
& Total equilibrium iterations over all accepted increments
& \\

\midrule

\(R_{\mathrm{comp}}\)
& Fraction of simulations that reach their endpoint
& \multirow{3}{0.20\textwidth}{Convergence robustness} \\

\(B_{95}\)
& 95th-percentile iterations per increment divided by the mean
& \\

\(B_{\max}\)
& Maximum iterations per increment divided by the mean
& \\

\midrule

\(\mathrm{Fo}_{95}\)
&  95th percentile of the local Fourier number
& Transport intensity \\

\bottomrule
\end{tabularx}
\end{table}

All simulations were run one at a time on a workstation with an Intel Xeon W-2223 processor and 64~GB of RAM, using four cores and a 72-hour wall-clock limit per simulation. Incomplete simulations are reported at their last converged increment and are included in the robustness assessment, with their cost measured up to termination.

\section{Numerical Benchmarks}
\label{sec:benchmarks}
This section first verifies the phase-field implementation on Miehe's benchmark. The strategies are then compared on the tensile, shear, and multiple-crack benchmarks using the metrics of Section~\ref{sec:evaluation_framework}, and the section closes with a cross-benchmark assessment.

\subsection{Verification: Miehe's Benchmark}
\label{sec:miehe}

\begin{figure}[ht]
  \centering
  \begin{subfigure}[b]{0.44\textwidth}
    \centering
    \includegraphics[width=\textwidth]{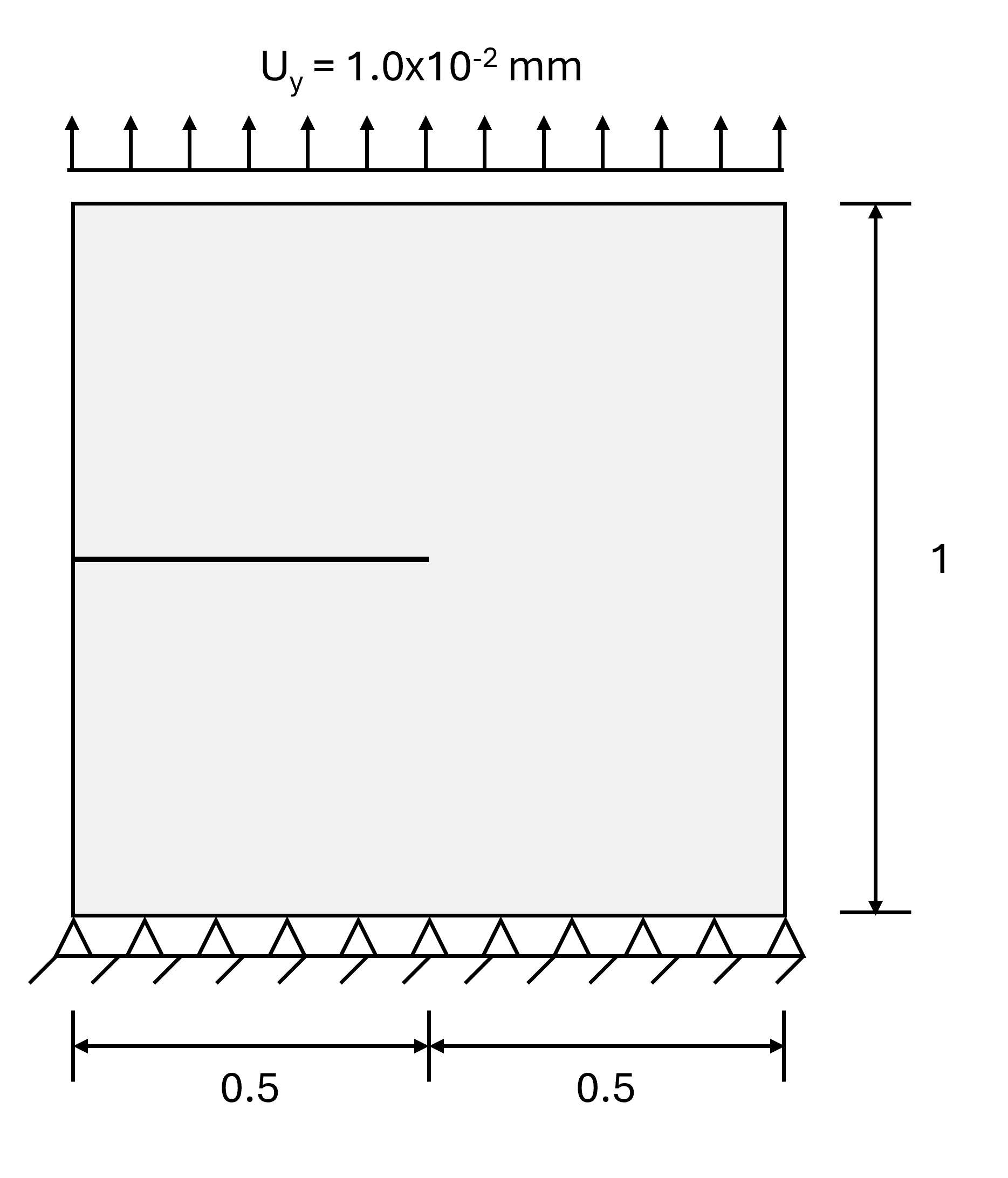}
    \caption{}
    \label{fig:miehe_geom}
  \end{subfigure}
  \hfill
  \begin{subfigure}[b]{0.55\textwidth}
    \centering
    \includegraphics[width=\textwidth]{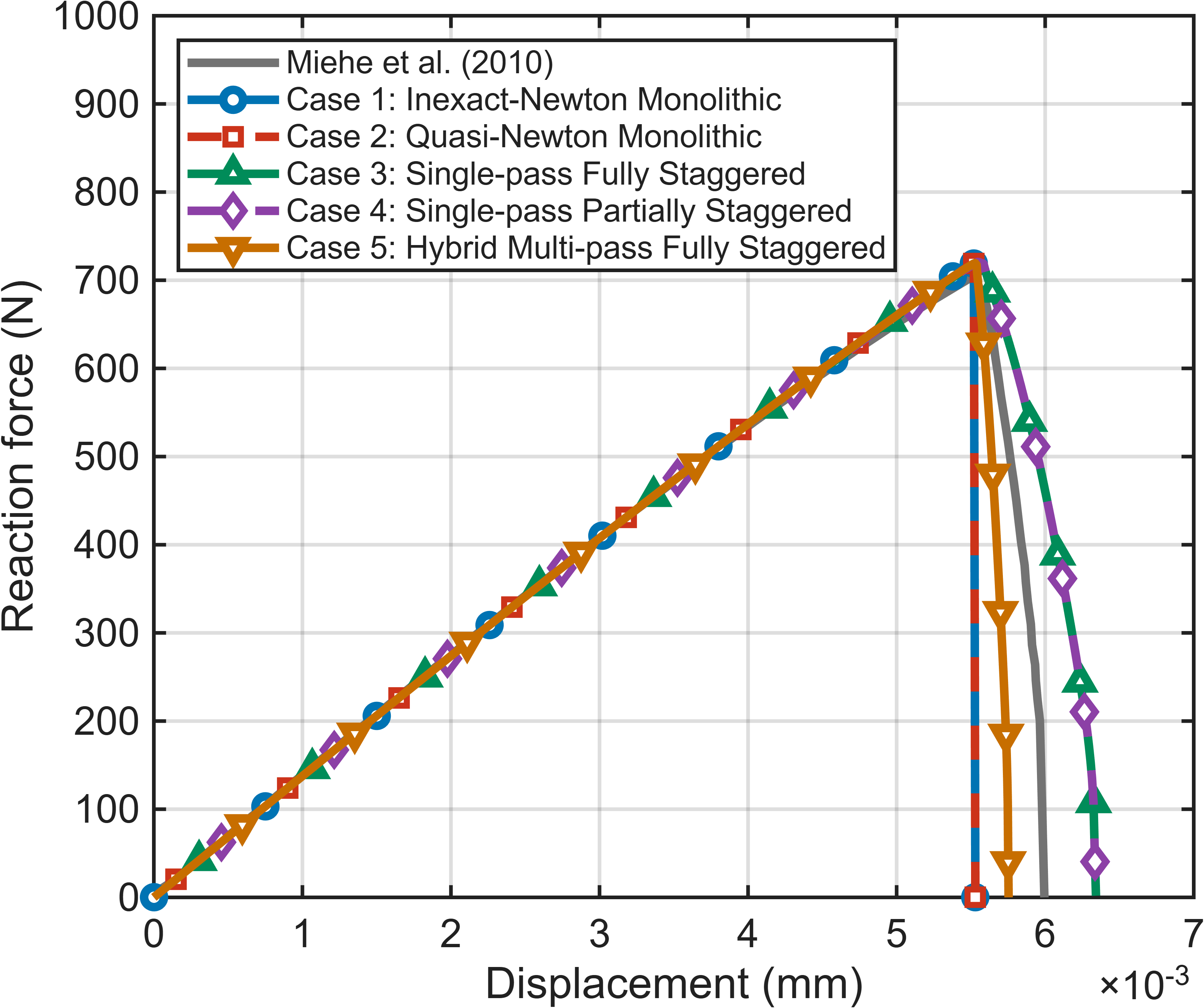}
    \caption{}
    \label{fig:miehe_fd}
  \end{subfigure}
  \caption{Miehe's single-edge-notched tension benchmark: (a)~geometry and boundary conditions (dimensions in millimeters) and (b)~force--displacement  responses of Cases~1--5 compared with~\citet{MieheHofacker2010}.}
  \label{fig:miehe}
\end{figure}

The phase-field implementation is first verified against the single-edge-notched tension test of~\citet{MieheHofacker2010}, with chemistry inactive. The $1\times1~\mathrm{mm}^2$ plate contains a horizontal crack from the left edge to its center (\Cref{fig:miehe}a) and is loaded by a vertical displacement $u_y$ on the top edge. The material parameters (Table~\ref{tab:miehe_params}) and the plane-strain AT2 model with $g(d)=(1-d)^2$ and no viscous regularization follow~\citet{MieheHofacker2010}. Displacement increments of $10^{-5}$~mm are applied for the first 500 increments and $10^{-6}$~mm thereafter. In this and every subsequent benchmark, all five cases use the same load and time stepping, except for the adaptive Case~4 analyses identified in Sections~\ref{sec:tensile} and~\ref{sec:multicracks}. The mesh is refined along the crack path to $h=10^{-3}$~mm ($l_0/h=7.5$). The model departs from the original benchmark in two respects: Q8 elements are used, and no tension--compression split is applied.

\begin{table}[h]
  \centering
  \caption{Material and numerical parameters for Miehe's single-edge-notched tension benchmark.}
  \label{tab:miehe_params}
  \begin{tabular}{ccccc}
    \toprule
    $E$ (MPa) & $\nu$ & $G_c$ (N/mm) & $l_0$ (mm) & $h$ (mm) \\
    \midrule
    $210\times10^{3}$ & 0.3 & 2.7 & 0.0075 & $1\times10^{-3}$ \\
    \bottomrule
  \end{tabular}
\end{table}

All five strategies reproduce the elastic response of the reference solution and its peak load of about 720~N (\Cref{fig:miehe}b). This agreement verifies the phase-field implementation. They differ only after the peak. The monolithic Cases~1 and~2 update all fields within the increment and give an abrupt load drop. The single-pass Cases~3 and~4 lag the damage, which spreads the release of elastic energy over several increments and gives a more gradual drop. Case~5 lies between the two groups. This post-peak dependence on the coupling update recurs in all subsequent benchmarks.
\subsection{Mode-I Tensile Test}
\label{sec:tensile}

\subsubsection{Problem Description}
\label{sec:tensile:setup}

\begin{figure}[htbp]
  \centering
  \includegraphics[width=0.9\textwidth]{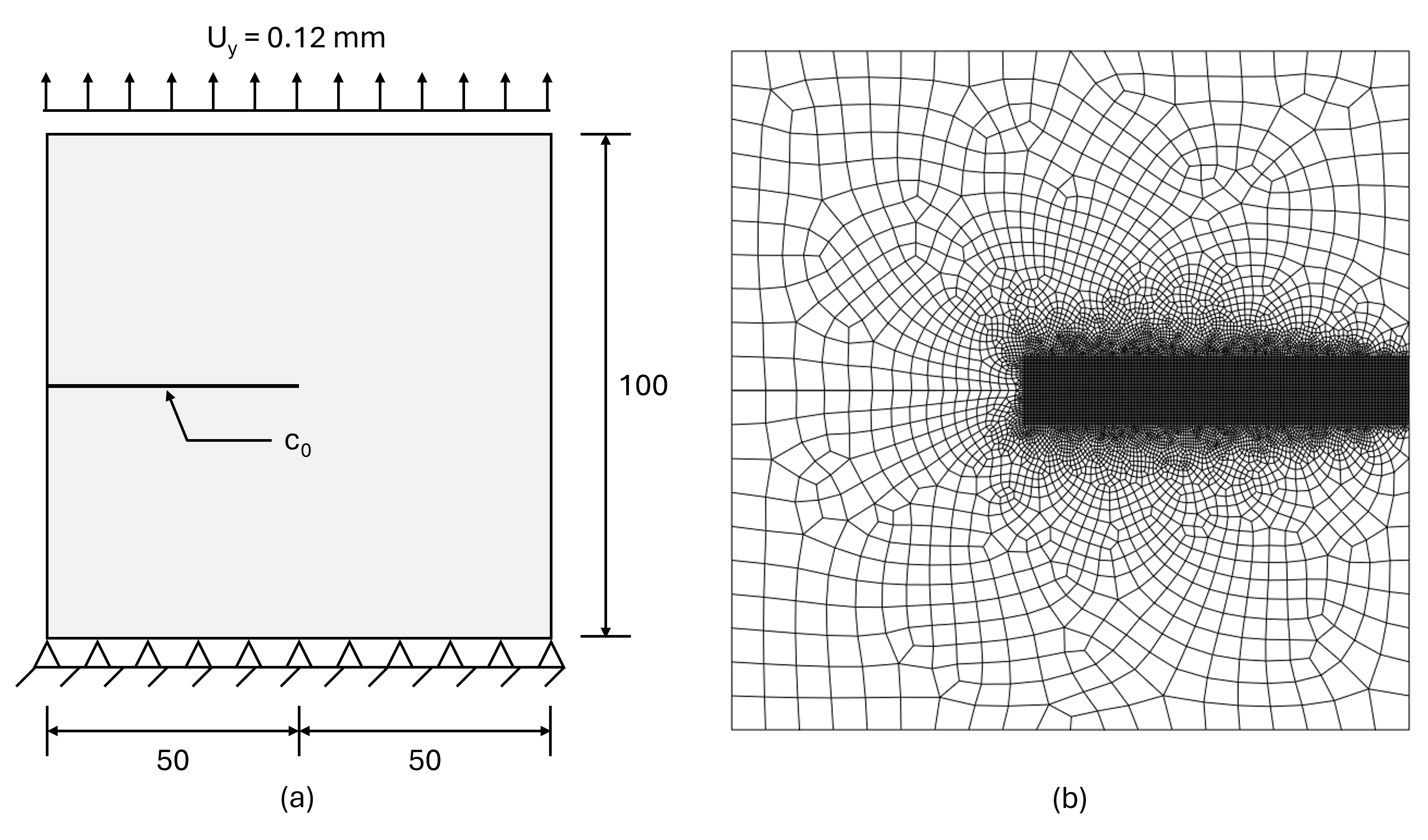}
  \caption{Mode-I single-edge-notched tensile benchmark: (a) geometry and boundary conditions and (b) representative locally refined mesh. Dimensions are in millimeters, and $c_0=0$ in the mechanical simulations.}
  \label{fig:tensile_geom}
\end{figure}

The single-edge-notched tension test is used to select the spatial and temporal discretizations and to compare the five cases under controlled loading. A $100\times100$~mm square contains a 50~mm horizontal pre-notch, modeled as a geometric slit, extending from the left edge to the center at mid-height (\Cref{fig:tensile_geom}a). The bottom edge is fixed, and a vertical displacement $U_y$ is applied to the top edge under plane strain. Mechanical simulations assess spatial convergence, and chemo-mechanical simulations assess temporal sensitivity, cost, and robustness. Both use the sandstone properties of Table~\ref{tab:tensile_props}.

\begin{table}[htbp]
  \centering
  \caption{Mechanical, phase-field, transport, and chemical parameters shared by the Mode-I tensile, single-edge-notched shear, and multiple-crack benchmarks.}
  \label{tab:tensile_props}
  \small
  \renewcommand{\arraystretch}{1.0}
  \begin{tabular}{@{} l c l l @{}}
    \toprule
    \textbf{Property} & \textbf{Symbol} & \textbf{Value} & \textbf{Ref.} \\
    \midrule
    \multicolumn{4}{@{}l}{\textit{Mechanical}} \\[2pt]
    Young's modulus & $E$ & 20\,000~MPa & \citep{Schuler2020} \\
    Poisson's ratio & $\nu$ & 0.2 & \citep{Schuler2020} \\
    Fracture energy & $G_c$ & 0.35~N/mm & \citep{Schuler2020} \\
    Phase-field length scale & $l_0$ & 1.0~mm & \citep{Schuler2020} \\
    Residual stiffness & $\kappa$ & $10^{-6}$ & \\
    \midrule
    \multicolumn{4}{@{}l}{\textit{Transport}} \\[2pt]
    Diffusion coefficient (undamaged) & $D_p$ & $1.25\times10^{-4}$~mm$^2$/s & \citep{Schuler2020} \\
    Diffusion coefficient in water & $D_0$ & $2\times10^{-3}$~mm$^2$/s & \citep{NavarreeSitchler2009} \\
    Damage-dependent diffusivity threshold & $d_{cr}$ & 0.4 & \citep{Gerard1998} \\
    Damage-dependent diffusivity exponent & $n_D$ & 5 & \citep{Gerard1998} \\
    Partial molar volume of CO$_2$ & $\bar{V}$ & 44\,000~mm$^3$/mol & \citep{Schuler2020} \\
    Temperature & $T$ & 363~K & \citep{Schuler2020} \\
    Gas constant & $R$ & 8314~N$\cdot$mm/(mol$\cdot$K) & \\
    \midrule
    \multicolumn{4}{@{}l}{\textit{Chemistry and porosity}} \\[2pt]
    Percolation threshold & $\Phi_c$ & 0.01 & \citep{Schuler2020} \\
    Initial calcite content & $n_0$ & $1.8\times10^{-6}$~mol/mm$^3$ & \citep{Schuler2020} \\
    Porosity connectivity parameters & $a$, $\beta$ & 2.4, 0.9 & \citep{NavarreeSitchler2009} \\
    Initial reactive surface area & $A_0$ & $2.71$~mm$^2$/mm$^3$ & \citep{Schuler2020} \\
    Molar volume of calcite & $M/\rho$ & 35\,000~mm$^3$/mol & \citep{Schuler2020} \\
    Chemical damage coefficient & $r$ & 10 & \citep{Schuler2020} \\
    Reactive surface area parameters & $\mathcal{C}$, $a_s$, $b_s$ & 37, 1.7, 0.69 & \citep{Khinast1996} \\
    \midrule
    \multicolumn{4}{@{}l}{\textit{pH model}} \\[2pt]
    Henry's law constant & $K_h$ & $1.73\times10^{-3}$ & \citep{Schuler2020} \\
    First dissociation constant & $K_{a1}$ & $2.5\times10^{-4}$~mol/L & \citep{Schuler2020} \\
    \bottomrule
  \end{tabular}
\end{table}

\subsubsection{Spatial Discretization and Mechanical Response}
\label{sec:tensile:mech}

The mesh is refined along the expected crack path, and three densities are characterized by $l_0/h$, where $h$ is the element size in the refined region (Table~\ref{tab:tensile_mesh}). Note that Case~5 uses three single-field elements per mesh cell, so it performs three times as many element evaluations on the same mesh. A total displacement $U_y=0.12$~mm is applied in increments of $1.2\times10^{-4}$~mm for the first 500 increments and $1.2\times10^{-6}$~mm thereafter to resolve the rapid crack propagation. Chemistry is switched off ($n_0=0$ and no CO$_2$ boundary condition), so $d_{\mathrm{chem}}=0$ and $c\equiv0$ while the three-field element is retained. These runs are therefore the chemistry-free limit of the same element rather than a separate two-field formulation.

\begin{table}[htbp]
  \centering
  \caption{Q8 mesh densities used in the mechanical Mode-I tensile spatial-discretization study.}
  \label{tab:tensile_mesh}
  \small
  \setlength{\tabcolsep}{8pt}
  \begin{tabular}{@{}ccc@{}}
    \toprule
    $l_0/h$ & $h$ (mm) & UEL elements \\
    \midrule
    2 & 0.5 & 5,317 \\
    4 & 0.25 & 14,657 \\
    8 & 0.125 & 48,170 \\
    \bottomrule
  \end{tabular}
\end{table}

Table~\ref{tab:tensile_mech_results} compares the peak loads with the Richardson-extrapolated value $F_{\mathrm{ref}}=743.59$~N. For every case, the error falls with refinement, from about $2\%$ at $l_0/h=2$ to below $0.6\%$ at $l_0/h=8$. The ranking of the cases is the same on all three meshes: Cases~1 and~2 are closest to $F_{\mathrm{ref}}$, Cases~3 and~4 are furthest from it, and Case~5 lies between them. At $l_0/h=4$, the error is already below $1\%$ for every case, whereas halving the element size again raises the cost by a factor of 3.3--5.3 without changing the crack path or the damage-band width. The $l_0/h=4$ mesh is therefore adopted for all subsequent simulations as a balance between accuracy and cost.

\begin{table}[h!]
  \centering
  \caption{Accuracy and computational cost of the mechanical Mode-I tensile spatial-discretization study. Peak-load errors are relative to the Richardson-extrapolated value $F_{\mathrm{ref}}=743.59$~N. CPU and wall-clock times are in hours.}
  \label{tab:tensile_mech_results}
  \small
  \begin{tabular}{@{}ccrrrr@{}}
    \toprule
    $l_0/h$ & Case & $F_{\max}$ (N) & Error (\%) & $t_{\mathrm{CPU}}$ & $t_{\mathrm{wall}}$ \\
    \midrule
    2 & 1 & 757.85 & 1.92 & 1.25 & 0.99 \\
      & 2 & 757.81 & 1.92 & 0.52 & 0.25 \\
      & 3 & 760.42 & 2.26 & 0.70 & 0.33 \\
      & 4 & 760.42 & 2.26 & 0.78 & 0.35 \\
      & 5 & 758.64 & 2.02 & 0.63 & 0.30 \\
    \midrule
    4 & 1 & 747.93 & 0.58 & 4.28 & 2.68 \\
      & 2 & 747.92 & 0.58 & 1.64 & 0.90 \\
      & 3 & 750.55 & 0.94 & 2.22 & 1.01 \\
      & 4 & 750.55 & 0.94 & 2.64 & 1.13 \\
      & 5 & 748.70 & 0.69 & 1.57 & 0.63 \\
    \midrule
    8 & 1 & 744.91 & 0.18 & 20.94 & 11.15 \\
      & 2 & 744.91 & 0.18 & 8.69 & 3.15 \\
      & 3 & 747.68 & 0.55 & 9.14 & 3.40 \\
      & 4 & 747.68 & 0.55 & 9.28 & 4.58 \\
      & 5 & 745.68 & 0.28 & 5.16 & 2.15 \\
    \bottomrule
  \end{tabular}
\end{table}

\begin{figure}[htbp]
  \centering
  \includegraphics[width=\textwidth]{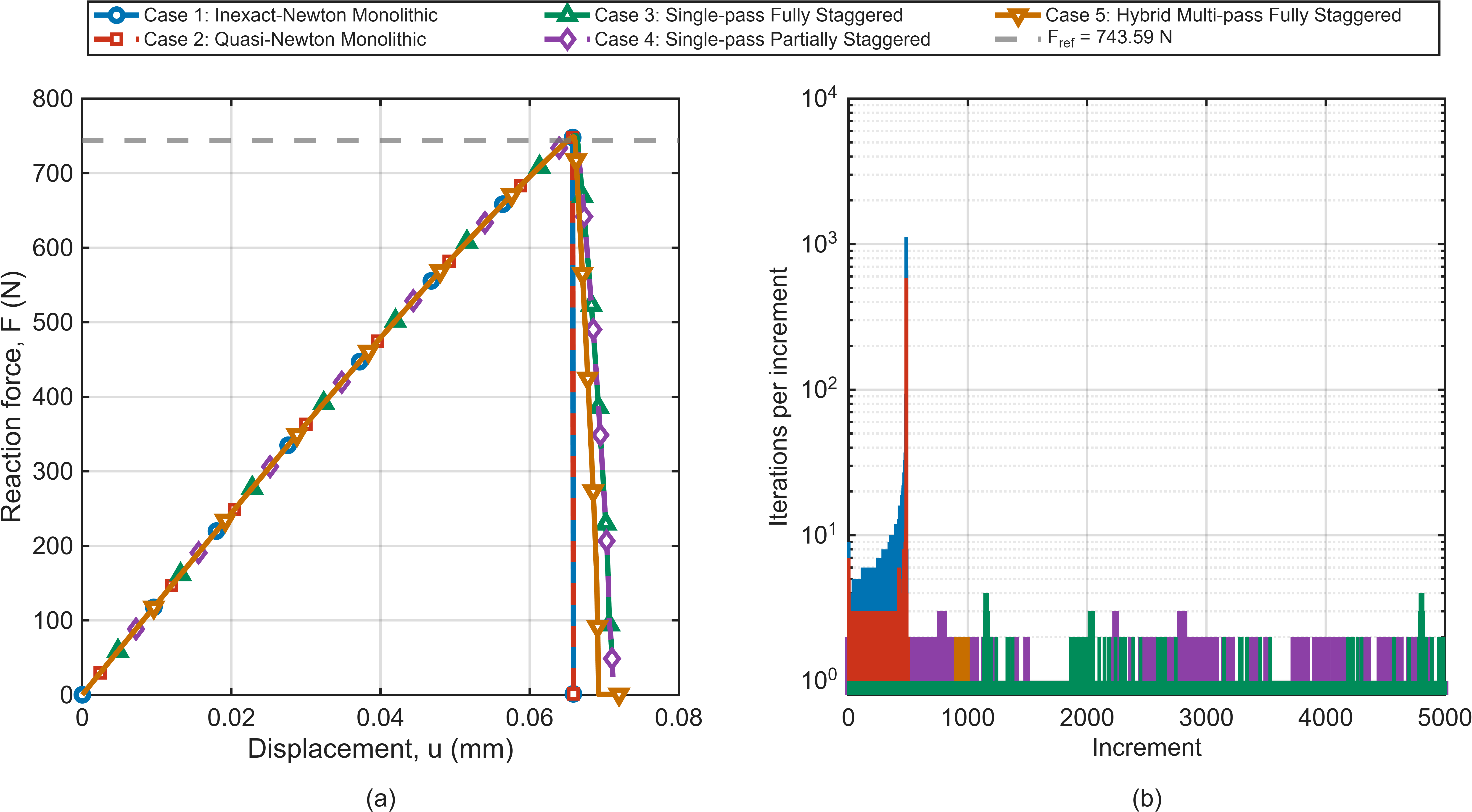}
  \caption{Mechanical Mode-I tensile benchmark on the $l_0/h=4$ mesh: (a) force--displacement responses and (b) equilibrium iterations per increment for Cases~1--5.}
  \label{fig:tensile_mech_response}
\end{figure}

\begin{figure}[htbp]
  \centering
  \includegraphics[width=0.9\textwidth]{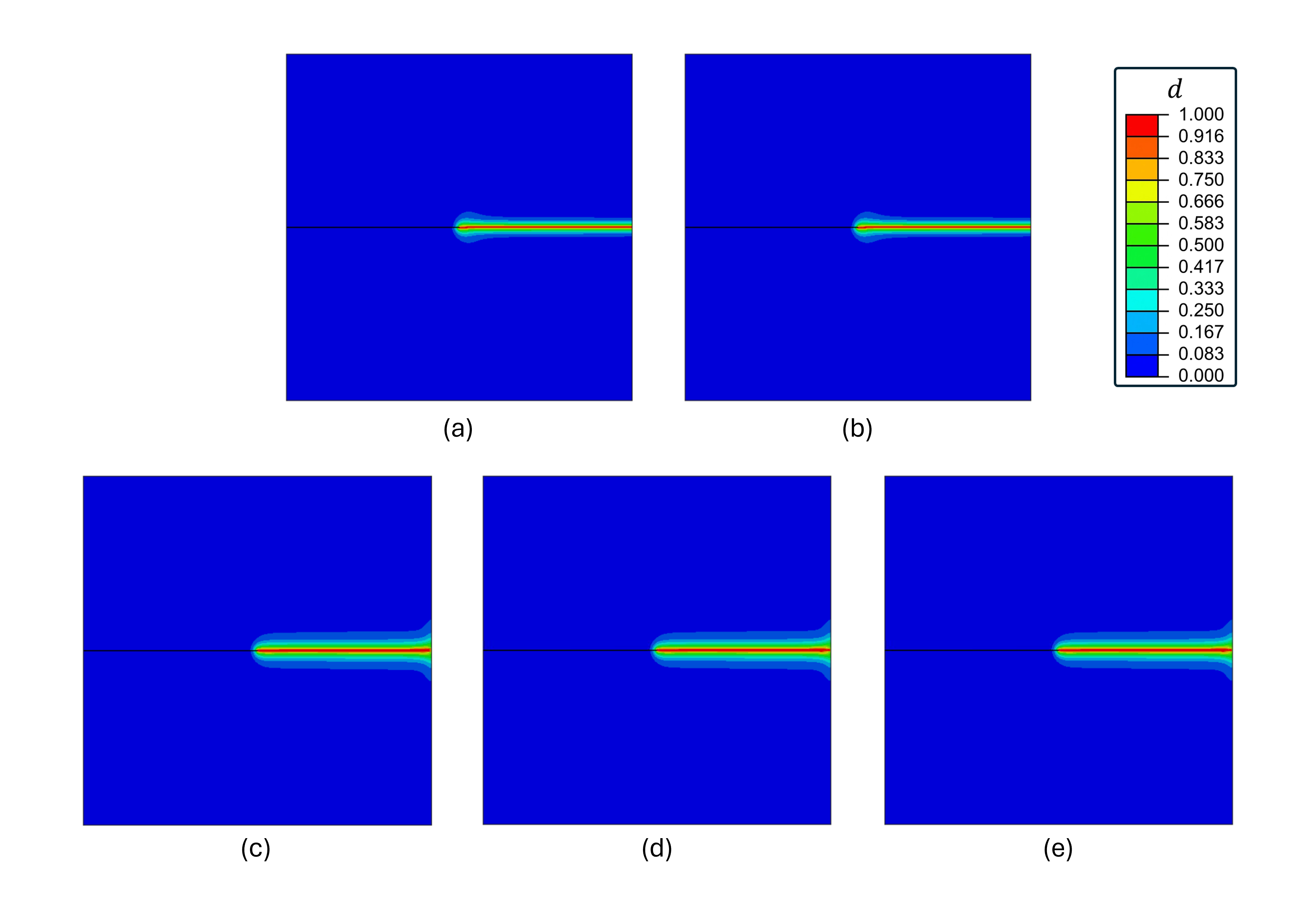}
  \caption{Phase-field damage $d$ at the final states of the mechanical Mode-I tensile benchmark on the $l_0/h=4$ mesh: (a) Case~1, (b) Case~2, (c) Case~3, (d) Case~4, and (e) Case~5.}
  \label{fig:tensile_mech_crack_schemes}
\end{figure}

On the selected mesh, the five cases agree closely up to the peak, with peak loads within $0.35\%$ of Case~1, and differ mainly after it (\Cref{fig:tensile_mech_response}a). Cases~1 and~2 drop abruptly, Cases~3 and~4 spread the loss of load over several increments, and Case~5 lies close to the monolithic solutions because it exchanges coupling quantities at every iteration (Section~\ref{sec:impl:summary}). In the iteration histories, the monolithic cases show isolated bursts near fracture, whereas the staggered cases keep low iteration counts (\Cref{fig:tensile_mech_response}b). Because $\bm{J}_{uc}$ and $\bm{J}_{cu}$ vanish without chemistry, Case~4 reduces to Case~3, and the two give identical responses. All five cases recover the same crack path and damage-band width (\Cref{fig:tensile_mech_crack_schemes}).

\subsubsection{Temporal Discretization and Accuracy}
\label{sec:tensile:chem}

The chemo-mechanical simulations use $l_0/h=4$, the chemical and transport parameters in Table~\ref{tab:tensile_props}, and a constant concentration $c_0=8\times10^{-7}$~mol/mm$^3$ on the crack faces. The specimen is exposed for $t_f=5\times10^6$~s while $U_y=0.12$~mm is applied linearly over the same period. Three fixed time steps, $\Delta t=1000$, 500, and 250~s, are compared. Because transport is integrated with the implicit backward-Euler method, this study assesses temporal accuracy and coupling error rather than stability. 

\begin{table}[h!]
  \centering
  \caption{Accuracy metrics for the chemo-mechanical Mode-I tensile temporal-discretization study. Fixed-step errors are relative to Case~1 at the same time step. Adaptive Case~4 errors are relative to Case~1 at $\Delta t=250$~s. A dash denotes an unavailable fixed-step result.}
  \label{tab:tensile_temporal_accuracy}
  \begin{tabular}{@{}lcrrrrrr@{}}
    \toprule
    $\Delta t$ (s) & Case & $F_{\max}$ (N) & $\varepsilon_{F_{\max}}$ (\%) & $\delta_{\mathrm{peak}}$ (mm) & $\varepsilon_{\delta_{\mathrm{peak}}}$ (\%) & $\delta_f$ (mm) & $\varepsilon_{\delta_f}$ (\%) \\
    \midrule
    1000 & 1 & 545.69 & 0.000 & 0.09000 & 0.000 & 0.09024 & 0.000 \\
         & 2 & 545.66 & 0.005 & 0.09000 & 0.000 & 0.09024 & 0.000 \\
         & 3 & 550.64 & 0.908 & 0.09144 & 1.600 & 0.10440 & 15.691 \\
         & 4 & -- & -- & -- & -- & -- & -- \\
         & 5 & 546.65 & 0.176 & 0.09012 & 0.133 & 0.09730 & 7.824 \\
    \midrule
    500 & 1 & 545.79 & 0.000 & 0.09012 & 0.000 & 0.09024 & 0.000 \\
        & 2 & 545.70 & 0.017 & 0.09012 & 0.000 & 0.09024 & 0.000 \\
        & 3 & 548.56 & 0.508 & 0.09084 & 0.799 & 0.09900 & 9.707 \\
        & 4 & -- & -- & -- & -- & -- & -- \\
        & 5 & 546.25 & 0.084 & 0.09024 & 0.133 & 0.09384 & 3.989 \\
    \midrule
    250 & 1 & 545.78 & 0.000 & 0.09006 & 0.000 & 0.09018 & 0.000 \\
        & 2 & 545.77 & 0.000 & 0.09006 & 0.000 & 0.09018 & 0.000 \\
        & 3 & 547.35 & 0.288 & 0.09054 & 0.533 & 0.09552 & 5.921 \\
        & 4 & -- & -- & -- & -- & -- & -- \\
        & 5 & 545.98 & 0.037 & 0.09018 & 0.133 & 0.09210 & 2.129 \\
    \midrule
    Adaptive & 4 & 546.17 & 0.071 & 0.09020 & 0.158 & 0.09158 & 1.553 \\
    \bottomrule
  \end{tabular}
\end{table}

\begin{figure}[htbp]
  \centering
  \includegraphics[width=\textwidth]{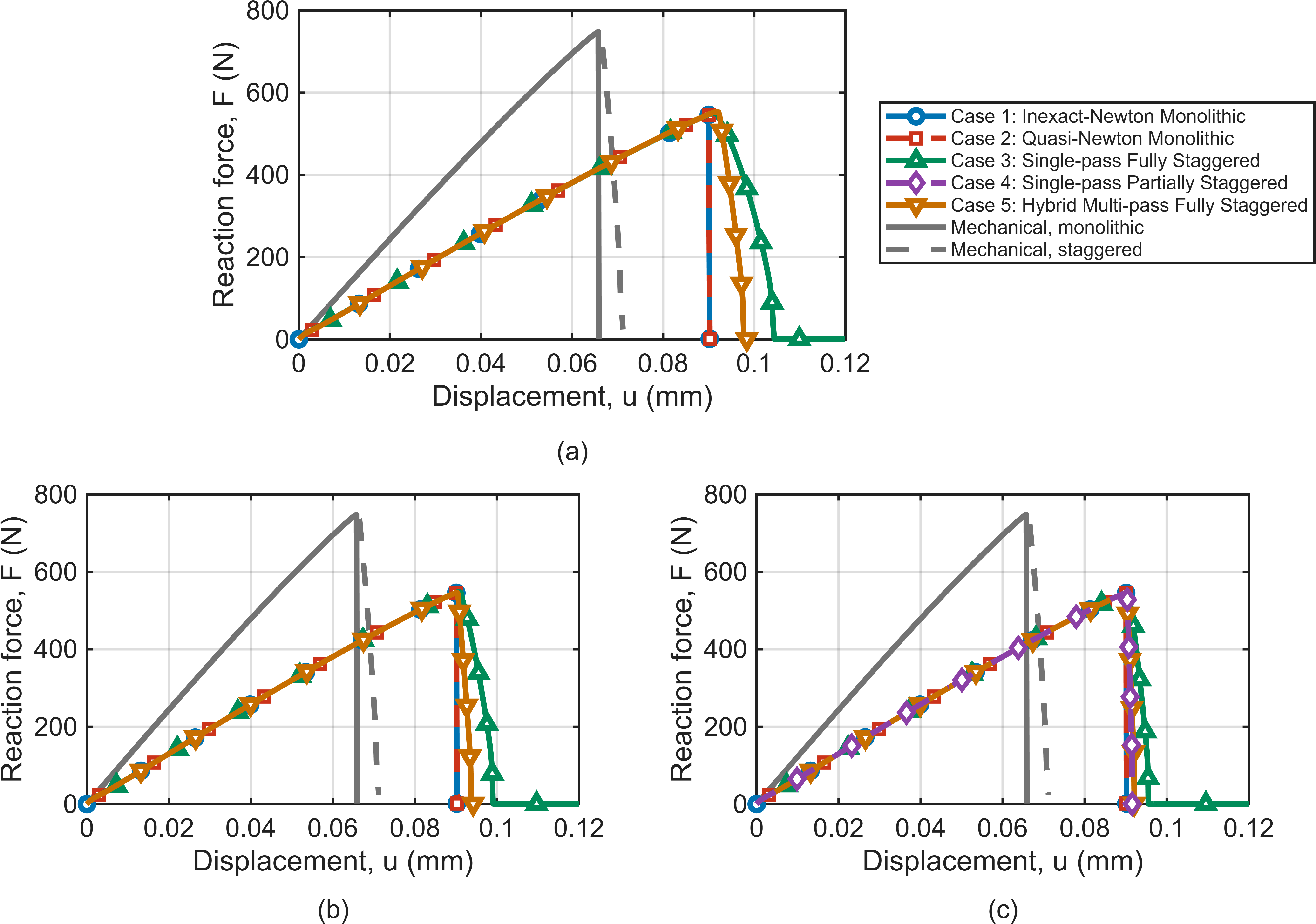}
  \caption{Chemo-mechanical Mode-I tensile benchmark: force--displacement responses at (a) $\Delta t=1000$~s, (b) $\Delta t=500$~s, and (c) $\Delta t=250$~s for Cases~1--3 and~5, with the adaptive Case~4 response in~(c). The gray curves correspond to the mechanical Case~1 and Case~3 responses.}
  \label{fig:tensile_chem_fd}
\end{figure}

Each case is measured against Case~1 at the same time step (Table~\ref{tab:tensile_temporal_accuracy}). Dissolution lowers the peak load by about $27\%$ relative to the mechanical problem and delays the peak. Three numerical observations stand out. First, Case~1 is practically insensitive to the time step over this range, with its peak load varying by less than $0.02\%$, and Case~2 coincides with it. Second, Case~3 is accurate at the peak but not after it. Its peak-load error stays below $1\%$, whereas its failure-displacement error falls only from about $16\%$ to $6\%$ as $\Delta t$ decreases from 1000 to 250~s. Peak quantities alone therefore understate the time-step sensitivity of single-pass schemes. Finally, Case~5 at least halves the failure-displacement error of Case~3 at every time step, because its repeated within-increment updates reduce the splitting error. As $\Delta t$ decreases, the post-peak branches of both staggered cases approach the abrupt monolithic load drop (\Cref{fig:tensile_chem_fd}), so the remaining differences reflect splitting error rather than a different fracture mode.

Case~4 is absent from this comparison because it fails at all three fixed time steps. An adaptive analysis (initial, minimum, and maximum increments of 10, $10^{-2}$, and 250~s) reaches fracture with a response close to that of Case~1 and a failure-displacement error of about $1.6\%$. Adaptive stepping can therefore rescue Case~4, but this does not establish fixed-step robustness.

The local concentration error $\varepsilon_{c,\mathrm{tip}}$ is evaluated in the elements ahead of the pre-notch tip, relative to Case~1 at $\Delta t=500$~s. At the pre-peak displacement $U_y=0.08$~mm, all errors are below $0.005\%$. Near the peak, they grow to $0.03\%$ for Case~2, $0.66\%$ for Case~3, $0.18\%$ for the adaptive Case~4, and $0.15\%$ for Case~5. Case~2 therefore remains closest to the reference, and Case~5 reduces the error of Case~3 by about three quarters as fracture is approached.

The step $\Delta t=500$~s is adopted for the transport-intensity study. It roughly halves the cost relative to $\Delta t=250$~s while keeping the peak-load and peak-displacement errors of Cases~3 and~5 below $1\%$. Their failure-displacement errors at this step, about $10\%$ for Case~3 and $4\%$ for Case~5, must be considered when interpreting the post-peak response.

\subsubsection{Computational Performance and Convergence Robustness}
\label{sec:tensile:performance}

\begin{figure}[htbp]
  \centering
  \includegraphics[width=\textwidth]{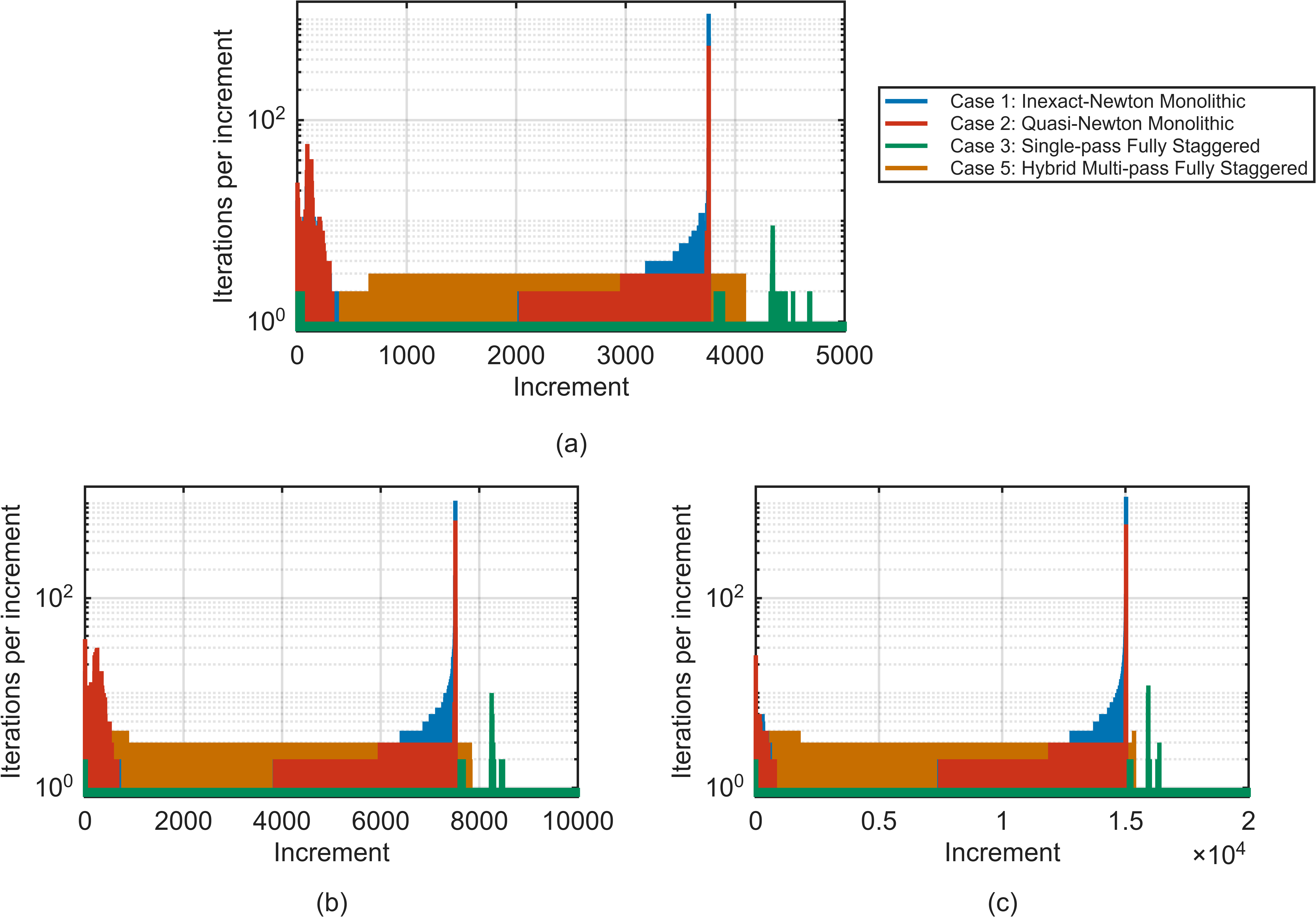}
  \caption{Chemo-mechanical Mode-I tensile benchmark: equilibrium iterations per increment for Cases~1--3 and~5 at (a) $\Delta t=1000$~s, (b) $\Delta t=500$~s, and (c) $\Delta t=250$~s. Case~4 is excluded because it does not complete at fixed time steps.}
  \label{fig:tensile_chem_iters}
\end{figure}

Table~\ref{tab:tensile_chem_performance} and \Cref{fig:tensile_chem_iters} show the cost and iteration behavior. Case~2 runs 1.7--2.5 times faster than Case~1 in CPU time with a practically identical response. Case~3 gives a similar speed-up, 1.7--2.3, and the lowest wall-clock time at each step, but at the cost of the post-peak errors identified above. Case~5 is slower than Cases~2 and~3 in CPU time but still about 1.5 times faster than Case~1, and its additional iterations buy accuracy. Case~4 completes only with adaptive stepping, at about 95 CPU hours, more than any fixed-step Case~1 analysis.

The burst factors show how the iteration demand is distributed, which runtime alone does not. Near fracture, Cases~1 and~2 need hundreds of times their mean iteration count in single increments. Cases~3 and~5 avoid such bursts, with $B_{\max}$ below 10 and about 2, respectively, but for different reasons. Case~3 performs a single pass per increment without checking convergence between the fields, so its counts stay low by construction. Case~5 repeats its field updates at every iteration and has the largest $N_{\mathrm{iter}}$ of all fixed-step cases, so its smooth history reflects regular work rather than less work. Iteration counts and burst factors must therefore be read together with runtime and completion.

\begin{table}[h!]
  \centering
  \caption{Computational performance and convergence robustness for the chemo-mechanical Mode-I tensile temporal-discretization study. CPU and wall-clock times are in hours. A dash denotes an unavailable comparison.}
  \label{tab:tensile_chem_performance}
  \small
  \setlength{\tabcolsep}{4pt}
  \begin{tabular}{@{}lclrrrrrr@{}}
    \toprule
    $\Delta t$ (s) & Case & Status & $t_{\mathrm{CPU}}$ & $t_{\mathrm{wall}}$ & $S_{\mathrm{CPU}}$ & $N_{\mathrm{iter}}$ & $B_{95}$ & $B_{\max}$ \\
    \midrule
    1000 & 1 & Completed & 9.08 & 5.04 & 1.00 & 8,411 & 2.24 & 509.17 \\
         & 2 & Completed & 5.42 & 2.99 & 1.68 & 7,421 & 1.55 & 277.66 \\
         & 3 & Completed & 5.25 & 2.19 & 1.73 & 4,456 & 0.98 & 8.79 \\
         & 4 & Failed & -- & -- & -- & -- & -- & -- \\
         & 5 & Completed & 5.90 & 2.20 & 1.54 & 12,315 & 1.29 & 1.93 \\
    \midrule
    500 & 1 & Completed & 17.31 & 9.70 & 1.00 & 16,086 & 2.80 & 498.81 \\
        & 2 & Completed & 7.75 & 5.16 & 2.23 & 13,025 & 1.73 & 380.47 \\
        & 3 & Completed & 8.69 & 3.44 & 1.99 & 8,332 & 0.99 & 7.92 \\
        & 4 & Failed & -- & -- & -- & -- & -- & -- \\
        & 5 & Completed & 11.88 & 4.49 & 1.46 & 24,267 & 1.29 & 1.93 \\
    \midrule
    250 & 1 & Completed & 36.15 & 21.23 & 1.00 & 27,106 & 2.22 & 650.42 \\
        & 2 & Completed & 14.64 & 8.76 & 2.47 & 19,860 & 2.27 & 454.08 \\
        & 3 & Completed & 15.66 & 6.10 & 2.31 & 16,026 & 0.99 & 9.93 \\
        & 4 & Failed & -- & -- & -- & -- & -- & -- \\
        & 5 & Completed & 23.98 & 9.14 & 1.51 & 47,681 & 1.29 & 1.93 \\
    \midrule
    Adaptive & 4 & Completed & 94.75 & 66.69 & -- & 73,590 & 1.48 & 9.62 \\
    \bottomrule
  \end{tabular}
\end{table}

\begin{figure}[htbp]
  \centering
  \includegraphics[width=\textwidth]{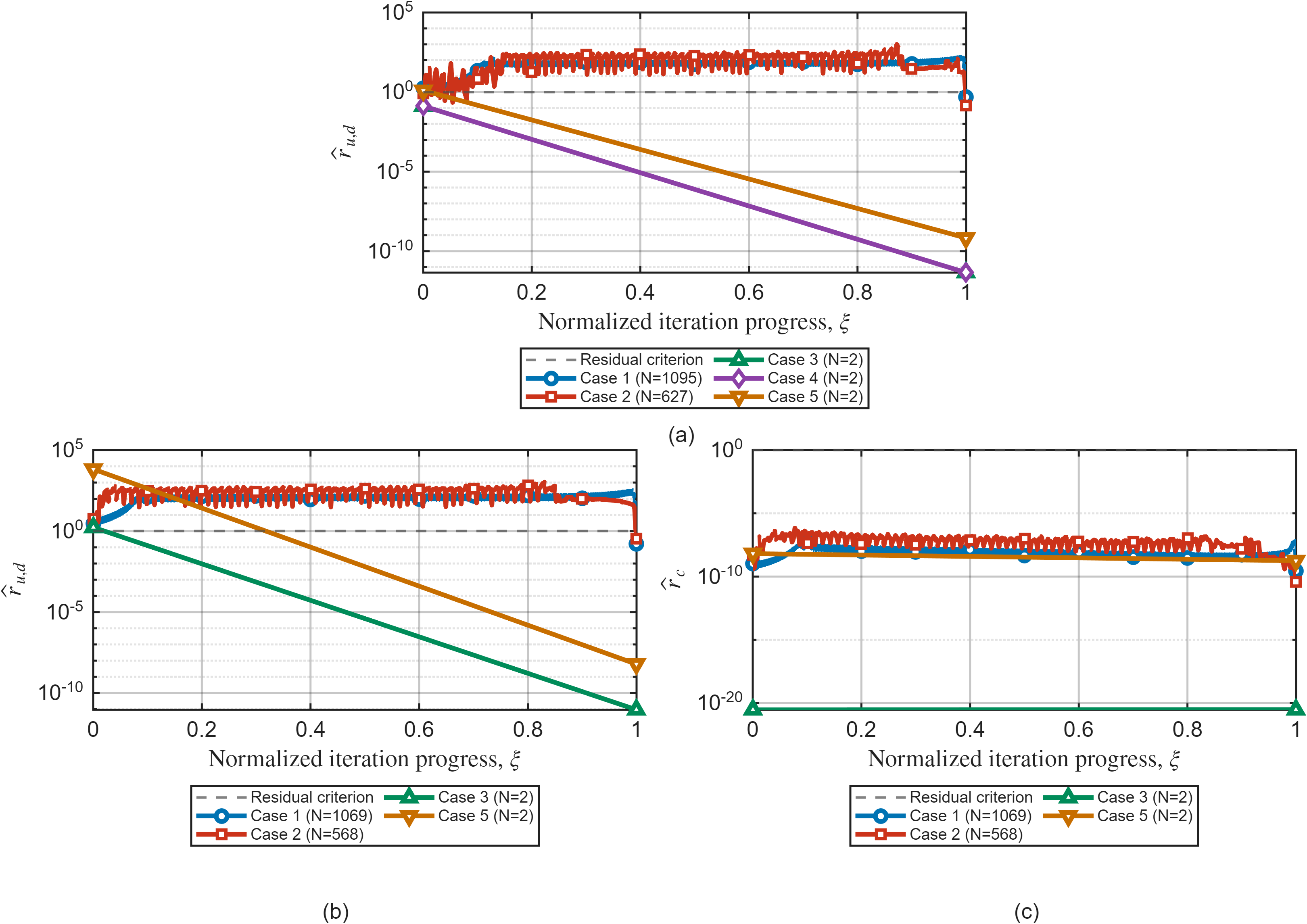}
  \caption{Criterion-normalized residual histories at increments during crack propagation in the Mode-I tensile benchmark: (a) displacement--phase-field residuals for the mechanical problem ($l_0/h=4$), (b) displacement--phase-field residuals for the chemo-mechanical problem ($\Delta t=500$~s), and (c) concentration residuals for the same chemo-mechanical problem. The dashed line marks the convergence criterion, and $N$ is the number of iterations in each increment. Case~4 is omitted from (b) and (c) because the fixed-step simulation did not complete.}
  \label{fig:tensile_residual_convergence}
\end{figure}

The residual histories near crack propagation (\Cref{fig:tensile_residual_convergence}) are consistent with the bursts. Cases~1 and~2 oscillate strongly over many iterations before satisfying the criterion, in both the mechanical and chemo-mechanical problems. In the mechanical problem, Cases~3--5 satisfy the displacement--phase-field criterion within two iterations. In the chemo-mechanical problem, Cases~3 and~5 finish below the criterion, and the concentration residual stays below its criterion throughout. For Case~3, however, these criteria are applied to equations with lagged coupling and therefore do not measure convergence between the fields.

\begin{figure}[htbp]
  \centering
  \includegraphics[width=0.8\textwidth]{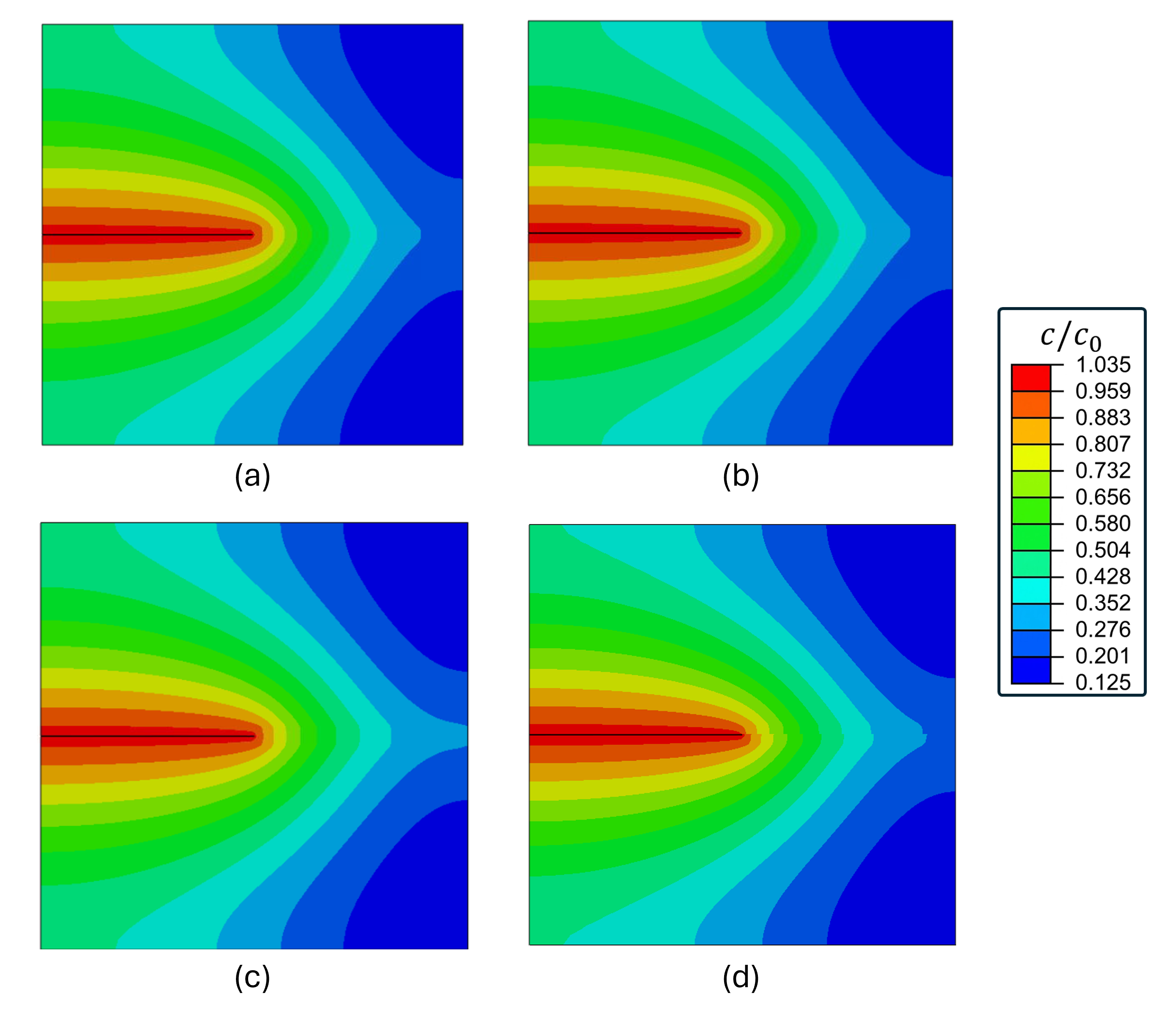}
  \caption{Normalized CO$_2$ concentration $c/c_0$ at the final states of the chemo-mechanical Mode-I tensile benchmark at $\Delta t=250$~s: (a) Case~1, (b) Case~2, (c) Case~3, and (d) Case~5.}
  \label{fig:tensile_chem_cratio}
\end{figure}

\begin{figure}[htbp]
  \centering
  \includegraphics[width=0.8\textwidth]{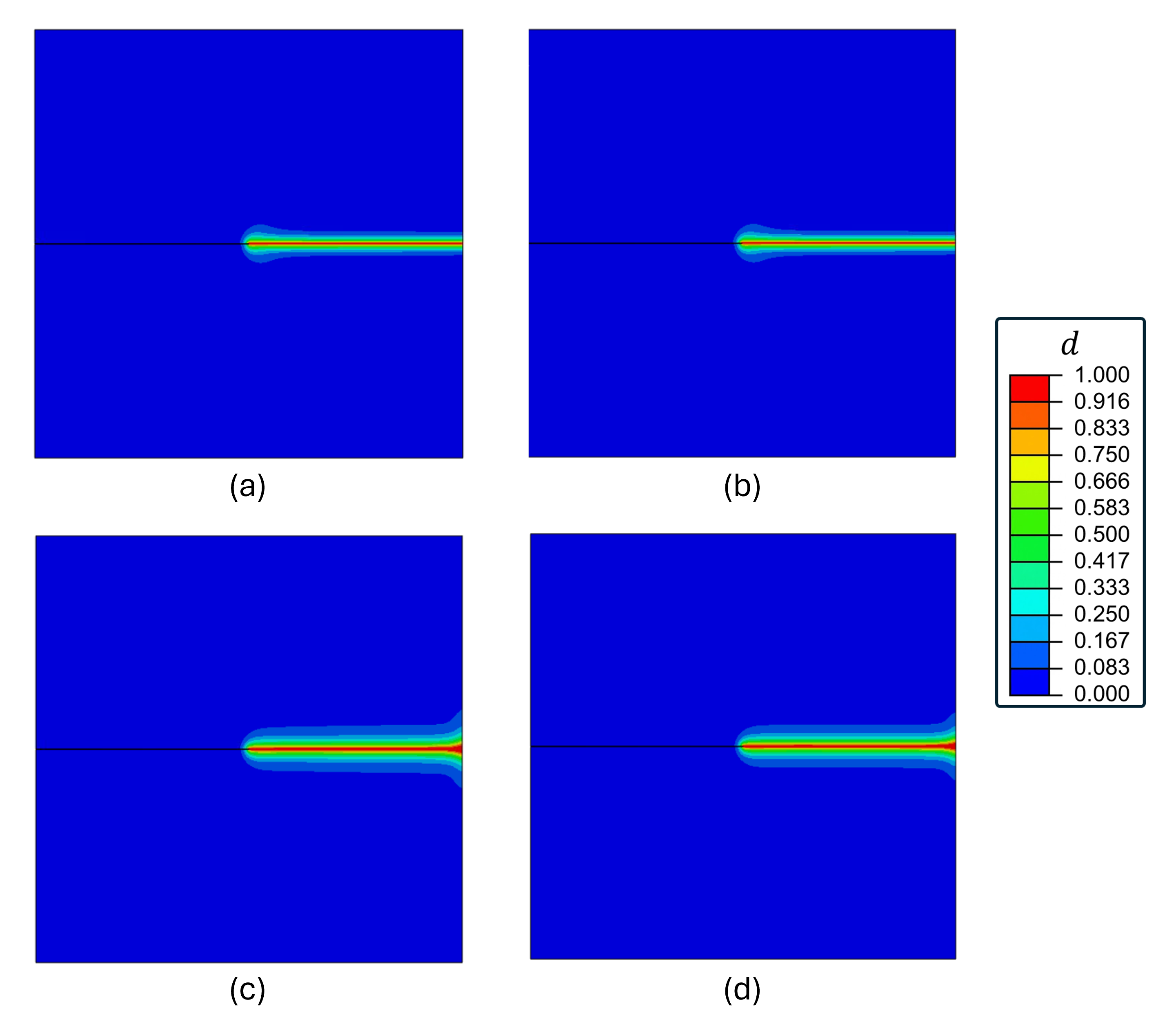}
  \caption{Phase-field damage $d$ at complete fracture in the chemo-mechanical Mode-I tensile benchmark at $\Delta t=250$~s: (a) Case~1, (b) Case~2, (c) Case~3, and (d) Case~5.}
  \label{fig:tensile_chem_crack}
\end{figure}

The concentration and phase-field contours (\Cref{fig:tensile_chem_cratio,fig:tensile_chem_crack}) show similar transport localization and crack paths for Cases~1--3 and~5.

\subsubsection{Influence of Transport Intensity}
\label{sec:tensile:transport}

Transport properties and dominant transport mechanisms vary across geotechnical and geomechanical applications. For example, fracture-mediated transport in hydraulic-fracturing problems can occur on markedly different time scales from the dissolved CO$_2$ transport considered in geological carbon storage. The present study focuses on the latter and therefore retains the baseline values of $D_0$ and $D_p$ listed in Table~\ref{tab:tensile_props} for the other benchmarks. To test how the strategies respond as transport becomes more severe, $D_0$ and $D_p$ are scaled together by factors of 10 and 1000 (denoted $10D$ and $1000D$), with $l_0/h=4$ and $\Delta t=500$~s. It should be noted that these scaled settings are not alternative material calibrations.
 
The measure $\mathrm{Fo}_{95}$ increases almost linearly with the scaling, from 1.5 for the baseline to 14.7 and 1468 (Table~\ref{tab:tensile_transport_accuracy}), confirming that it measures transport intensity as expected from its definition. The scaling also enlarges the transport contributions to $\bm{J}_{cc}$ and $\bm{J}_{cu}$, which may affect the convergence behavior, although this is not quantified separately.

Between the $10D$ and $1000D$ settings, the peak response is practically unchanged. Diffusivity affects the mechanics only indirectly, through dissolution damage. Over the loading and dissolution time scale considered, the peak response is therefore insensitive to further increases in transport rate. The comparison thus tests solver robustness under increasing transport intensity while the mechanical response remains broadly comparable.

\begin{table}[htbp]
  \centering
  \caption{Accuracy and completion metrics for the Mode-I tensile transport-intensity study. Errors are relative to Case~1 at the same transport setting, and the baseline problem has $\mathrm{Fo}_{95}^{(1)}=1.47$. A dash denotes a failed run or an inapplicable comparison.}
  \label{tab:tensile_transport_accuracy}
  \resizebox{\textwidth}{!}{%
  \begin{tabular}{@{}ccclrrrrrr@{}}
    \toprule
    Scale & $\mathrm{Fo}_{95}^{(1)}$ & Case & Status & $F_{\max}$ (N) & $\varepsilon_{F_{\max}}$ (\%) & $\delta_{\mathrm{peak}}$ (mm) & $\varepsilon_{\delta_{\mathrm{peak}}}$ (\%) & $\delta_f$ (mm) & $\varepsilon_{\delta_f}$ (\%) \\
    \midrule
    $10D$ & 14.68 & 1 & Completed & 545.78 & 0.000 & 0.09012 & 0.000 & 0.09024 & 0.000 \\
          &       & 2 & Completed & 545.74 & 0.007 & 0.09012 & 0.000 & 0.09024 & 0.000 \\
          &       & 3 & Completed & 547.33 & 0.285 & 0.09060 & 0.533 & 0.09552 & 5.851 \\
          &       & 4 & Failed    & --     & --    & --      & --    & --      & --    \\
          &       & 5 & Completed & 546.14 & 0.065 & 0.09024 & 0.133 & 0.09372 & 3.856 \\
    \midrule
    $1000D$ & 1467.92 & 1 & Completed & 545.79 & 0.000 & 0.09012 & 0.000 & 0.09024 & 0.000 \\
            &         & 2 & Failed    & --     & --    & --      & --    & --      & --    \\
            &         & 3 & Completed & 547.36 & 0.289 & 0.09060 & 0.533 & 0.09552 & 5.851 \\
            &         & 4 & Completed & 547.32 & 0.280 & 0.09060 & 0.533 & 0.09552 & 5.851 \\
            &         & 5 & Completed & 546.25 & 0.084 & 0.09024 & 0.133 & 0.09384 & 3.989 \\
    \bottomrule
  \end{tabular}%
  }
\end{table}

\begin{table}[htbp]
  \centering
  \caption{Computational performance and convergence robustness for the Mode-I tensile transport-intensity study. CPU and wall-clock times are in hours.}
  \label{tab:tensile_transport_performance}
  \begin{tabular}{@{}ccrrrrrr@{}}
    \toprule
    Scale & Case & $t_{\mathrm{CPU}}$ & $t_{\mathrm{wall}}$ & $S_{\mathrm{CPU}}$ & $N_{\mathrm{iter}}$ & $B_{95}$ & $B_{\max}$ \\
    \midrule
    $10D$ & 1 & 16.20 & 9.20 & 1.00 & 14,167 & 2.18 & 653.67 \\
          & 2 & 7.56 & 4.60 & 2.14 & 10,219 & 1.47 & 415.59 \\
          & 3 & 8.25 & 3.33 & 1.96 & 8,903 & 1.79 & 53.65 \\
          & 4 & 0.016 & 0.006 & -- & -- & -- & -- \\
          & 5 & 10.07 & 3.94 & 1.61 & 8,327 & 1.89 & 2.83 \\
    \midrule
    $1000D$ & 1 & 15.05 & 8.52 & 1.00 & 13,760 & 2.19 & 585.86 \\
            & 2 & 8.94 & 5.52 & -- & -- & -- & -- \\
            & 3 & 7.83 & 3.24 & 1.92 & 8,898 & 1.79 & 53.68 \\
            & 4 & 9.00 & 3.58 & 1.67 & 8,901 & 1.79 & 53.66 \\
            & 5 & 9.24 & 3.71 & 1.63 & 8,563 & 1.83 & 2.75 \\
    \bottomrule
  \end{tabular}
\end{table}

Tables~\ref{tab:tensile_transport_accuracy} and~\ref{tab:tensile_transport_performance} show three trends. First, the completed staggered cases stay close to Case~1 at the peak. Their post-peak errors, however, persist as in the baseline problem. Second, completion depends on the scheme and does not change monotonically with transport intensity. Case~4 fails at $10D$ but completes at $1000D$, and Case~2 does the opposite. A plausible cause of the Case~2 failure is that, at high diffusivity, the nonsymmetric stress-assisted part of $\bm{J}_{cc}$ gains weight, and the symmetric BFGS approximation cannot represent it (Section~\ref{sec:impl:monolithic}). Cases~1, 3, and~5 complete under both settings. The measure $\mathrm{Fo}_{95}$ is therefore an indicator of transport intensity, not a convergence threshold. Finally, the cases differ in their iteration bursts. Case~1 retains large bursts, and those of Case~3 grow several-fold relative to the baseline. Case~5 keeps $B_{\max}$ below 3. It needs fewer iterations than Case~3 but more CPU time, because each iteration evaluates three elements per cell.

For this benchmark, Case~2 gives the most accurate low-cost solution when it converges. Case~3 gives the lowest fixed-step wall-clock time and suits rapid screening followed by time-step refinement. Case~5 is an intermediate choice that improves the post-peak accuracy of Case~3 and suppresses iteration bursts at a higher cost. Case~1 remains the reference, and Case~4 is not a reliable fixed-step option.
\subsection{Shear Test}
\label{sec:shear}

\subsubsection{Problem Description}
\label{sec:shear:setup}

\begin{figure}[htbp]
  \centering
  \includegraphics[width=0.9\textwidth]{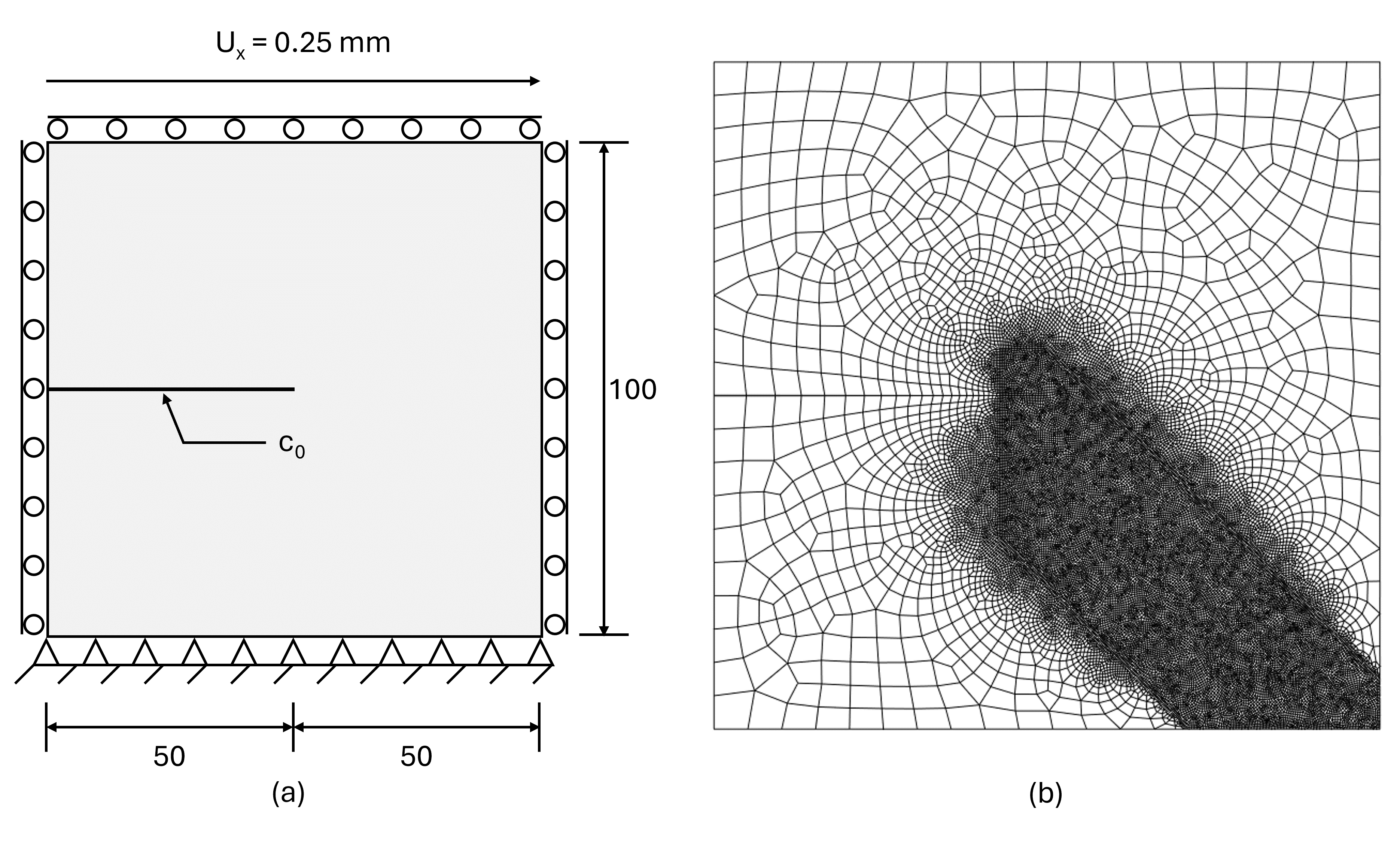}
  \caption{Shear test: geometry, boundary conditions, and locally refined mesh. Dimensions are in millimeters.}
  \label{fig:shear_geom}
\end{figure}

The second benchmark is the single-edge-notched shear test. Unlike in the tensile test, the rotating principal-stress field produces a curved crack that grows from the notch tip toward the bottom-right corner~\citep{Miehe2010a,Ambati2015}. The benchmark therefore tests the strategies under a nonsymmetric hydrostatic-stress field and a curved crack path. The specimen, material properties, and discretization of the tensile test are retained. The bottom edge is fixed, a horizontal displacement $U_x$ is applied to the top edge, and the vertical displacement is constrained on the top and side edges (\Cref{fig:shear_geom}). The mesh ($l_0/h=4$) is refined in the bottom-right quadrant along the expected crack path and contains 24,660 elements. In the mechanical analysis, $U_x$ is increased to 0.015~mm and then to 0.25~mm in a second step. In the chemo-mechanical analysis, $c_0=8\times10^{-7}$~mol/mm$^3$ is prescribed on the crack faces, and the displacement and concentration are applied together over $t_f=5\times10^6$~s with $\Delta t=500$~s.

\subsubsection{Mechanical Response}
\label{sec:shear:mech}

\begin{figure}[htbp]
  \centering
  \includegraphics[width=\textwidth]{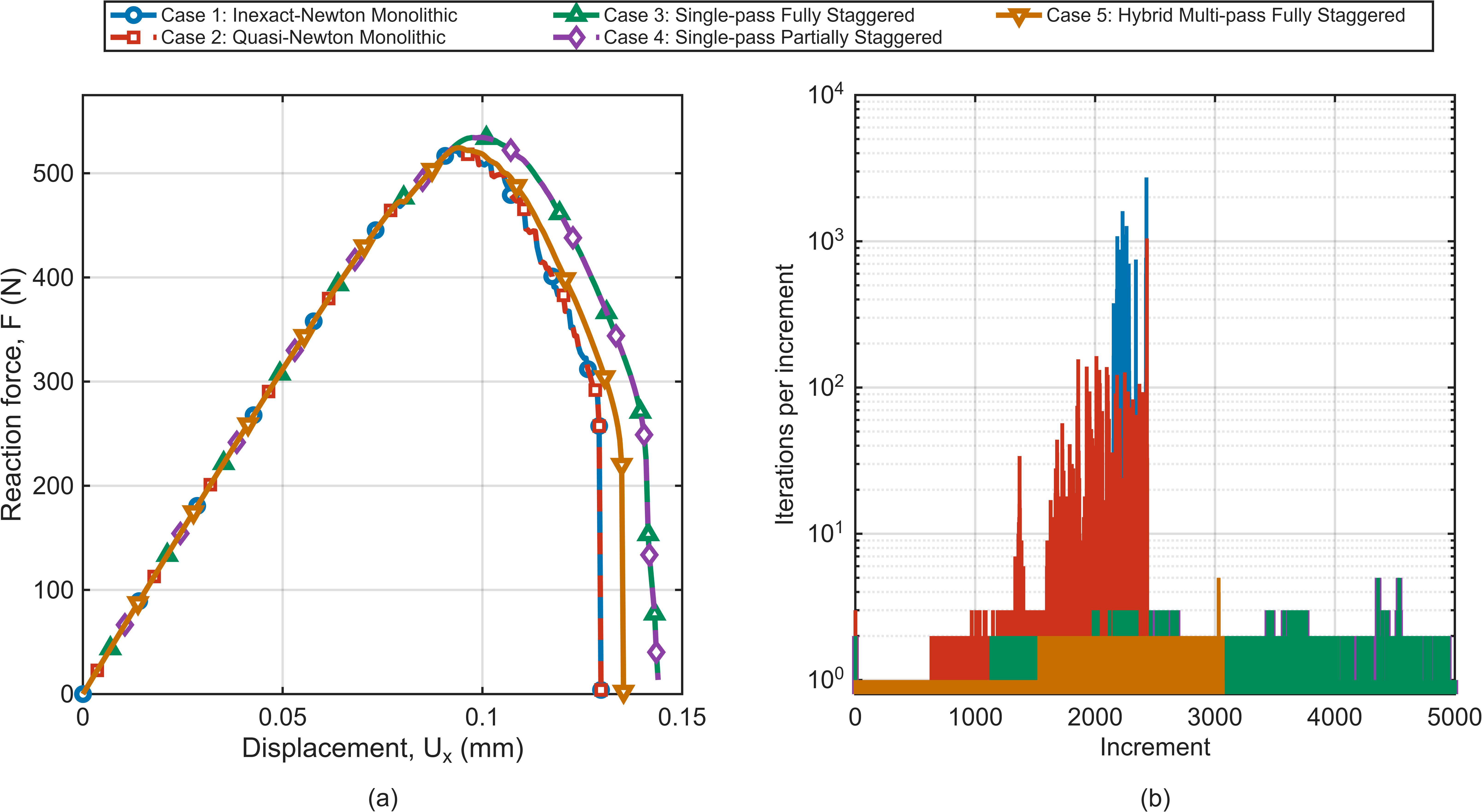}
  \caption{Mechanical shear test: (a) force--displacement responses and (b) equilibrium iterations per increment for Cases~1--5.}
  \label{fig:shear_mech_response}
\end{figure}

\Cref{fig:shear_mech_response}(a) compares the force--displacement responses. Unlike the abrupt load drop of the tensile test, the shear response has a rounded peak and a gradual post-peak decline, because the curved crack grows progressively under the rotating stress field. Cases~1 and~2 coincide, as do Cases~3 and~4. The single-pass cases overestimate the peak load by about $2.5\%$ and delay the peak and failure by about $7\%$ and $11\%$, the largest errors of any mechanical benchmark (Table~\ref{tab:shear_accuracy}). Case~5 recovers most of this loss of accuracy, reducing these errors to about $0.6\%$, $1\%$, and $4\%$, although it does not coincide with the monolithic reference.

\Cref{fig:shear_mech_response}(b) shows the corresponding equilibrium iterations. The monolithic cases show pronounced iteration bursts during crack growth, whereas the staggered cases keep regular counts. Case~1 has the largest absolute spike, but Case~2 has the larger $B_{\max}$ because its mean iteration count is lower. Among the staggered strategies, Case~5 needs slightly fewer iterations than Cases~3 and~4. Cases~3--5 also have the lowest and nearly equal costs (Table~\ref{tab:shear_performance_robustness}). Case~5 is the cheapest, about eight times faster than Case~1 in CPU time, even though it evaluates three elements per cell. The number of elements per cell therefore does not by itself determine the cost.

\begin{figure}[htbp]
  \centering
  \includegraphics[width=0.9\textwidth]{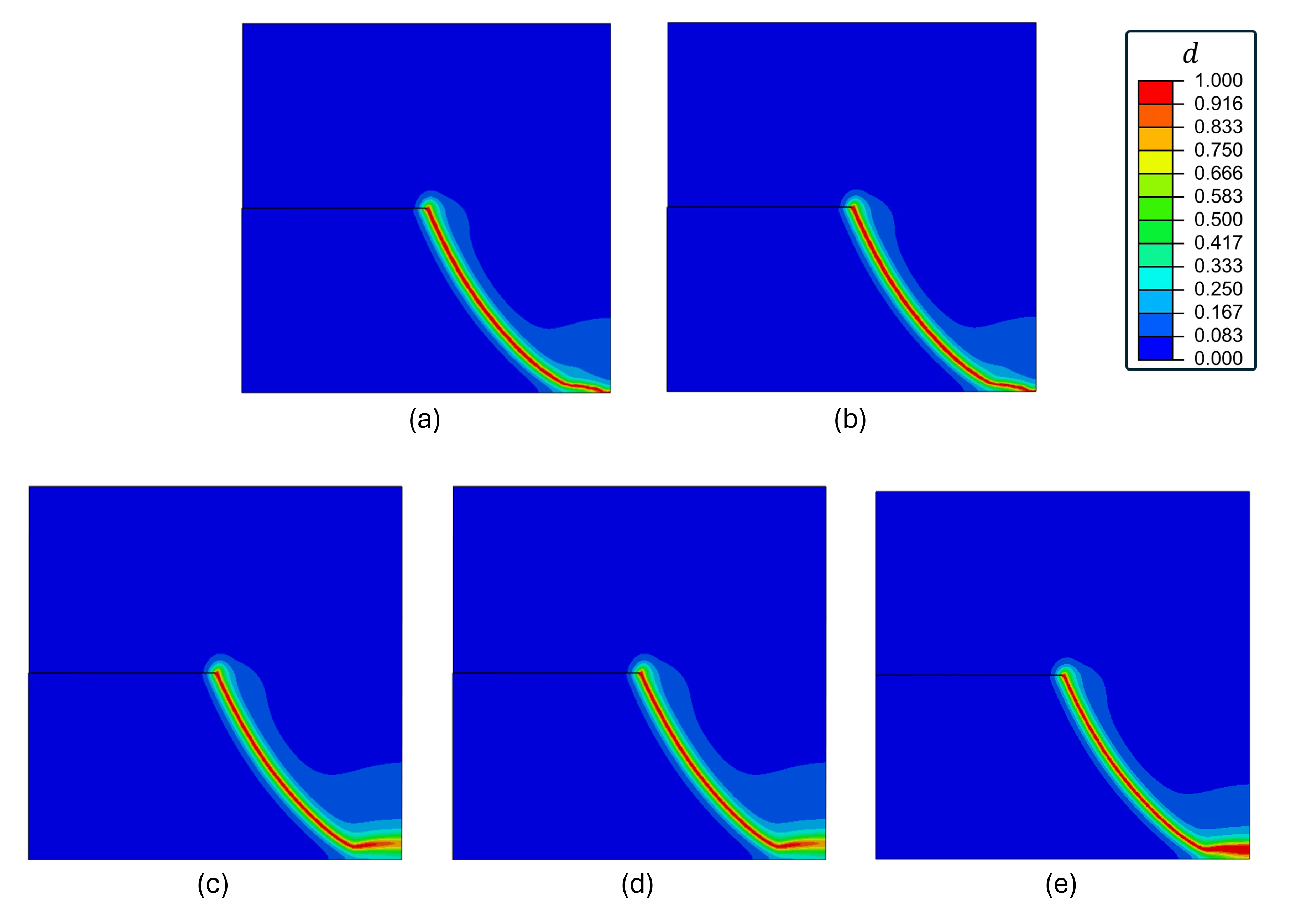}
  \caption{Phase-field damage $d$ at the final states of the mechanical shear test: (a) Case~1, (b) Case~2, (c) Case~3, (d) Case~4, and (e) Case~5.}
  \label{fig:shear_mech_crack}
\end{figure}

Despite these differences in accuracy and cost, all five cases predict a crack that initiates at the notch tip and curves toward the bottom-right corner, with similar damage-band widths (\Cref{fig:shear_mech_crack}). This path is consistent with the crack paths reported for this benchmark~\citep{Miehe2010a,Ambati2015}.

\subsubsection{Chemo-mechanical Response}
\label{sec:shear:chem}

\begin{figure}[htbp]
  \centering
  \includegraphics[width=\textwidth]{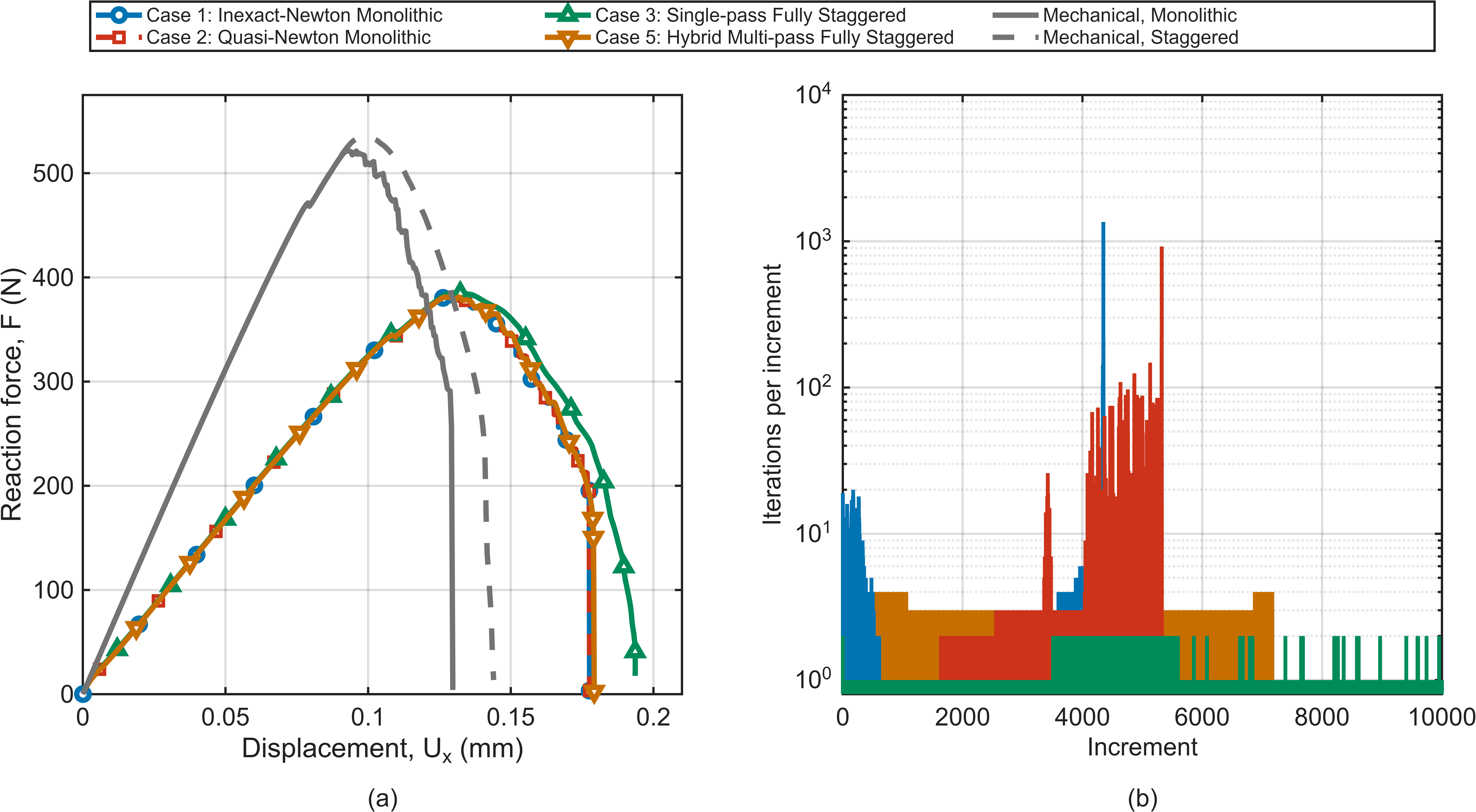}
  \caption{Chemo-mechanical shear test: (a) force--displacement responses and (b) equilibrium iterations per increment for Cases~1--3 and~5. Case~4 terminates before the comparable response range. Gray curves are the mechanical Case~1 and Case~3 responses.}
  \label{fig:shear_chem_response}
\end{figure}

\Cref{fig:shear_chem_response}(a) compares the force--displacement responses of Cases~1--3 and~5 with the mechanical curves. Dissolution lowers the peak load by about $27\%$ relative to the mechanical problem and shifts it to a larger displacement for every completed case, reflecting the progressive degradation around the growing crack. Case~2 again coincides with Case~1. Case~3 overestimates the peak load and peak displacement by about $1\%$ and $2\%$, and its failure-displacement error is about $9\%$ (Table~\ref{tab:shear_accuracy}). Case~5 reduces these errors to below $0.4\%$ at the peak and below $1\%$ at failure, so its largest improvement over Case~3 is in the post-peak response. Case~4 terminates at $U_x\approx1.4\times10^{-3}$~mm, before the peak, so it is assessed for robustness only.

The local concentration error $\varepsilon_{c,\mathrm{tip}}$ follows the same ranking. It is evaluated before the peak, at $U_x=0.12$~mm, in the elements directly below the pre-notch tip along the initial crack path. The error is $0.009\%$ for Case~2, $0.123\%$ for Case~3, and $0.066\%$ for Case~5. Case~2 therefore coincides with Case~1 in this region, and Case~5 halves the error of Case~3.

\begin{table}[htbp]
  \centering
  \caption{Accuracy metrics for the mechanical and chemo-mechanical shear test. A dash denotes a failed run or an inapplicable comparison.}
  \label{tab:shear_accuracy}
  \begin{tabular}{@{}lcrrrrrr@{}}
    \toprule
    Response & Case & $F_{\max}$ (N) & $\varepsilon_{F_{\max}}$ (\%) & $\delta_{\mathrm{peak}}$ (mm) & $\varepsilon_{\delta_{\mathrm{peak}}}$ (\%) & $\delta_f$ (mm) & $\varepsilon_{\delta_f}$ (\%) \\
    \midrule
    Mechanical & 1 & 521.59 & 0.000 & 0.09302 & 0.000 & 0.12968 & 0.000 \\
               & 2 & 521.59 & 0.000 & 0.09302 & 0.000 & 0.12968 & 0.000 \\
               & 3 & 534.50 & 2.475 & 0.09960 & 7.074 & 0.14398 & 11.025 \\
               & 4 & 534.50 & 2.475 & 0.09960 & 7.074 & 0.14398 & 11.025 \\
               & 5 & 524.48 & 0.553 & 0.09396 & 1.011 & 0.13532 & 4.349 \\
    \midrule
    Chemo-mechanical & 1 & 381.09 & 0.000 & 0.12775 & 0.000 & 0.17763 & 0.000 \\
                     & 2 & 381.09 & 0.000 & 0.12775 & 0.000 & 0.17763 & 0.000 \\
                     & 3 & 385.08 & 1.049 & 0.13048 & 2.139 & 0.19359 & 8.990 \\
                     & 4 & -- & -- & -- & -- & -- & -- \\
                     & 5 & 381.26 & 0.046 & 0.12825 & 0.391 & 0.17925 & 0.912 \\
    \bottomrule
  \end{tabular}
\end{table}

\Cref{fig:shear_chem_response}(b) shows the corresponding equilibrium iterations. Cases~1 and~2 again develop large bursts during fracture, with $B_{\max}$ in the hundreds, whereas Cases~3 and~5 keep regular histories with similar totals (Table~\ref{tab:shear_performance_robustness}). Case~5 therefore improves on the accuracy of Case~3 without increasing the iteration count. Case~3 is the cheapest completed strategy. Case~5 needs slightly more CPU time than Case~2 but less wall-clock time, because of its different parallel utilization and additional element evaluations. Case~4 fails despite accumulating tens of thousands of iterations with moderate burst factors, so moderate bursts do not by themselves indicate completion.

The chemo-mechanical Case~1 analysis reaches failure, but not the end of the prescribed loading history, within the 72-hour limit. It therefore provides a valid reference response up to failure. Its cost at the failure endpoint is estimated in two steps. The elapsed time is scaled by $\delta_f/\delta_{\mathrm{last}}=0.88$ (the ratio of the failure displacement to the displacement at the last converged increment) and the CPU time then follows from the CPU-to-wall-time ratio of the mechanical Case~1 analysis. The resulting speed-ups, about 5.4, 9.1, and 5.0 for Cases~2, 3, and~5, are approximate because the iteration demand varies during fracture.

\begin{table}[htbp]
  \centering
  \caption{Computational performance and convergence robustness for the mechanical and chemo-mechanical shear test. CPU and wall-clock times are in hours. The chemo-mechanical Case~1 times and corresponding speed-ups are displacement-ratio estimates. A dash denotes an unavailable comparison.}
  \label{tab:shear_performance_robustness}
  \small
  \setlength{\tabcolsep}{3.5pt}
  \begin{tabular}{@{}lclrrrrrr@{}}
    \toprule
    Response & Case & Status & $t_{\mathrm{CPU}}$ & $t_{\mathrm{wall}}$ & $S_{\mathrm{CPU}}$ & $N_{\mathrm{iter}}$ & $B_{95}$ & $B_{\max}$ \\
    \midrule
    Mechanical & 1 & Completed & 55.36 & 38.81 & 1.00 & 59,552 & 3.01 & 134.83 \\
               & 2 & Completed & 10.71 & 5.99 & 5.17 & 14,228 & 3.51 & 216.14 \\
               & 3 & Completed & 7.52 & 2.90 & 7.36 & 3,986 & 1.63 & 2.45 \\
               & 4 & Completed & 8.01 & 3.13 & 6.91 & 3,986 & 1.63 & 2.45 \\
               & 5 & Completed & 7.06 & 2.89 & 7.85 & 3,823 & 1.60 & 4.00 \\
    \midrule
    Chemo-mechanical & 1 & Incomplete & $\approx90.70$ & $\approx63.59$ & 1.00 & 15,482 & 3.47 & 427.65 \\
                     & 2 & Completed & 16.73 & 10.22 & $\approx5.42$ & 21,310 & 3.00 & 307.07 \\
                     & 3 & Completed & 10.00 & 6.52 & $\approx9.07$ & 8,436 & 1.84 & 1.84 \\
                     & 4 & Failed & -- & -- & -- & 38,911 & 2.19 & 9.50 \\
                     & 5 & Completed & 18.17 & 7.52 & $\approx4.99$ & 8,291 & 1.73 & 2.59 \\
    \bottomrule
  \end{tabular}
\end{table}

\begin{figure}[htbp]
  \centering
  \includegraphics[width=0.8\textwidth]{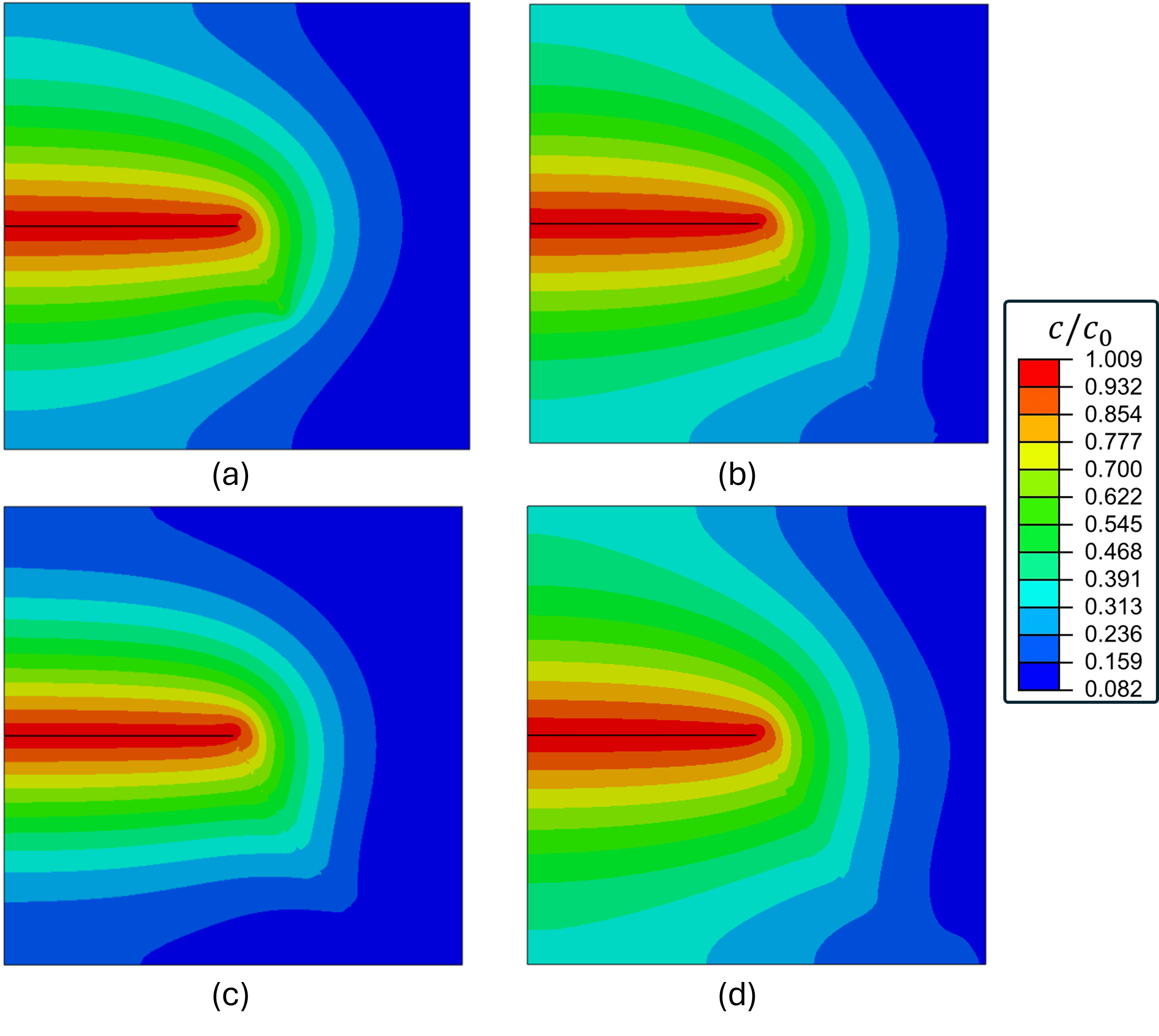}
  \caption{Normalized CO$_2$ concentration $c/c_0$ at the final stored states of the chemo-mechanical shear test: (a) Case~1, (b) Case~2, (c) Case~3, and (d) Case~5.}
  \label{fig:shear_chem_cratio}
\end{figure}

The concentration contours at the final states (\Cref{fig:shear_chem_cratio}) show similar nonsymmetric transport patterns for Cases~1--3 and~5, shaped by diffusion, the curved crack, and the hydrostatic-stress field. Case~4 is not shown because it terminates before a comparable state.

The shear benchmark separates the practical roles of the five strategies. Case~2 provides the most efficient tightly coupled solution and reproduces the Case~1 response. Case~3 is the fastest option for an initial chemo-mechanical screening analysis, but its single-pass coupling gives the largest errors among the completed solutions, particularly in $\varepsilon_{\delta_f}$. Case~5 provides a more accurate staggered alternative. It reduces the chemo-mechanical $\varepsilon_{\delta_f}$ from $8.990\%$ to $0.912\%$ and maintains low $B_{95}$ and $B_{\max}$, but at the cost of a higher $t_{\mathrm{CPU}}$ than Cases~2 and~3. Case~1 remains the reference when tightly coupled accuracy is required, whereas Case~4 is not a reliable default because it does not complete the chemo-mechanical response.

\subsection{Multiple Pre-Existing Cracks}
\label{sec:multicracks}

\subsubsection{Problem Description}
\label{sec:multicracks:setup}

\begin{figure}[htbp]
  \centering
  \includegraphics[width=0.85\textwidth]{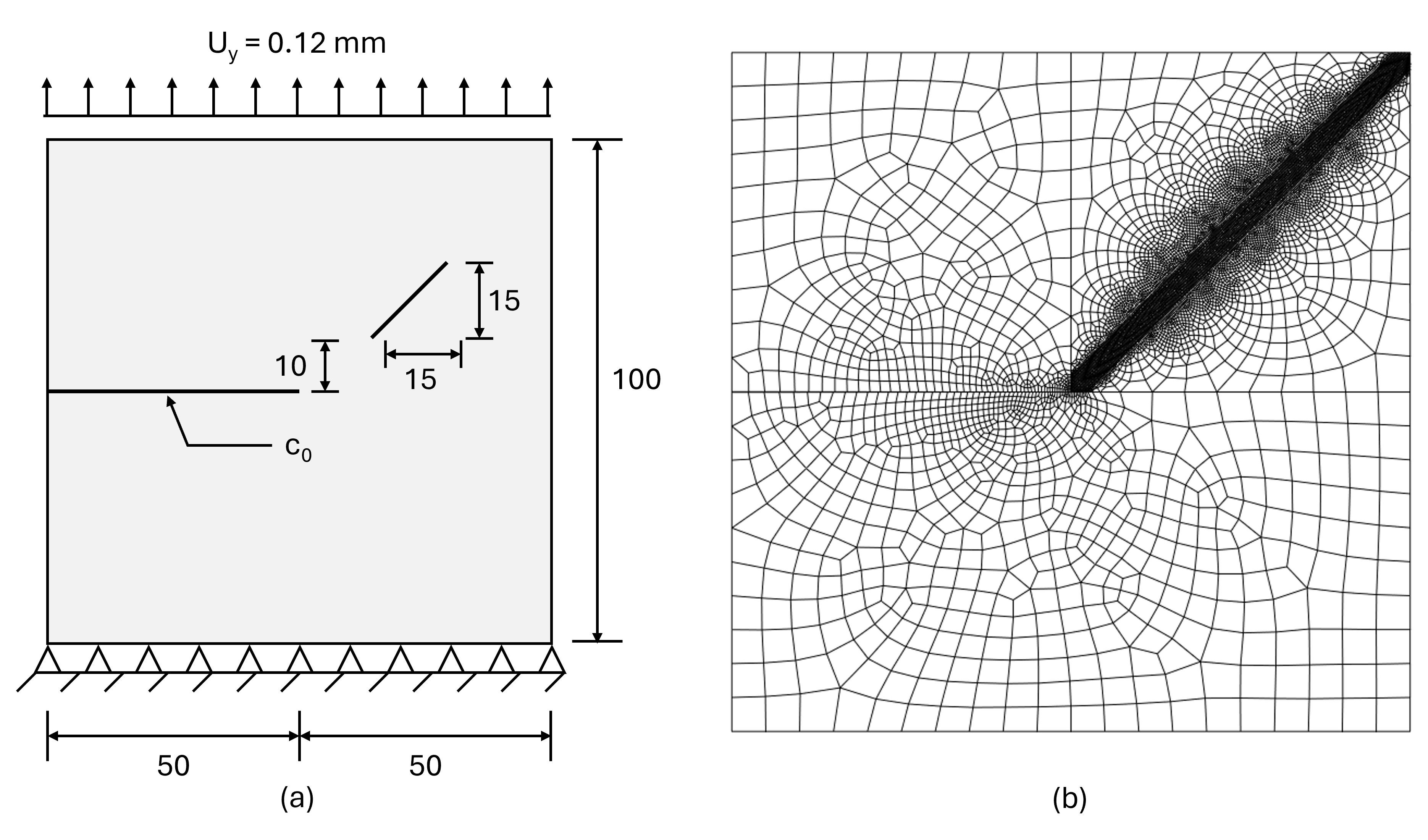}
  \caption{Multiple-crack benchmark: geometry, boundary conditions, and locally refined mesh. The secondary crack is inclined at $45^{\circ}$ and extends from $(65,60)$ to $(80,75)$~mm. Dimensions are in millimeters.}
  \label{fig:multicracks_geom}
\end{figure}

The third benchmark adds an inclined crack ahead of the main crack tip in the tensile configuration (\Cref{fig:multicracks_geom}). The two cracks interact through the mechanical and chemical fields, and the ligament between them allows crack initiation, temporary arrest, reloading, and coalescence~\citep{Nguyen2016,Bourdin2000}. The benchmark is therefore more demanding than the single-crack test, because the solver must follow two fracture events and their interaction. The geometry, boundary conditions, loading, and material properties are those of the tensile benchmark, and the time step is $\Delta t=500$~s. In the chemo-mechanical simulations, $c_0$ is applied only on the faces of the main crack, so CO$_2$ reaches the inclined crack only by diffusion and stress-assisted flux. The mesh ($l_0/h=4$) is refined around both crack tips and the coalescence region and contains 10,946 elements.

\subsubsection{Mechanical Response}
\label{sec:multicracks:mech}

\begin{figure}[htbp]
  \centering
  \includegraphics[width=\textwidth]{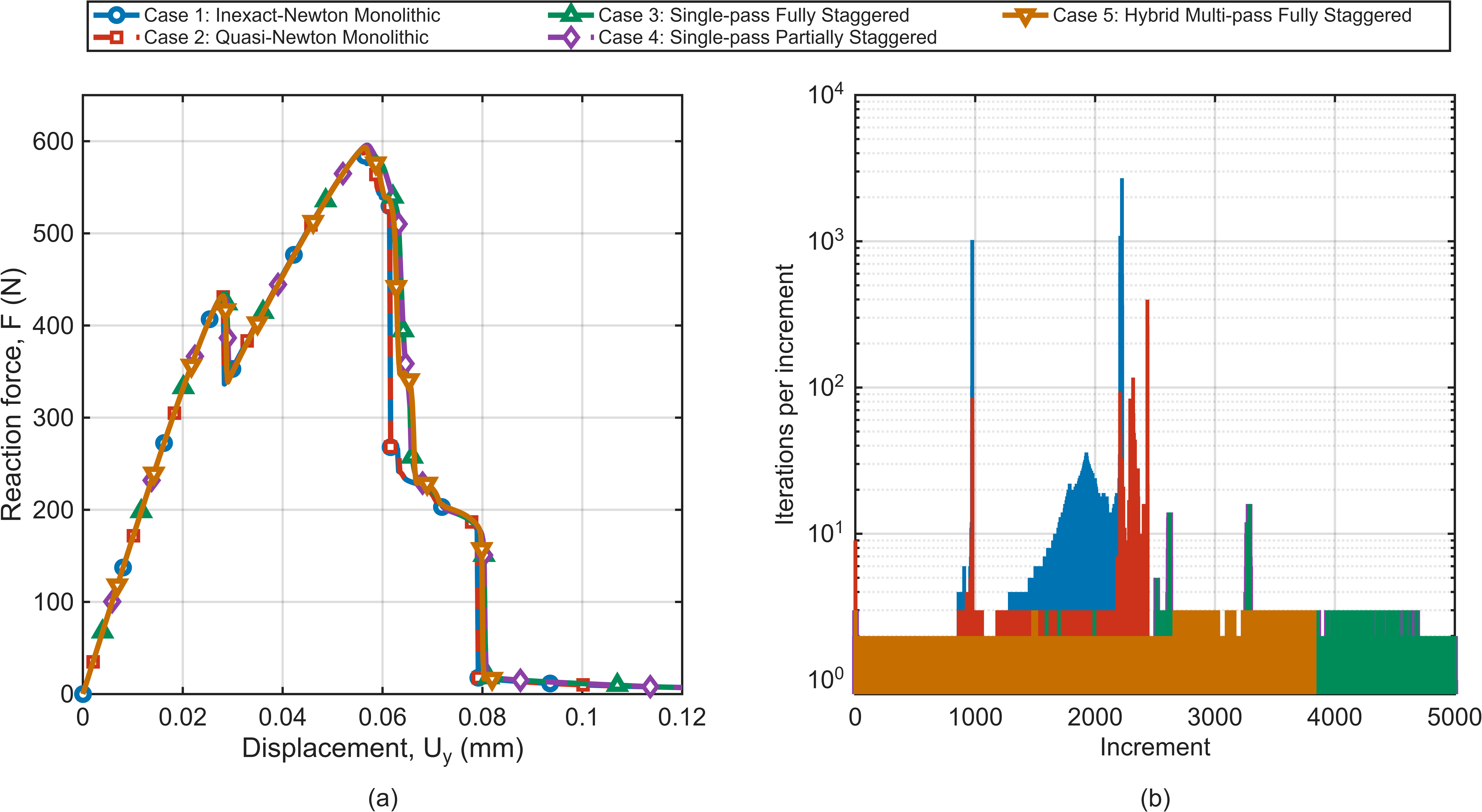}
  \caption{Mechanical multiple-crack benchmark: (a) force--displacement responses and (b) equilibrium iterations per increment for Cases~1--5. The two peaks correspond to crack initiation and coalescence.}
  \label{fig:multicracks_mech_response}
\end{figure}

\Cref{fig:multicracks_mech_response}(a) compares the force--displacement responses. All five cases reproduce the two-peak response. The first peak is followed by a partial load drop as the crack from the main tip advances and temporarily arrests. Reloading then leads to the global peak, associated with renewed propagation and failure of the ligament. Accuracy follows the pattern of the previous benchmarks, with smaller errors. Cases~3 and~4 again coincide, with peak-load and failure-displacement errors of about $0.5\%$ and $2.0\%$, and Case~5 reduces them to about $0.1\%$ and $1.4\%$ (Table~\ref{tab:multicracks_accuracy}). Case~2 remains closest to Case~1.

\Cref{fig:multicracks_mech_response}(b) shows the corresponding equilibrium iterations. The monolithic cases develop iteration bursts at both fracture events, with $B_{\max}$ above 100, whereas Case~5 has the smoothest history. Case~5 is also the fastest, about eight times faster than Case~1 in CPU time, although it needs more iterations than Cases~3 and~4 (Table~\ref{tab:multicracks_performance_robustness}). 

\begin{figure}[htbp]
  \centering
  \includegraphics[width=0.9\textwidth]{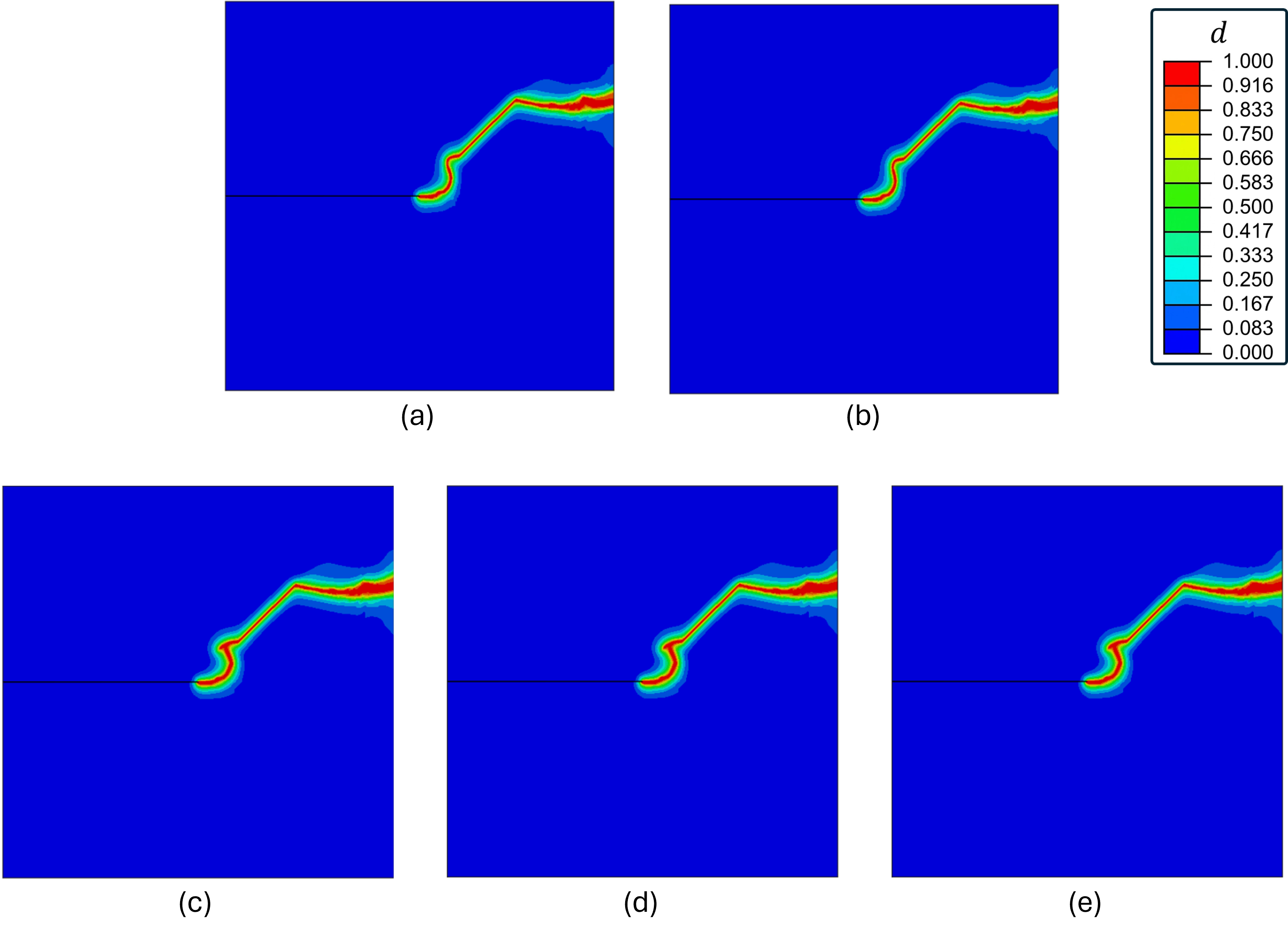}
  \caption{Phase-field damage $d$ at the final states of the mechanical multiple-crack benchmark: (a) Case~1, (b) Case~2, (c) Case~3, (d) Case~4, and (e) Case~5.}
  \label{fig:multicracks_mech_crack}
\end{figure}

Despite these differences in iteration behavior and cost, all five cases predict the same crack path (\Cref{fig:multicracks_mech_crack}). The crack initiates at the tip of the main crack, kinks upward through a short sigmoidal segment, and propagates toward the inclined crack, with which it coalesces. The damage-band width is also similar across the cases.

\subsubsection{Chemo-mechanical Response}
\label{sec:multicracks:chem}

\begin{figure}[htbp]
  \centering
  \includegraphics[width=\textwidth]{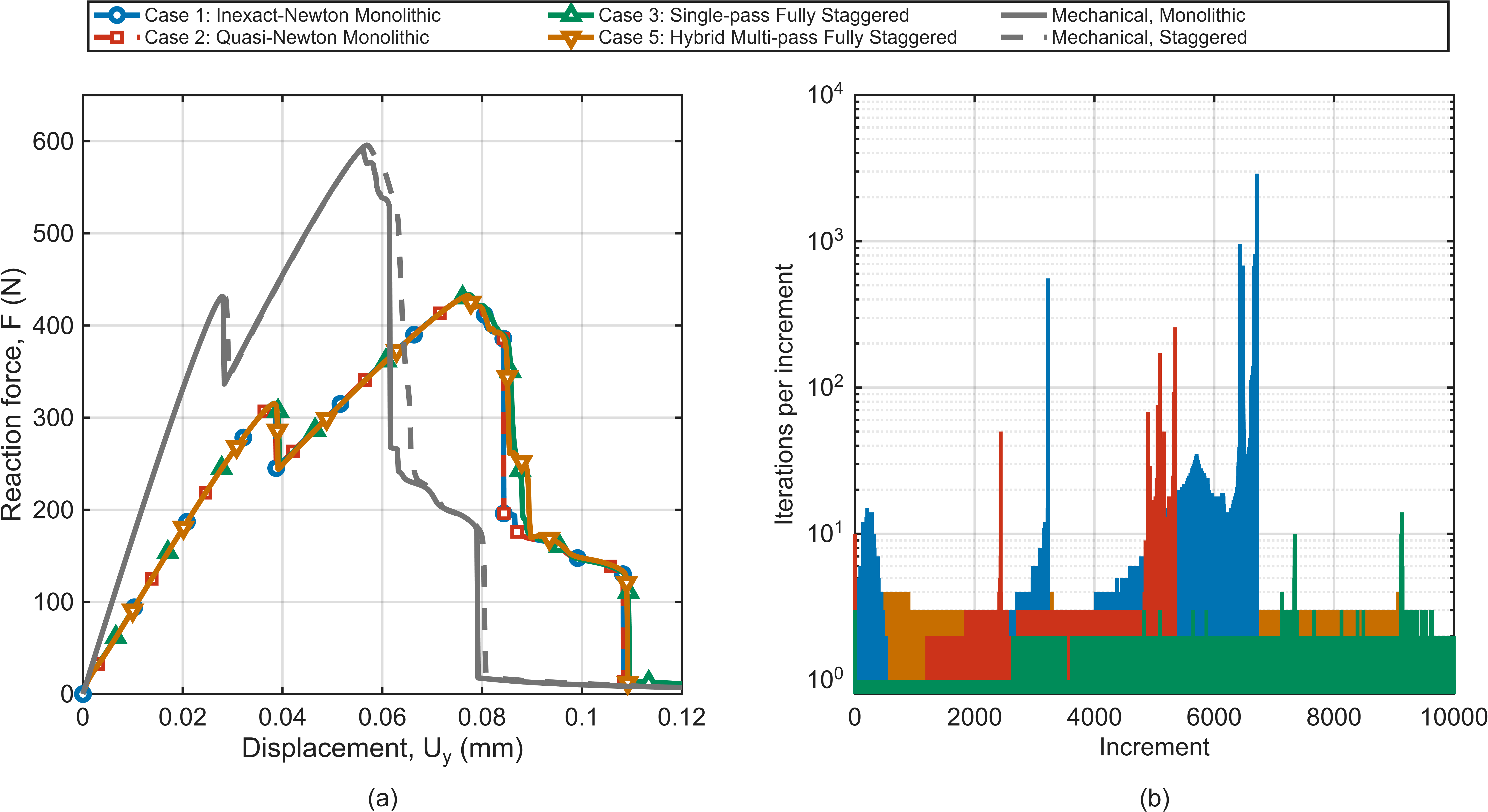}
  \caption{Chemo-mechanical multiple-crack benchmark: (a) force--displacement responses and (b) equilibrium iterations per increment for Cases~1--3 and~5. Case~4 terminates before the comparable response range. Gray curves are the mechanical Case~1 and Case~3 responses.}
  \label{fig:multicracks_chem_response}
\end{figure}

\Cref{fig:multicracks_chem_response}(a) compares the force--displacement responses of Cases~1--3 and~5, which reach complete fracture, with the mechanical curves. The two-peak response is retained. Dissolution lowers both peaks by about $27\%$ relative to the mechanical problem and delays the global peak from about 0.056 to 0.077~mm. Both effects arise because dissolution progressively reduces the stiffness and load-carrying capacity before coalescence. Case~2 coincides with Case~1, and the errors of Case~3 are small, at most about $1.1\%$ in the failure displacement. Case~5 reduces them further, most strongly for the peak load, from about $0.19\%$ to $0.01\%$ (Table~\ref{tab:multicracks_accuracy}).

The local concentration error $\varepsilon_{c,\mathrm{tip}}$ follows the same ranking. It is evaluated in the elements ahead of the pre-notch tip of the main crack, before and at the global peak. Case~2 stays within $0.015\%$ of Case~1 at both states. The errors of Case~3 are about $0.84\%$ and $0.89\%$, and Case~5 reduces them to about $0.17\%$ and $0.18\%$, roughly one fifth of the Case~3 values.

Case~4 is excluded from these comparisons because it does not reach the first peak. Its fixed-step analysis fails. An adaptive rerun reaches only $U_y\approx6\times10^{-4}$~mm within the 72-hour limit, despite about 73,000 iterations over more than 46,000 increments. Its burst factors are low, but a low iteration demand per increment does not establish robustness when the analysis cannot reach the response of interest.

\begin{table}[htbp]
  \centering
  \caption{Accuracy metrics for the mechanical and chemo-mechanical multiple-crack benchmark. A dash denotes an unavailable comparison because chemo-mechanical Case~4 terminates before the peak response.}
  \label{tab:multicracks_accuracy}
  \begin{tabular}{@{}lcrrrrrr@{}}
    \toprule
    Response & Case & $F_{\max}$ (N) & $\varepsilon_{F_{\max}}$ (\%) & $\delta_{\mathrm{peak}}$ (mm) & $\varepsilon_{\delta_{\mathrm{peak}}}$ (\%) & $\delta_f$ (mm) & $\varepsilon_{\delta_f}$ (\%) \\
    \midrule
    Mechanical & 1 & 592.57 & 0.000 & 0.05616 & 0.000 & 0.07919 & 0.000 \\
               & 2 & 592.35 & 0.038 & 0.05616 & 0.000 & 0.07919 & 0.000 \\
               & 3 & 595.67 & 0.523 & 0.05684 & 1.218 & 0.08078 & 2.015 \\
               & 4 & 595.67 & 0.523 & 0.05684 & 1.218 & 0.08078 & 2.015 \\
               & 5 & 593.42 & 0.143 & 0.05639 & 0.406 & 0.08033 & 1.440 \\
    \midrule
    Chemo-mechanical & 1 & 432.47 & 0.000 & 0.07704 & 0.000 & 0.10836 & 0.000 \\
                     & 2 & 432.47 & 0.000 & 0.07704 & 0.000 & 0.10836 & 0.000 \\
                     & 3 & 433.29 & 0.189 & 0.07740 & 0.467 & 0.10956 & 1.107 \\
                     & 4 & -- & -- & -- & -- & -- & -- \\
                     & 5 & 432.43 & 0.009 & 0.07716 & 0.156 & 0.10920 & 0.775 \\
    \bottomrule
  \end{tabular}
\end{table}

\Cref{fig:multicracks_chem_response}(b) shows the corresponding equilibrium iterations. Case~1 develops the largest bursts and is by far the most expensive (Table~\ref{tab:multicracks_performance_robustness}). Case~5 has the lowest burst factors and is the fastest completed strategy, about ten times faster than Case~1 in CPU time. This holds even though it needs more iterations than Cases~2 and~3 and evaluates three elements per cell. Case~2 needs less CPU time than Case~3 but more wall-clock time.

\begin{table}[htbp]
  \centering
  \caption{Computational performance and convergence robustness for the mechanical and chemo-mechanical multiple-crack benchmark. CPU and wall-clock times are in hours. A dash denotes an unavailable comparison.}
  \label{tab:multicracks_performance_robustness}
  \small
  \setlength{\tabcolsep}{3.5pt}
  \begin{tabular}{@{}lclrrrrrr@{}}
    \toprule
    Response & Case & Status & $t_{\mathrm{CPU}}$ & $t_{\mathrm{wall}}$ & $S_{\mathrm{CPU}}$ & $N_{\mathrm{iter}}$ & $B_{95}$ & $B_{\max}$ \\
    \midrule
    Mechanical & 1 & Completed & 28.17 & 15.50 & 1.00 & 67,309 & 2.65 & 148.88 \\
               & 2 & Completed & 3.92 & 2.24 & 7.19 & 8,872 & 1.67 & 166.85 \\
               & 3 & Completed & 4.38 & 2.39 & 6.43 & 4,854 & 1.56 & 10.90 \\
               & 4 & Completed & 4.92 & 2.62 & 5.73 & 4,854 & 1.56 & 10.90 \\
               & 5 & Completed & 3.50 & 1.49 & 8.05 & 8,030 & 1.40 & 1.40 \\
    \midrule
    Chemo-mechanical & 1 & Completed & 103.14 & 66.11 & 1.00 & 96,940 & 2.79 & 270.23 \\
                     & 2 & Completed & 14.41 & 10.89 & 7.16 & 21,658 & 2.08 & 107.57 \\
                     & 3 & Completed & 17.33 & 9.17 & 5.95 & 11,571 & 1.58 & 7.89 \\
                     & 4 & Failed & -- & -- & -- & 73,218 & 1.90 & 6.34 \\
                     & 5 & Completed & 10.63 & 4.04 & 9.70 & 28,217 & 1.29 & 1.61 \\
    \bottomrule
  \end{tabular}
\end{table}

\begin{figure}[htbp]
  \centering
  \includegraphics[width=0.8\textwidth]{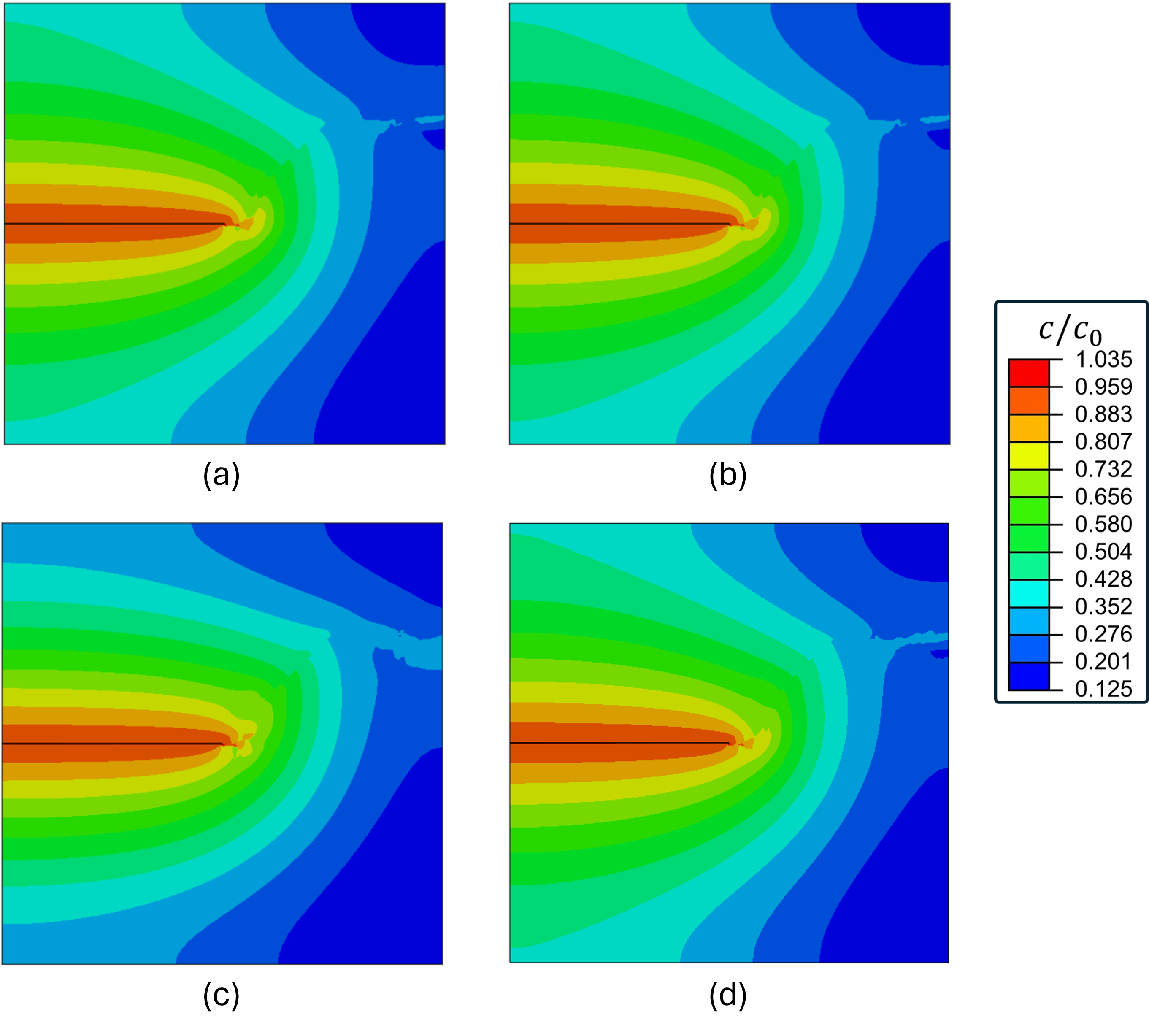}
  \caption{Normalized CO$_2$ concentration $c/c_0$ at the final stored states of the chemo-mechanical multiple-crack benchmark for: (a) Case~1, (b) Case~2, (c) Case~3, and (d) Case~5.}
  \label{fig:multicracks_cratio}
\end{figure}

The concentration contours at complete fracture (\Cref{fig:multicracks_cratio}) show similar large-scale transport patterns for Cases~1--3 and~5. The concentration is highest near the faces of the main crack, and the front extends toward the coalescence region. The completed cases also recover the same coalescence path as the mechanical analyses, so dissolution lowers and delays the peaks without redirecting the crack.

In this benchmark, Case~2 agrees most closely with Case~1 in both the response and the local concentration. Case~5 gives the best overall balance of accuracy, cost, completion, and iteration regularity, with chemo-mechanical response errors below $0.8\%$ and concentration errors below $0.2\%$. Case~3 remains a viable single-pass option but is slower and less accurate than Case~5. Case~1 remains the reference for coupled verification, and Case~4 is not a reliable default because its chemo-mechanical analysis does not reach the first peak.
\subsection{Cross-benchmark Assessment}
\label{sec:discussion}

The central question of this investigation is which strategy an analyst should try first for a new coupled dissolution--fracture problem. The three benchmarks show that no single metric identifies the best strategy. A strategy that requires few equilibrium iterations is not necessarily the least expensive, and a strategy that is inexpensive when it converges is not necessarily the most reliable. The strategies must therefore be assessed on solution accuracy, computational cost, and robustness together.

\subsubsection{Solution Accuracy}

Case~2 is practically coincident with Case~1 in every completed analysis, with chemo-mechanical peak-load and local concentration errors of at most $0.03\%$. Case~3 matches the chemo-mechanical peak load to within about $1\%$ but has much larger post-peak errors, up to about $16\%$ in failure displacement. These errors decrease with time-step refinement. Peak agreement alone is therefore insufficient to verify a single-pass solution. Case~5 reduces the post-peak error of Case~3 in every chemo-mechanical comparison, by up to an order of magnitude. It also keeps the chemo-mechanical peak-load error below $0.2\%$ and generally improves the local concentration agreement. Without chemistry, Case~4 coincides with Case~3. Most of its fixed-step chemo-mechanical analyses fail, so they inform robustness rather than accuracy. Adaptive stepping, however, recovers the tensile response.

\subsubsection{Computational Performance}

The monolithic strategies show iteration bursts near fracture, in which single increments need up to about 650 (Case~1) and 450 (Case~2) times their mean iteration count. In the baseline chemo-mechanical analyses, Cases~3 and~5 keep $B_{\max}$ below 10 and 3, respectively. Under increased transport intensity, the bursts of Case~3 grow several-fold, to about 54, whereas Case~5 stays below 3. Regular iteration demand, however, implies neither low cost nor completion.

Completion is the complementary measure. The comparison set comprises the seven fixed-step chemo-mechanical analyses: three tensile time steps, two increased-transport settings, the shear benchmark, and the multiple-crack benchmark. Mechanical and adaptive analyses are excluded. The completion ratios for Cases~1--5 are:
\begin{equation}
R_{\mathrm{comp}}=
\left\{
\tfrac{6}{7},\,
\tfrac{6}{7},\,
\tfrac{7}{7},\,
\tfrac{1}{7},\,
\tfrac{7}{7}
\right\}
=
\left\{
0.86,\,
0.86,\,
1.00,\,
0.14,\,
1.00
\right\}
\label{eq:cross_benchmark_completion}
\end{equation}
The Case~1 shear analysis reaches failure but is counted as incomplete because it does not finish the prescribed loading history within the 72-hour limit. It is therefore a time-limit stop rather than a convergence failure. These ratios apply only to this set and are not general convergence probabilities. The transport-intensity study further adds that completion is scheme-dependent and non-monotonic in $\mathrm{Fo}_{95}$, since Cases~2 and~4 fail at opposite settings.

\subsubsection{Practical Solver Guidance}

Two workflows are recommended. When several strategies can be tested, Case~3 is used for inexpensive screening of the model setup, crack path, and concentration field. Its post-peak response is then checked by time-step refinement or against Case~2, which is the most efficient tightly coupled strategy when it converges. Case~1 is reserved for selected verification. When a single strategy must be chosen, Case~5 is the most balanced choice. It completes every fixed-step analysis, is more accurate than Case~3, and has regular iteration histories at moderate cost. Case~4 completes only one of the seven fixed-step chemo-mechanical analyses and is not recommended as a default for the present implementation. This hierarchy converts the benchmark comparison into practical guidance for implementing strongly coupled multiphysics fracture models in a commercial finite-element environment.

\section{Conclusions}
\label{sec:conclusions}
This study assessed five solution strategies for a three-field model of dissolution-driven phase-field fracture, implemented in user elements in Abaqus. The strategies were the block-diagonal inexact-Newton monolithic (Case~1), quasi-Newton (BFGS) monolithic (Case~2), single-pass fully staggered (Case~3), single-pass partially staggered $[\bm{u},c]$ (Case~4), and hybrid multi-pass fully staggered (Case~5) schemes. All strategies share the same formulation. They were compared for accuracy, computational cost, and robustness on mechanical and chemo-mechanical tension, shear, and multiple-crack benchmarks, including increased-transport conditions. The main conclusions are:

\begin{enumerate}[label=(\roman*)]
  \item Case~2 agrees most closely with the Case~1 reference, with chemo-mechanical peak-load and concentration errors of at most $0.03\%$ and CPU-time speed-ups of 1.7--7.2. It is the preferred tightly coupled strategy, but its failure at the highest transport intensity shows that convergence is not guaranteed.

  \item Case~3 is an inexpensive screening strategy that completes every fixed-step chemo-mechanical analysis. Its post-peak response is, however, sensitive to the time step, with failure-displacement errors of up to about $16\%$. This response should therefore be verified by time-step refinement or against a tightly coupled solution.

  \item Case~5 also completes every fixed-step analysis and reduces the post-peak error of Case~3 by up to an order of magnitude. Its chemo-mechanical peak-load and concentration errors remain below $0.2\%$. Its multi-pass updates reduce, but do not eliminate, the splitting error, and the cost of Case~5 relative to Case~3 is problem-dependent.

  \item Robustness cannot be inferred from runtime or iteration counts alone. Over the fixed-step chemo-mechanical analyses, the completion ratios are 0.86, 0.86, 1.00, 0.14, and 1.00 for Cases~1--5, respectively. Case~4 is therefore not recommended as a default for this study, although adaptive stepping recovers its tensile response. Under increased transport, Cases~2 and~4 fail at opposite settings. This shows that $\mathrm{Fo}_{95}$ measures transport intensity but does not define a convergence threshold.
\end{enumerate}

Two practical workflows follow from these conclusions. When several strategies can be tested, Case~3 provides screening and Case~2 provides a tightly coupled solution. When only one strategy can be used, Case~5 is the most balanced choice. These conclusions apply to the present two-dimensional implementation, the constitutive model without a tension--compression split, and the discretizations and transport ranges investigated. Future work should examine fully converged multi-pass variants, alternative field partitions, three-dimensional problems, broader transport regimes, and coupled hydraulic processes.

\section*{CRediT authorship contribution statement}
\textbf{Maithah M. Alaleeli:} Methodology, Software, Validation, Formal analysis, Investigation, Data curation, Visualization, Writing -- original draft, Writing -- review \& editing. \textbf{Mohamed Elkhodbia:} Investigation, Data curation, Visualization, Writing -- original draft, Writing -- review \& editing. \textbf{Mostafa E. Mobasher:} Conceptualization, Methodology, Supervision, Funding acquisition, Resources, Writing -- review \& editing.

\section*{Declaration of competing interest}
The authors declare that they have no conflict of interest.

\section*{Data availability}
Data will be made available on request.

\section*{Acknowledgements}
This work was partially supported by the Sand Hazards and Opportunities for Resilience, Energy, and Sustainability (SHORES) Center, funded by Tamkeen under the NYU Abu Dhabi Research Institute Award CG013, and by the NYUAD Kawader Program.

\appendix
\section{Derivation of the Strong Form}
\label{app:strong_form}

This appendix derives the strong form in Eq.~\eqref{eqn:strong_forms} from the energy functional and from mass conservation. For a given chemical damage $d_{\mathrm{chem}}$, the first variation of the functional in Eq.~\eqref{eq:mechanical_fracture_functional} with respect to $\bm{u}$ and $d$ is:
\begin{align}
  \delta\Pi={}&\int_\Omega\left[
  g\,\frac{\partial\Psi_0}{\partial\bm{\varepsilon}}:\delta\bm{\varepsilon}
  +\frac{\partial g}{\partial d}\Psi_0\,\delta d
  +G_c\left(\frac{d}{l_0}\delta d+l_0\nabla d\cdot\nabla\delta d\right)
  \right]\mathrm{d}\Omega
  \nonumber\\
  &-\int_\Omega\bm{b}\cdot\delta\bm{u}\,\mathrm{d}\Omega
  -\int_{\partial\Omega_h}\bm{h}\cdot\delta\bm{u}\,\mathrm{d}\Gamma .
  \label{eq:app_first_variation}
\end{align}
Using $\bm{\sigma}=g\,\partial\Psi_0/\partial\bm{\varepsilon}$ and $\delta\bm{\varepsilon}=\operatorname{sym}\nabla\delta\bm{u}$, the divergence theorem gives
\begin{align}
  \int_\Omega\bm{\sigma}:\nabla\delta\bm{u}\,\mathrm{d}\Omega
  &=\int_{\partial\Omega}(\bm{\sigma}\cdot\bm{n})\cdot\delta\bm{u}\,\mathrm{d}\Gamma
  -\int_\Omega(\nabla\cdot\bm{\sigma})\cdot\delta\bm{u}\,\mathrm{d}\Omega,
  \nonumber\\
  \int_\Omega\nabla d\cdot\nabla\delta d\,\mathrm{d}\Omega
  &=\int_{\partial\Omega}(\nabla d\cdot\bm{n})\,\delta d\,\mathrm{d}\Gamma
  -\int_\Omega\nabla^2d\,\delta d\,\mathrm{d}\Omega .
\end{align}
Substituting these identities into Eq.~\eqref{eq:app_first_variation} with $\delta\bm{u}=\bm{0}$ on $\partial\Omega_u$ gives:
\begin{align}
  \delta\Pi={}&-\int_\Omega(\nabla\cdot\bm{\sigma}+\bm{b})\cdot\delta\bm{u}\,\mathrm{d}\Omega
  +\int_{\partial\Omega_h}(\bm{\sigma}\cdot\bm{n}-\bm{h})\cdot\delta\bm{u}\,\mathrm{d}\Gamma
  \nonumber\\
  &+\int_\Omega\left[G_c\left(\frac{d}{l_0}-l_0\nabla^2d\right)+\frac{\partial g}{\partial d}\Psi_0\right]\delta d\,\mathrm{d}\Omega
  +\int_{\partial\Omega}G_cl_0(\nabla d\cdot\bm{n})\,\delta d\,\mathrm{d}\Gamma .
\end{align}
Stationarity, $\delta\Pi=0$ for arbitrary $\delta\bm{u}$ and $\delta d$, gives Eq.~\eqref{eqn:strong_forms_a} with the traction condition in Eq.~\eqref{eqn:bc_b}, together with
\begin{equation}
  G_c\left(\frac{d}{l_0}-l_0\nabla^2d\right)+\frac{\partial g}{\partial d}\Psi_0=0
  \quad\text{in }\Omega,
  \qquad
  \nabla d\cdot\bm{n}=0\quad\text{on }\partial\Omega .
\end{equation}
With $\partial g/\partial d=-2(1-d_{\mathrm{chem}})(1-d)$ from Eq.~\eqref{eq:degradation_combined}, and with $\Psi_0$ replaced by the history field $H$ of Eq.~\eqref{eq:history} to prevent crack healing~\citep{MieheHofacker2010}, this becomes Eq.~\eqref{eqn:strong_forms_b}.

The transport equation is obtained directly from mass conservation. For an arbitrary fixed control volume $V\subseteq\Omega$, conservation of dissolved CO$_2$ without a reaction sink requires
\begin{equation}
  \frac{\mathrm{d}}{\mathrm{d}t}\int_V c\,\mathrm{d}V=-\int_{\partial V}\bm{j}_c\cdot\bm{n}\,\mathrm{d}A
  =-\int_V\nabla\cdot\bm{j}_c\,\mathrm{d}V .
\end{equation}
Because $V$ is arbitrary, Eq.~\eqref{eqn:strong_forms_c} follows.

\section{Implementation in Abaqus}
\label{app:abaqus}

This appendix gives the Abaqus-specific details of the implementation, namely the element layout, the solution controls, and the way Abaqus reports the residuals.

\paragraph{Element layout} In Cases~1--4, each node of a Q8 user element carries two displacement degrees of freedom (DOFs~1--2), one phase-field DOF (DOF~3), and one concentration DOF (DOF~4), which gives 32 DOFs per element. The solution strategy is selected at run time by a flag passed through the element property array. Case~5 represents each mesh cell by three coincident user elements that share its nodes: a 16-DOF mechanical element, an 8-DOF phase-field element, and an 8-DOF transport element. Because Abaqus/CAE cannot display user-element state variables directly, each cell is also overlaid with a stiffness-free \texttt{CPE8} element that shares its nodes and contributes no residual or tangent. This element is used only for visualization.

\paragraph{Solution controls} Cases~1, 3, 4, and~5 use \texttt{*Solution Technique, type=NEWTON} within a \texttt{*Static} step, with nonsymmetric matrix storage (\texttt{*Step, unsymm=YES}). Case~2 uses \texttt{*Solution Technique, type=QUASI-NEWTON, reform kernel=8}, which reassembles the tangent after every eight BFGS updates, with the default symmetric matrix storage. The quasi-Newton technique in Abaqus is available only for symmetric systems. For Case~2, Abaqus therefore replaces the element matrix by its symmetric part, $\tfrac{1}{2}(\bm{A}+\bm{A}^T)$, where $\bm{A}$ is the matrix returned by the UEL. Line search and increment cutback are activated through \texttt{*Controls}. The default Abaqus tolerances are used for all cases: $5\times10^{-3}$ for the residual and $10^{-2}$ for the solution correction.

\paragraph{Residual reporting} Abaqus/Standard reports residuals for DOFs~1--3 under its force category and for DOF~4 under its moment category. In the present elements, the moment-category residual is therefore the residual of the concentration equation, and the force-category residual combines the displacement and phase-field equations. The normalized residuals in Eq.~\eqref{eq:criterion_normalised_residual} use the corresponding time-averaged force and moment measures computed by Abaqus, which are denoted $\bar f_{u,d}^{(k)}$ and $\bar f_c^{(k)}$ in Section~\ref{sec:evaluation_framework}.

\newpage
\bibliography{references}
\end{document}